\documentclass[]{aastex631}
\usepackage{amsmath}
\usepackage{natbib}
\usepackage{url}
\usepackage{ulem}
\usepackage{footnote}
\usepackage{IEEEtrantools}
\usepackage{cases}
\usepackage[overload]{empheq}
\usepackage{cleveref} 
\usepackage[version=3]{mhchem}
\usepackage{empheq}

\begin{document}
\title{Thermal Evolution and magnetic history of rocky planets}

\correspondingauthor{Jisheng Zhang}
\email{zhangjis@uchicago.edu, larogers@uchicago.edu}

\author{Jisheng Zhang}
\affil{Department of Astronomy and Astrophysics, the University of Chicago, 5640 S. Ellis Ave, Chicago, IL 60637, USA}

\author{Leslie A. Rogers}
\affiliation{Department of Astronomy and Astrophysics, the University of Chicago, 5640 S. Ellis Ave, Chicago, IL 60637, USA}

\begin{abstract}
We present a thermal evolution model coupled with a Henyey solver to study the circumstances under which a rocky planet could potentially host a dynamo in its liquid iron core and/or magma ocean. We calculate the evolution of planet thermal profiles by solving the energy balance equations for both the mantle and the core. We use a modified mixing length theory to model the convective heat flow in both the magma ocean and solid mantle. In addition, by including the Henyey solver, we self-consistently account for adjustments in the interior structure and heating (cooling) due to planet contraction (expansion). We evaluate whether a dynamo can operate using the critical magnetic Reynolds number. We run simulations to explore how planet mass ($M_{pl}$), core mass fraction (CMF) and equilibrium temperature ($T_{eq}$) affect the evolution and lifetime of possible dynamo sources. We find that the $T_{eq}$ determines the solidification regime of the magma ocean, and only layers with melt fraction greater than a critical value of 0.4 may contribute to the dynamo source region in the magma ocean. We find that the mantle mass, determined by $M_{pl}$ and CMF, controls the thermal isolating effect on the iron core. In addition, we show that the liquid core last longer with increasing planet mass. For a core thermal conductivity of 40$\ \mathrm{Wm^{-1}K^{-1}}$, the lifetime of the dynamo in the iron core is limited by the lifetime of the liquid core for 1$M_{\oplus}$ planets, and by the lack of thermal convection for 3$M_{\oplus}$ planets. 
\end{abstract}


\section{Introduction \label{Intro}}
Planetary magnetic fields are ubiquitous in the solar system and the explanation for their existence has been linked to planetary interiors. Strong planet-scale magnetic fields have been detected for Mercury, Earth, Ganymede, and all of the gas giants in the solar system \citep[]{Ness75,ness1986, ness1989,Finlay2010,kivelson1996,Bur55,smith1980}. These magnetic fields have been attributed to dynamo currents, which are generated by thermal convection, compositional convection, differential rotation, or a combination of several processes within the planetary interior. Consequently, the presence or absence of strong planet-scale magnetic fields constrains the planet's thermal evolution history and interior dynamics. For example, the discovery of Jupiter's magnetic field indicates a convectively-driven dynamo action in the electrically conductive metallic hydrogen layer \citep{sta10}, whereas the lack of a magnetic field for Venus indicates an inefficient core cooling rate due to the lack of plate tectonics \citep{stevenson2010}.

The observational detection of an exoplanet's magnetic field would open an unprecedented window into exoplanet interiors and add a new dimension to exoplanet characterization. To date, the characterization of exoplanets has focused on their shrouding atmospheres and mass-radius relationships. Radius and mass measurements derived from transit photometry and radial velocity data constrain planetary average densities and bulk compositions. Additionally, transmission and emission spectroscopy provide information about the atmospheric thermal budget and chemical composition. However, exoplanet interior structure and dynamics remain ambiguous. The interpretation of planet mass-radius measurements suffers from degeneracies;  a single mass-radius measurement can be consistent with various bulk compositions and interior structures \citep[e.g.,][]{adams2008,rogers2010}. Eventually, observational measurements of exoplanet magnetic fields may extend the current mass-radius relationship to a mass-radius-magnetic field strength diagram and help reduce the degeneracy in the interpretation. Most generally, a strong planetary scale magnetic field requires an electrically conductive convecting fluid region within the planet. In the case of a rocky exoplanet, a magnetic field detection would imply the existence of a convecting liquid iron core or a layer of molten silicate in the mantle \citep{soubiran2018}.

Detection of planetary-scale magnetic fields on planets with a range of atmosphere compositions enables studies of the influence of magnetic fields on atmosphere preservation. The comparison among the atmospheres of Earth, Venus and Mars is often used to argue that a dynamo-generated magnetic field could protect the atmosphere of a rocky planet from losing its water \citep{lundin2007}. However, the mechanism of such protection remains unclear. Moreover, despite of Earth having a global magnetic field, its atmosphere loss rate is comparable to other rocky planets in the solar system \citep{gunell2018}. Ideally, a large sample of planets with a range of atmospheric compositions and magnetic field properties will allow studies of the interactions between magnetic fields and atmosphere loss rate. The results would have implications for the assessment of the habitability of exoplanets, and complement the current focus on biosignatures in exoplanetary spectra. 

Given the importance of dynamo-generated magnetic fields, their detection will be a frontier in exoplanetary science in the coming decades. So far, various groups have looked for observational signatures of magnetic fields from exoplanets \citep[e,g.,][]{winter2006,zarka2007, hallinan2013}. \citet{kao18} determined the magnetic field strength of a brown dwarf by detecting of its radio aurora emission. The recently commissioned Low Frequency Array \citep[LOFAR,][]{haarlem2013} and Long Wavelength Array \citep[LWA,][]{ellingson2009} reach $\sim$mJy sensitivities at frequencies below 100MHz, enabling searches for the magnetic fields of gas giants. Using LOFAR, \citet{turner2020} detected radio signals from $\tau$ Bo\"{o}tis and $\nu$ Andromedae systems, which are attributed to the self-sustaining magnetic fields of Jovian-sized planets. Due to their weaker expected magnetic field strength \citep[$\lesssim3G$,][]{Bonati2020}, detections of radio signals from sub-Neptunes and super-Earths may require observations at frequency below 10 MHz, which are inaccessible to ground-based telescopes due to Earth's ionospheric cutoff. The Sun Radio Interferometer Space Experiment \citep[SunRISE,][]{sunrise} is a spaced-based telescope designed to study the sun at frequencies below 15 MHz. SunRISE will not be sensitive enough to detect exoplanets, but may observe auroral emission from Saturn, and thus prove the concept of using space-based telescopes to detect planetary radio aurora in a bandpass below 15 MHz. To guide future surveys for exoplanetary magnetic fields and to eventually translate detected signals into constraints on the planets' interior structure and dynamics, extensive modeling of the thermal evolution history of exoplanets across a range of masses, compositions, and stellar incident flux is needed. 

Multiple groups have investigated the thermal and magnetic evolution of rocky planets using a box model \citep[e.g.,][]{stevenson1983, papuc2008, dri14}. In their models, the mantle and core are treated as two boxes, and the thermal history is described by the time evolution of the average mantle and core temperatures. The heat flow throughout the planet is evaluated by modeling conduction across the thermal-boundary layers at the core-mantle boundary (CMB) and the planetary surface. Under the assumption of whole mantle convection, the heat flow across the thermal boundary layers are set by the critical Rayleigh number, a dimensionless number describing the onset of convection flow. Boundary layer theory and box models are effectively zero-dimensional energy-balance models. The thermophysical properties of mantle and core materials (e.g., heat capacity, viscosity, and thermal expansion coefficient) are modeled with an effective value that is constant throughout each ``box'' (upper mantle, lower mantle, core, and thermal boundary layers). They are not true 1-D treatments wherein thermophysical and structure quantities are calculated as a function of radius. In addition, with constant pressure and density profiles adopted for the core and mantle, the work done by planet expansion/contraction is not self-consistently included. 

We have developed a new fully 1-D thermal evolution model for rocky planets to explore the possibility of molten silicate or liquid iron serving as dynamo source regions. Our code solves for the radial distribution of temperature, melt fraction, pressure, density, and conductive and convective heat fluxes, and evolves these profiles forward in time. Thermophysical properties that affect the planet's structure and evolution, such as viscosity and thermal expansion coefficient, are evaluated with local pressure and temperature. 
We use a generalized  Schwarzschild criterion based on density gradient to determine where convection occurs in the mantle and a modified mixing length theory to model the convective heat flow. The mixing length theory is traditionally used to estimate the convective heat flow of inviscid fluid \citep[e.g. heat transport in stars,][]{vitense53}. The modified formulation \citep[e.g.,][]{sas86, abe1995, tachinami2011, wagner2019} accounts for the viscous drag force induced by viscous fluid with a small Reynolds number, such as solid phase mantle silicate. We further extend the modified mixing length theory to estimate the convective heat flow in both single- and multi-phase regions (Appendix~\ref{section:mlt_x}). Additionally, we include a Henyey solver (section~\ref{Structure}) to adjust the pressure and density profiles within the model as the planet cools. This feature allows us to include heating (cooling) due to planetary contraction (expansion) self-consistently. 
The final result is to obtain the time evolution of the convective heat flow throughout the planet, which we use to determine the location and duration of possible dynamo source regions. 

In this paper, we present our numerical model for the interior structure and evolution of rocky planets. We apply the model to simulate a sparse grid of 12 planets with 2 planetary masses ($M_{pl}=$1 and 3 $M_{\oplus}$), 2 core mass fractions (CMF$=$0.326 and 0.7), and 3 equilibrium temperatures ($T_{eq}=$255~K, 1350~K and 2500~K), as a first demonstration of the model's applicability. We aim to explore the potential influence of $M_{pl}$, CMF and $T_{eq}$ on the thermal and magnetic history. Our paper is organized as follows. We describe the numerical model in section~\ref{methodology}. We validate the model through comparison to the Earth and other evolution models in section~\ref{validation}. We present the thermal evolution results for the grid of 12 planet case studies in section~\ref{result}. Finally, we provide a discussion and conclusion in section~\ref{section:conclusion}.

\section{Method \label{methodology}}

\begin{figure}
\begin{center}
\includegraphics[scale=0.55]{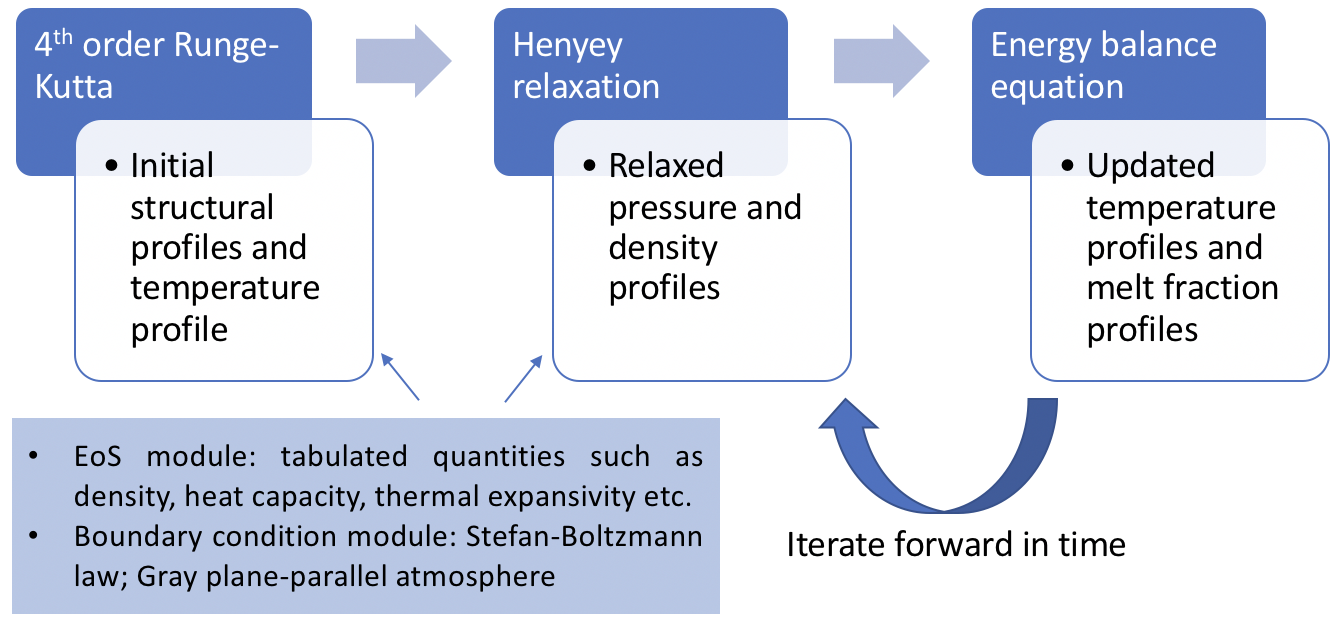}
\end{center}
\caption{Algorithm of the model. Initial profiles are prepared step 1. Subsequent updates in temperature, melt fraction and structural profilesare carried out by iterating step 2 and 3. \label{fig:algorithm}}
\end{figure}

We have developed a 1-D spherically symmetric discretized model to simulate the thermal evolution of rocky planets as they cool from an initial post-formation hot state with a liquid iron core and molten silicate mantle. We apply the model to explore the impact of planetary mass ($M_{pl}$), core mass fraction (CMF), and equilibrium temperature ($T_{eq}$) on the thermal history of rocky planets and the timing and duration of possible dynamos in both the liquid core and magma ocean. 

In this paper, we consider planets consisting of a silicate mantle surrounding a pure iron core. Future work will explore the effect of different chemical compositions in the core and mantle as well as the level of differentiation between the iron and silicates in the planets.

The overall algorithm is briefly summarized in figure~\ref{fig:algorithm} and here:
\begin{enumerate}
\item The planet is discretized radially into multiple cells with equal mass in the core and mantle depending on $M_{pl}$ and CMF. An initial structure profile, including radius ($r$), pressure ($P$) and density ($\rho$) as a function of mass interior ($m$), is prepared by integrating the structural equations (section~\ref{Structure}) with appropriate equations of state (section~\ref{EoS}) using a fourth order Runge-Kutta method. A hot adiabatic initial temperature profile ($T$) is assumed in both the core and the mantle. 
\item The solution to the interior structure ($r$, $P$ and $\rho$ as a function of $m$) --- be it the initial structure from step 1 or a subsequent structure updated after a thermal timestep from step 3 --- is updated by a Henyey solver (section~\ref{Structure}) so that it satisfies the boundary conditions at both the center and the surface. The relative tolerance in the Henyey relaxation for both radius and pressure is set to be $10^{-4}$.
\item A new thermal profile at the new thermal timestep, $t+\Delta t$, is obtained by solving the energy balance equation in both the core and the mantle (sections~\ref{core_cooling} and~\ref{section:mantle heat transport}). The surface heat flow is set by gray-body radiation. The cooling rate of the core is determined by heat conduction through the core-mantle boundary as well as the rate of inner solid core nucleation. Radial profiles of thermophysical properties, such as viscosity, are updated based on the new thermal profile.  
\end{enumerate}
Time evolution is calculated by iterating steps 2 and 3 forward in time.

Compared to a parameterized model \citep[e.g.,][]{papuc2008,dri14} that estimates the convective flux under the assumption of whole mantle convection using a global Rayleigh number, our approach has two key advantages. First, our model takes variations of thermophysical properties with depth in the planet into account. Second, instead of assuming whole mantle convection, we directly evaluate which layers are unstable to convection with a generalized Schwarzschild criterion based on the entropy gradient. 

In the rest of the section, we explain each component in more detail. We discuss the interior structure calculation in section~\ref{Structure}, choice of bulk composition and the corresponding equation of state (EoS) in section~\ref{EoS}, energy transport in the core and the mantle in sections~\ref{core_cooling} and~\ref{section:mantle heat transport}, mantle viscosity in section~\ref{section:viscosity}, internal heat production in section~\ref{section:Qrad}, boundary conditions in section~\ref{section:boundary condition}, and the dynamo criterion in section~\ref{section:dynamo_criterion}. 

\subsection{Interior Structure \label{Structure}}
We consider a spherically symmetric planet that evolves quasi-statically in hydrostatic equilibrium. In our interior structure model, the iron core and silicate mantle are each divided into hundreds to thousands of cells with equal mass of $dm$ (a Langrangian formulation), depending on $M_{pl}$ and CMF. For example, the iron core and silicate mantle of an Earth-like planet ($M_{pl}=1M_{\oplus}$ and CMF=0.326) are each divided in to 1000 cells. At each instant in time, the interior structure is described by the following coupled differential equations for radius $r(m)$ and pressure $P(m)$:

\begin{equation}
\frac{\partial r}{\partial m}=\frac{1}{4\pi r^2 \rho},
\label{eq:dm}
\end{equation}

\begin{equation}
\frac{\partial P}{\partial m}=-\frac{Gm}{4\pi r^4},
\label{eq:dP}
\end{equation}

\noindent where $r$ is the radial distance from the center of the planet, $G$ is the gravitational constant, and $\rho$ is the local density, calculated based on the EoS (described in section \ref{EoS}) for a given material at the local pressure and temperature level.

The coupled differential equations~\ref{eq:dm} and \ref{eq:dP} have boundary conditions at both the center and the surface. The radius at the center ($m=0$) reduces to 0, while the pressure level at the surface ($m=M_{pl}$) is set by the atmospheric boundary condition (herein chosen to be 1 bar). A detailed description of the surface boundary conditions can be found in section~\ref{section:boundary condition}.

To start the evolution, the initial structure profile is obtained by numerically solving the above differential equations using a fourth order Runge-Kutta technique integrating from the planet center to the surface (a shooting technique). We set the initial temperature at the planet center to be 500~K above the melting temperature. The initial temperature distribution in the core and the mantle is assumed to be adiabatic. Our choice of the initial temperature distribution is hot enough such that both the core and the mantle are fully molten. We reiterate the shooting technique to search for an appropriate central pressure corresponding to the chosen $M_{pl}$ and CMF until the surface pressure satisfies the surface boundary condition with a relative error less than $10^{-4}$. The resulting solution is fed into the Henyey solver (discussed below) to further reduce the relative error in the surface pressure.  

When applying shooting techniques (like the well-known Runga-Kutta method) to  model the planet interior structure, the boundary conditions are only fully satisfied at either the planet surface (when integrating from the outside in) or at the center (when integrating from the inside out), not both. Iteration is needed to reach a solution that satisfies the second boundary condition within a preset tolerance. For example, when setting the initial starting structure, we iterate to find the central pressure corresponding to the desired total planet mass, $M_{pl}$, and other models that apply a shooting method to integrate from the outside in \citep[e.g.][]{val2006,rogers2010} iterate to solve for the total planet radius that leads to $r\approx0$ at $m=0$. To model the detailed time evolution and energy transport throughout the planet, it is necessary to simultaneously resolve the boundary conditions at the center and the surface.
 
We use a Henyey solver \citep{hen59}, an implicit iterative Newton-Raphson integration scheme, to solve equations~\ref{eq:dm} and \ref{eq:dP}, while simultaneously satisfying the specified boundary conditions at both the planet center and the surface. The Henyey solver approximates the spatial derivatives in equations~\ref{eq:dm} and \ref{eq:dP} at the boundary between two adjacent mass cells using the finite difference method, resulting in $2(n-1)$ difference equations for $n$ mass cells. Together with 2 boundary conditions at planet center and surface, there are $2n$ equations governing $2n$ unknowns ($P$ and $r$ at each cell). Provided with an approximated solution to the unknowns, an improved solution can be found with the Newton-Raphson method. The Henyey solver reiterates the procedure a few times until a converged solution is reached (i.e. the maximum change in the radius and pressure profiles is less than the relative tolerance of $10^{-4}$). We run the Henyey solver after each thermal timestep to update the radius, pressure, and volume of each cell based on the new temperature and melt fraction profiles. A major advantage of the method is that it can reach convergence rapidly given a reasonable starting point  (which, in our case, is the solution for the pressure and radius profiles at the previous timestep).  On top of this, adjustments in the interior structure due to phase transitions in the mantle and core are taken into account self-consistently. 

\subsection{Bulk Composition and Equation of State \label{EoS}}

The density within each layer is calculated based on the composition and the appropriate choice of EoS, which is a unique function that relates density to temperature and pressure for a given material. Our code is modular and could be readily extended to include EoS options in addition to the ones described in this section, enabling sensitivity analyses over the uncertainty of various EoSs and assumed compositions (for example, MAGARATHEA compiles a list of EoSs appropriate for mantles and cores of super-Earths \citep{huang2022}). However, this is beyond the scope of this paper. For this paper, we have chosen a set of EoSs that are semi-empirical fits to experimental data by high-pressure experiments or \emph{ab initio} calculations. In this section, we will discuss the bulk composition of the planets and the choice of EoS for the individual materials. 

\subsubsection{Composition and Phase Structure}
In this study, we consider terrestrial planets with a pure iron core and a silicate mantle. The mantle is molten while the local temperature is above the melting point of Mg-perovskite. The transition between liquid and solid phases of mantle silicate is given in a Simon form melting curve \citep{belonoshko2005},

\begin{equation}\label{eqn:pv_melt}
    T_{si,m}=T_{si,0}\left(\frac{P_{\mathrm{GPa}}}{4.6}+1\right)^{0.33},
\end{equation}

\noindent where $T_{si,0}=1831~K$ and $P_{\mathrm{GPa}}$ is the pressure in GPa. The latent heat release upon melting of mantle silicate, $L$, is $7.322\times10^5~\mathrm{J\,kg^{-1}}$\citep{hess1990}. Upon solidification, the mantle is further divided into a lower mantle consisting of Mg-perovskite (\ce{MgSiO3}) and an upper mantle consisting of olivine (\ce{Mg2SiO4}). There is a pressure-induced phase transition from olivine to Mg-perovskite with a Clapeyron slope of $-0.0013$~MPa~K$^{-1}$ and a phase boundary of $P_0=25$~GPa at the ambient temperature of $T_0=300$~K \citep{fei04}. A study exploring other choices for the Mg/Si ratios in the mantle will be pursued in the future. 

The iron core is divided into a solid inner core and a liquid outer core. The solid-liquid iron phase transition is determined by the melting curve \citep{zhang2015},

\begin{equation}\label{eqn:iron_melt}
T_{Fe,m}=T_{Fe,0}\left(\frac{P_{\mathrm{GPa}}}{31.3}+1\right)^{1/1.99},
\end{equation}
where $T_{Fe,0}=1900K$. The latent heat of fusion of iron due to core nucleation, $L_{Fe}$, is $1.2\times10^6\mathrm{Jkg^{-1}}$\citep{anderson1997}.
 
 Seismological data and high-pressure experiments have indicated the presence of light-element impurities in the iron core and iron pollution in the mantle silicate \citep[e.g.,][]{prem,andy2005, cote2008}. The influence of the presence of impurities in the iron core and silicate mantle is left for future work. 

\subsubsection{EoS of Molten Silicate}
For molten silicate, we adopt the recently developed EoS for liquids under high temperature and pressure conditions \citep{wolf2018}. The EoS comprises an isothermal component and a thermal perturbation described by the generalized Rosenfeld-Tarazona model \citep{rosenfeld1998}. The EoS reads

\begin{equation}
    P(\rho,T)=P(\rho,T_0)+\Delta P_E(\rho,T)+\Delta P_S(\rho,T),
\end{equation}
where the first term on the right is the isothermal component and the second and the third terms are the energetic and entropic contributions to the thermal perturbation. The isothermal component is given by the Vinet EoS \citep{vinet1989},

\begin{equation}
    P(\rho,T_0)=3K_0\left(\frac{\rho}{\rho_0}\right)^{2/3}\left[1-\left(\frac{\rho}{\rho_0}\right)^{-1/3}\right]\exp\left[\frac{3}{2}\left(K_0'-1\right)\left(1-\left(\frac{\rho}{\rho_0}\right)^{-1/3}\right)\right],
\end{equation}
where $T_0$ is the reference temperature of 3000~K, and $\rho_0$, $K_0$, and $K_0'$ are the density, isothermal bulk modulus, and derivative of isothermal bulk modulus at the reference temperature and zero pressure. The thermal perturbation terms, which are described by the generalized Rosenfeld-Tarazona model, are written as:

\begin{equation}
    \begin{split}
        \Delta P_E(\rho,T)&=-b'(\rho)[f_T(T)-f_T(T_0)],\\
        \Delta P_S(\rho,T)&=\frac{b'(\rho)}{m-1}[T(f_T'(T)-f_T'(T_{0S}))-T_0(f_T'(T)-f_T'(T_{0S}))]+\gamma_{0S}\rho c_{V,0S}(T-T_0).
    \end{split}
\end{equation}
In these two equations, $b'(\rho)$ is the derivative of thermal coefficients, which is given as 

\begin{equation}
    b(\rho)=\sum_nb_n\left(\frac{\rho_0}{\rho}-1\right)^n,
\end{equation}
where $b_n$ are fitted polynomial parameters, and $f_T$ measures the deviation from the reference temperature,

\begin{equation}
    f_T=\left(\frac{T}{T_0}\right)^m-1,
\end{equation}
where $m$ is a power law exponent with a theoretically expected value of 0.6 \citep{rosenfeld1998}. $T_{0S}$, $c_{V,0S}$ and $\gamma_{0S}$ are the temperature, specific heat capacity at constant volume and Gr\"uneisen parameter along the reference adiabat, whose expressions are given in \citet{wolf2018}.

Model parameters, including $\rho_0$, $K_0$, $K_0'$ and $b_n$, are fitted to molecular dynamic simulations of \citet{spera2011} with data over a range of temperatures ($\sim$2500~K-5000~K) and pressures (0-150~GPa). The resulting EoS accurately predicts densities of liquid \ce{MgSiO3} from shock-wave experiments on \ce{MgSiO3} glass \citep[roughly in the range of 50-150~GPa and 3000-6000~K,][]{mosenfelder2009}.

Thermophysical properties, such as specific heat capacity at constant volume and pressure per unit mass, $c_V$ and $c_P$, and Gr{\"u}neisen parameter, $\gamma=(\partial lnT/\partial lnV)_S$, are also directly output by the EoS, and expressions are provided in the Appendix of \citet{wolf2018}. The isothermal bulk modulus, $K_T$, and thermal expansion coefficient, $\alpha$, are calculated using their definitions and quantities output by the EoS, 

\begin{equation}\label{eqn:alpha}
    \begin{split}
        K_T&=-\rho\left.\frac{\partial P}{\partial \rho}\right\vert_T,\\
        \alpha&=\frac{\gamma\rho c_V}{K_T},
    \end{split}
\end{equation}
where $V$ is the volume at pressure, $P$, and temperature, $T$. 

\subsubsection{EoS for Iron and Solid Phase Silicate}
Similar to the EoS of \ce{MgSiO3} melt, the EoSs for Mg-perovskite, olivine as well as liquid- and solid-phase iron all consist of an isothermal component and a thermal perturbation. In this section, we provide a brief overview on the choices of EoSs. 

For the upper mantle silicate (olivine) and liquid iron, we choose the third order Birch-Murnaghan EoS (BM3) to predict the density at a given pressure level and a constant reference temperature, $T_0=300K$: 

\begin{equation}
    P(\rho,T_0)=\frac{3}{2}K_0\left[\left(\frac{\rho}{\rho_0}\right)^{\frac{7}{3}}-\left(\frac{\rho}{\rho_0}\right)^{\frac{5}{3}}\right]\left[1+\frac{3}{4}\left(K_0'-4\right)\left(\left(\frac{\rho}{\rho_0}\right)^{\frac{3}{2}}-1\right)\right],
\end{equation}
where $K_0$ and $K_0'$ are the isothermal bulk modulus at reference state and the derivative with respect to the pressure. 

The BM3 EoS results from the expansion of Eulerian finite strain and has been shown to be adequate for modeling the upper mantle of the Earth \citep{bir52}, however extrapolation beyond $100$~GPa, a pressure level common for the interior of super-Earth-sized planets, becomes uncertain \citep{stacey04}. Therefore, for the lower mantle (perovskite) and the solid inner iron core ($\epsilon$-Fe), we choose Keane's equation of state, which is based on the importance of the derivative of bulk modulus in the limit of infinite pressure $K'_{\infty}$ \citep{stacey08}. (Note that BM3 is used for the outer liquid iron core, as the liquid phase does not exist at infinite pressure and current \textit{ab initio} calculations do not provide a fitted value for $K'_{\infty}$). The Keane's EoS is proven to be consistent with the thermodynamics in the high-pressure limit and compatible with data obtained by high-pressure experiments \citep{sta11}. It is given by the following:

\begin{equation}
P(\rho,T_0)=K_0\left\{\frac{K^{\prime}_0}{K^{\prime2}_{\infty}}\left[\left(\frac{\rho}{\rho_0}\right)^{K^{\prime}_{\infty}}-1\right]-\left(\frac{K^{\prime}_0}{K^{\prime}_{\infty}}-1\right)ln\left(\frac{\rho}{\rho_0}\right)\right\}.
\end{equation}
Fitted values for $\rho_0, K_0, K'_0$ and $K'_{\infty}$ for the various materials and EoSs are summarized in table~\ref{EoS parameter}.

\begin{deluxetable}{lcccccccccccc}\label{EoS parameter}
\tabletypesize{\footnotesize}
\tablewidth{0.99\textwidth}
\tablecaption{Fitted EoS parameters}
\tablehead{     
  \colhead{Material} &\colhead{EoS} &\colhead{$\rho_0(\mathrm{{kg\,m^{-3}}}$)} & \colhead{$K_0 (\mathrm{GPa}$)} & \colhead{$K_0'$} &\colhead{$K_{\infty}'$} & \colhead{$\gamma_0$} & \colhead{$\gamma_{\infty}$} &\colhead{$\theta_0(\mathrm{K})$} & \colhead{$\lambda$} & \colhead{n} & \colhead{$T_0$ ($\mathrm{K}$)} & \colhead{Ref.}}
  
\startdata
Olivine    & BM3 & 3213.7    & 127.4    & 4.2 & N/A & 1.31 & N/A & 760 & 3.2 & 49.7542 & 300 & 1\\
Perovskite & Keane & 4105.9  & 267.7 & 4.04 & 2.63 & 1.506 & 1.148 & 1114 & 7.025 & 49.8058 & 300 & 2\\
Liquid Fe  & BM3 & 7037.8  & 83.7 & 5.97 & N/A & 2.033 & N/A & 263 & 1.168 & 17.907 & 1181 & 3\\
$\epsilon-\mathrm{Fe}$ & Keane & 8269.4  & 164.7 & 5.65  & 2.94 & 1.875 & 1.305 & 430 & 3.289 & 17.907 & 300 & 4\\
\enddata
\tablerefs{(1) \cite{Katsura2009}; (2) \cite{Oganov2001}; (3) \cite{Dorogokupets2017}; (4) \cite{Dewaele2006}}
\end{deluxetable}

To incorporate the effect of temperature on density of perovskite, olivine and liquid iron, we add a Debye thermal pressure term, which accounts for the lattice vibrational energy with a cut-off frequency corresponding to the Debye frequency. The Debye thermal pressure is given as:
\begin{equation}
    P_{De}(\rho,T)=\gamma \rho E_{in}(\rho, T),
\end{equation}
where $\gamma$ is the Gr{\"u}neisen parameter, and $E_{in}$ is the internal energy per unit mass. $E_{in}$ is calculated based on the Debye model:
\begin{equation}
    E_{in}=9nR_gT\left(\frac{T}{\theta}\right)^3\int^{\frac{\theta}{T}}_0 \frac{t^3}{e^t-1}dt,
\end{equation}
where $n$ is the atomic molar density in (mol~kg$^{-1}$), $R_g$ is the molar gas constant,and $\theta$ is the Debye temperature.
. The complete equation of state is then: 
\begin{equation}
    P\left(\rho, T\right)=P\left(\rho, T_0\right)+ [P_{De}\left(\rho, T\right)-P_{De}\left(\rho, T_0\right)].
\end{equation}

In addition to the effect of Debye thermal pressure, the density of solid iron is subject to an additional effect of anharmonic and electronic thermal pressures. The complete EoS for solid iron is
\begin{equation}
    P(\rho,T)=P\left(\rho, T_0\right)+[P_{De}\left(\rho, T\right)-P_{De}\left(\rho, T_0\right)]+[P_{an}(\rho,T)-P_{an}(\rho,T_0)]+[P_{el}(\rho,T)-P_{el}(\rho,T_0)],
\end{equation}
where $P_{an}$ and $P_{el}$ are anharmonic and electronic thermal pressures, and given as:
\begin{equation}
    P_{an}=\frac{3}{2}R_g\rho ma_0\left(\frac{\rho}{\rho_0}\right)^{-m}T^2,
\end{equation}
\begin{equation}
    P_{el}=\frac{3}{2}R_g\rho ge_0\left(\frac{\rho}{\rho_0}\right)^{-g}T^2.
\end{equation}
Parameters $a_0, m, e_0$ and $g$ are given by \citet{Dewaele2006}: $a_0=3.7\times10^{-5}\text{K}^{-1}$, m=1.87, $e_0=1.95\times10^{-4}\text{K}^{-1}$ and g=1.339.

The Gr{\"u}neisen parameter, $\gamma$, and the Debye temperature, $\theta$, vary with density. For olivine and liquid iron, these values are calculated as

\begin{equation}
    \begin{split}
        \gamma&=\gamma_0\left(\frac{\rho}{\rho_0}\right)^{-\lambda},\\
        \theta&=\theta_0\exp\left[\frac{\gamma_0-\gamma}{\lambda}\right].
    \end{split}
\end{equation}
For perovskite and $\epsilon$-Fe, these values are given in the Al'Tshuler form \citep{altshuler1987},

\begin{equation}
    \begin{split}
        \gamma&=\gamma_{\infty}+\left(\gamma_0-\gamma_{\infty}\right)\left(\frac{\rho}{\rho_0}\right)^{-\lambda},\\
        \theta&=\theta_0\left(\frac{\rho}{\rho_0}\right)^{\gamma_{\infty}}\exp\left[\frac{\left(1-(\rho/\rho_0)^{-\lambda}\right)\left(\gamma_0-\gamma_{\infty}\right)}{\lambda}\right],
    \end{split}
\end{equation}
where subscript $0$ denotes values evaluated at reference state and $\infty$ at infinite large pressure. Parameters relevant to the thermal correction are summarized in table~\ref{EoS parameter}.

The thermal conductivity, $k$, and specific heat capacity at constant pressure per unit mass, $c_P$, are approximately constant for mantle silicate within the pressure range ($\lesssim500$~GPa) in the mantle of the studied planets in this paper \citep{sta11}.
We use average Earth mantle and core values, which are summarized in table~\ref{thermophysical properties}. The isothermal bulk modulus, $K_T$, and the thermal expansion coefficient, $\alpha$, are calculated using the expression in eqn~\ref{eqn:alpha}, where $c_V$ is

\begin{equation}
    c_V=\frac{c_P}{1+\gamma \alpha T}.
\end{equation}

\begin{deluxetable}{lccc}\label{thermophysical properties}
\tabletypesize{\footnotesize}
\tablewidth{0.99\textwidth}
\tablecaption{Thermophysical properties of mantle and core}
\tablehead{     
  \colhead{Material} &\colhead{k($\mathrm{Wm^{-1}K^{-1}}$)} & \colhead{$\mathrm{c_P}$ ($\mathrm{JK^{-1}kg^{-1}}$)}  &\colhead{Ref.}
}
  
\startdata
Olivine     & 4    & 1250  &  1 \\
Perovskite & 10  & 1260 & 1 \\
Molten silicate  & 4  & N/A & 2 \\
Fe  & $40^{(3)}$  & $840^{(1)}$  & 1, 3 \\
\enddata
\tablerefs{(1) \citet{yukutake2000};  (2) \citet{bower2018}}; (3) \citet{kon2016}
\end{deluxetable}

\subsection{Core Cooling \label{core_cooling}}
The temperature distribution in the liquid iron core is assumed to be adiabatic, and the adiabatic temperature gradient is given as 

\begin{equation} \label{adiabat}
\left(\frac{\partial T}{\partial P}\right)_S=\frac{\alpha T}{\rho c_P},
\end{equation}
where $\alpha$ and $c_P$ denote the thermal expansion coefficient and specific heat per unit mass at constant pressure. The inner core starts solidifying once the adiabat of the liquid iron core intersects with the melting curve of iron. We assume the solid inner core to be conductive and solve the heat conduction equation to calculate its cooling process,
\begin{equation}
    \rho c_P\frac{\partial T}{\partial t}=\frac{1}{r^2}\frac{\partial}{\partial r}\left(r^2k_{c}\frac{\partial T}{\partial r}\right)+\alpha T\frac{\partial P}{\partial t},
\end{equation}
where $k_{c}$ is the thermal conductivity of iron. The second term on the right side of the equation is rate of heating due to adiabatic compression. Pressure changes in the planet are calculated using the Henyey solver.

The cooling process of the liquid core is described by its energy balance equation,
\begin{equation}\label{eqn:core_energy}
    \int_{M_{ic}}^{M_c}c_P\dot{T}dm-\int_{M_{ic}}^{M_c}\frac{\alpha T}{\rho} \dot{P}dm=4\pi R_c^2k_{LM}\left.\frac{\partial T}{\partial r}\right\vert_{\text{CMB}}-4\pi R_{\text{ICB}}^2k_c\left.\frac{\partial T}{\partial r}\right\vert_{\text{ICB}}+L_{\text{Fe}}\dot{M}_{ic},
\end{equation}
where $M_{ic}$ and $M_c$ are masses of the solid inner core and the entire iron core, $R_{\text{ICB}}$ and $R_c$ are radii of the solid inner core and the entire core, $k_{LM}$ and $k_c$ are the thermal conductivity of the lower mantle at CMB and iron at the inner core boundary (ICB), and $\left(\partial T/\partial r\right)_{\text{CMB}}$ and $\left(\partial T/\partial r\right)_{\text{ICB}}$ are temperature gradients at the base of the mantle and the base of the liquid core. The left hand side of the equation is the internal energy change rate of the iron core. The first and second terms on the right hand side of equation~\ref{eqn:core_energy} are the heat loss rate by conduction across the CMB and rate of heat gain across the ICB by conduction. The third term is the latent heat release rate due to inner core solidification.

The thermal history of the liquid core can be described by a potential temperature, $T_{Pot}$, which is the temperature of the liquid iron core if it were adiabatically decompressed to a fixed pressure reference point, $P_{Pot}$ \citep{unterborn2019}. For each planet we choose an arbitrary reference point at a pressure level $\sim$ 5 GPa lower than $P_{\text{CMB}}$ at the beginning of the evolution. For example, the $P_{Pot}$ for the planet with 1$M_{\oplus}$ and CMF=0.326 is 120 GPa. The first term on the left side of equation~\ref{eqn:core_energy} is related to $T_{Pot}$ by the following equation

\begin{equation}\label{eqn:core_internal}
    \int_{M_{ic}}^{M_c}c_P\dot{T}dm=\int_{M_{ic}}^{M_c}c_P\left.\frac{\partial T}{\partial T_{Pot}}\right\vert_{P}\dot{T}_{Pot}dm+\int_{M_{ic}}^{M_c}c_P\left.\frac{\partial T}{\partial P}\right\vert_{T_{Pot}}\dot{P}dm.
\end{equation}
$T_{Pot}$ is a proxy for entropy, and $(\partial T/\partial P)_{T_{Pot}}$ is the adiabatic temperature gradient, given by equation~\ref{adiabat}. The second term on the right side of equation~\ref{eqn:core_internal} becomes
\begin{equation}\label{eqn: adiabatic_compression_liquid_core}
    \int_{M_{ic}}^{M_c}c_P\left.\frac{\partial T}{\partial P}\right\vert_{T_{Pot}}\dot{P}dm=\int_{M_{ic}}^{M_c}\frac{\alpha T}{\rho}\dot{P}dm.
\end{equation}
We can rewrite equation~\ref{eqn:core_energy} in terms of $T_{Pot}$ using equations~\ref{eqn:core_internal} and~\ref{eqn: adiabatic_compression_liquid_core}, and the adiabatic compression term (the second term) on the left side of equation~\ref{eqn:core_energy} is cancelled by equation~\ref{eqn: adiabatic_compression_liquid_core}:
\begin{equation}\label{eqn:core_energy_Tpot}
    \dot{T}_{Pot}\int_{M_{ic}}^{M_c}c_P\left.\frac{\partial T}{\partial T_{Pot}}\right\vert_{P}dm=4\pi R_c^2k_{LM}\left.\frac{\partial T}{\partial r}\right\vert_{\text{CMB}}-4\pi R_{\text{ICB}}^2k_c\left.\frac{\partial T}{\partial r}\right\vert_{\text{ICB}}+L_{\text{Fe}}\dot{M}_{ic}.
\end{equation}
This is because heating due to adiabatic compression is accounted for when parameterizing the P-T profile by $T_{Pot}$.

$\dot{T}_{Pot}$ is expressed in terms of the change in conditions at the CMB by:

\begin{equation}\label{eqn:dT_an}
    \dot{T}_{Pot}=\left.\frac{\partial T_{Pot}}{\partial T_{\text{CMB}}}\right\vert_{P_{\text{CMB}}}\dot{T}_{\text{CMB}}+\left.\frac{\partial T_{Pot}}{\partial P_{\text{CMB}}}\right\vert_{T_{\text{CMB}}}\dot{P}_{\text{CMB}}.
\end{equation}
The partial derivatives with respect to $T_{Pot}$, $T_{\text{CMB}}$ and $P_{\text{CMB}}$ in equations~\ref{eqn:core_internal} and~\ref{eqn:dT_an} are calculated numerically using the pre-calculated table of core adiabats. 

The inner core nucleation decelerates the cooling rate of the core by releasing latent heat. Given a potential temperature, we can determine $M_{ic}$, and the pressure and radius at the boundary of inner and outer cores, $P_{\text{ICB}}$ and $R_{\text{ICB}}$, by the intersection of the core adiabat and the melting curve (equation~\ref{eqn:iron_melt}). Consequently, we can express $\dot{M}_{ic}$ as 

\begin{equation}\label{eqn:dMic}
   \dot{M}_{ic}=\left\{
\begin{array}{ll}
      0 & P_{\text{ICB}}>P_{\text{center}}\text{ or }P_{\text{ICB}}<P_{\text{CMB}}\\
       \frac{dM_{ic}}{dP_{\text{ICB}}}\frac{dP_{\text{ICB}}}{dT_{Pot}}\dot{T}_{Pot}
       & P_{\text{CMB}}<P_{\text{ICB}}<P_{\text{center}}.\\
\end{array}
\right.
\end{equation}
$dP_{\text{ICB}}/dT_{Pot}$ is evaluated numerically and $dM_{ic}/dP_{\text{ICB}}$ is obtained from equation~\ref{eq:dP},

\begin{equation}
    \frac{dM_{ic}}{dP_{\text{ICB}}}=-\frac{4\pi R_{\text{ICB}}^4}{GM_{ic}}.
\end{equation}
Combining equations~\ref{eqn:core_energy} to~\ref{eqn:dMic}, we obtain the energy balance equation of the core as a function of a single temperature, $T_{\text{CMB}}$.

\subsection{Mantle Heat Transport}\label{section:mantle heat transport}
The basic equation for calculating the energy transport in the mantle is provided by \cite{sas86}. Here, we express the energy transport equation in terms of entropy instead of temperature,

\begin{equation}\label{eqn:mantle_energy}
    \rho T\frac{\partial s}{\partial t}=-\frac{1}{r^2}\frac{\partial}{\partial r}\left(r^2F\right)+\rho H,
\end{equation}
where $s$ is specific entropy per unit mass, $H$ is energy generation rate per unit mass, and $F$ is the sum of conductive and convective flux. We write equation~\ref{eqn:mantle_energy} in terms of entropy because it is a natural coordinate for convecting systems of both pure solid/liquid and partially molten aggregates.

Thermal conduction is given by Fourier's law, which is split into an adiabatic and a super/sub-adiabatic part \citep{bower2018}, 

\begin{equation}
    F_{cond}=-\rho c_P \kappa \left.\frac{\partial T}{\partial r}\right|_s-\rho T\kappa\frac{\partial s}{\partial r},
\end{equation}
where $\kappa$ is the thermal diffusivity, and $\left(\partial T/\partial r\right)_s$ is the adiabatic temperature gradient, given by $\left(\partial T/\partial r\right)_s=\rho g\left(\partial T/\partial P\right)_s$ . 

Thermal convection occurs when the entropy gradient, $\partial s/\partial r$, is negative. We use the modified mixing length theory \citep{sas86,abe1995,senshu2002} to evaluate the convective heat flow, which considers the energy transport by the vertical motion by fluid parcels. This method allows us to use local values of physical quantities to evaluate heat transport within each zone. $F_{\text{conv}}$, the convective heat flux, is given as 

\begin{equation}
    F_{\text{conv}}=-\rho T\kappa_h\frac{\partial s}{\partial r},
    \label{eq:Fconv_kh}
\end{equation}
where $\kappa_h$ is eddy diffusivity. 

In a fully solid/liquid case, convection is driven by the temperature gradient, and $\kappa_h$ is 

\begin{subequations}\label{eddy_T}
\begin{empheq}[left={\kappa_h=\ \empheqlbrace}]{align}
\sqrt{-\frac{\alpha gTl^4}{16c_P}\frac{\partial s}{\partial r}} & \text{\ \ \ \ \ for\ }-\frac{\alpha gTl^4}{18c_P\nu^2}\frac{\partial s}{\partial r}>\frac{9}{8} \label{eddy_T_invisc}\\
-\frac{\alpha gTl^4}{18c_P\nu}\frac{\partial s}{\partial r} & \text{\ \ \ \ \ for\ }0<-\frac{\alpha gTl^4}{18c_P\nu^2}\frac{\partial s}{\partial r}<\frac{9}{8}, \label{eddy_T_visc}
\end{empheq}
\end{subequations}
where $g$ gravitational acceleration, $l$ mixing length, and $\nu$ kinematic viscosity. The eddy diffusivity for inviscid fluid (equation~\ref{eddy_T_invisc}) is derived in \cite{vitense53} and for viscous fluid (equation~\ref{eddy_T_visc}) is adapted from \cite{sas86}. Both equations in the original papers are expressed in terms of the temperature gradient, and the ones in terms of the entropy gradient are provided in \citep{bower2018}. A detailed derivation is given in Appendix~\ref{section:mlt}. The transition between the inviscid and viscous regimes is determined by the velocity of fluid parcels. In equation~\ref{eddy_T_invisc}, the velocity is estimated based on the exchange between the kinetic and gravitational potential energy of parcels; in equation~\ref{eddy_T_visc}, the velocity is estimated by the Stokes velocity, which considers the force balance between buoyancy force and viscous drag force exerted on fluid parcels. 

In the case of a solid and liquid mixture, $\rho$ is determined from the densities of pure solid, $\rho_s$, and pure liquid, $\rho_m$, at $T_{si,m}$ at a specified pressure by the additive volume mixing rule. The temperature profile is fixed on the melting curve of mantle silicate for the idealized cases of a single pure compound, and convection is then driven by the melt fraction gradient rather than the temperature gradient. We derived $\kappa_h$ in such a case (see Appendix~\ref{section:mlt_x} for details), 

\begin{subequations}\label{eddy_x}
\begin{empheq}[left={\kappa_h=\ \empheqlbrace}]{align}
\sqrt{-\frac{\alpha_xgT_{si,m}l^4}{16L}\frac{\partial s}{\partial r}} & \text{\ \ \ \ \ for\ }-\frac{\alpha_xgT_{si,m}l^4}{18 L\nu^2}\frac{\partial s}{\partial r}>\frac{9}{8} \label{eddy_x_invisc}\\
-\frac{\alpha_xgT_{si,m}l^4}{18 L\nu}\frac{\partial s}{\partial r} & \text{\ \ \ \ \ for\ }0<-\frac{\alpha_xgT_{si,m}l^4}{18 L\nu^2}\frac{\partial s}{\partial r}<\frac{9}{8}, \label{eddy_x_visc}
\end{empheq}
\end{subequations}
where $\alpha_x$ is the expansion coefficient due to changes in melt fraction, and defined as 

\begin{equation}\label{eqn:drhodx}
    \alpha_x\equiv-\frac{1}{\rho}\left.\frac{\partial \rho}{\partial x}\right\vert_P=\rho\left(\frac{1}{\rho_m}-\frac{1}{\rho_s}\right).
\end{equation}
The melt fraction $x$ is related to the entropy by

\begin{subequations}\label{melt_frac}
\begin{empheq}[left={x(s,P)=\ \empheqlbrace}]{align}
0 & \text{\ \ \ \ \ for\ }s\leq s_{s}(P) \\
\frac{s-s_{s}(P)}{s_{m}(P)-s_{s}(P)} & \text{\ \ \ \ \ for\ }s_{s}(P)<s<s_{m}(P) \\
1 & \text{\ \ \ \ \ for\ } s\geq s_{m}(P),
\end{empheq}
\end{subequations}
where $s_{m}$ and $s_{s}$ are the specific entropies of pure liquid and pure solid at $T_{si,m}(P)$. Similarly to the single phase case, the transition between formula~\ref{eddy_x_invisc} and~\ref{eddy_x_visc} is determined by the velocity of the convecting parcels. 

To improve the numerical stability of the code, we employ a transition function, $z(y)$, to ensure that $\kappa_h$ varies smoothly and continuously when the mantle silicate goes through phase transitions from pure liquid phase to a liquid and solid mixture, and from the mixture to pure solid phase. $\kappa_h$ is then 

\begin{subequations}
\begin{empheq}[left={\kappa_h=\ \empheqlbrace}]{align}
z(y_m)\kappa_{h,\ref{eddy_T_invisc}}+(1-z(y_m))\kappa_{h,\ref{eddy_x_invisc}} & \text{\ \ \ \ \ for\ inviscid\ fluid} \label{eqn:transition_inviscid} \\
(1-z(y_s))\kappa_{h,\ref{eddy_T_visc}}+z(y_s)\kappa_{h,\ref{eddy_x_visc}} & \text{\ \ \ \ \ for\ viscous\ fluid }, \label{eqn:transition_viscid}
\end{empheq}
\end{subequations}
where $\kappa_{h,\ref{eddy_T_invisc}}$,$\kappa_{h,\ref{eddy_T_visc}}$,$\kappa_{h,\ref{eddy_x_invisc}}$ and $\kappa_{h,\ref{eddy_x_visc}}$ are given by corresponding formula for $\kappa_h$. The transition function, $z(y)$, approaches 0 as $y\to-\infty$, and 1 as $y\to\infty$, and is given as,

\begin{equation}\label{transition function}
    z(y)=\frac{1}{2}+\frac{1}{2}\text{tanh}(y),
\end{equation}
where $y$ is 

\begin{equation}
    \begin{split}
        y_m&=\frac{s-s_{m}}{s_w}\text{\ \ \ \ \ for\ inviscid\ fluid }\\
        y_s&=\frac{s-s_{s}}{s_w}\text{\ \ \ \ \ for\ viscous\ fluid. }
    \end{split}
\end{equation}
with $s_w$ being the transition width. We choose $s_w$ to be $s_w=0.02\times(s_{m}-s_{s})$, which corresponds to 0.02 in melt fraction. $y_m$ and $y_s$ appear in equations~\ref{eqn:transition_inviscid} and \ref{eqn:transition_viscid} separately because the transition from fully liquid to the 2-phase mixture happens in the inviscid regime while the transition from the 2-phase mixture to fully solid happens in the viscid regime. 

The mixing length, $l$, describes the characteristic length that a fluid parcel can travel due to the thermal buoyancy force before it merges with the surroundings. \cite{abe1995} set $l$ to be the distance from the nearest boundary of convection ($D$). Several groups \citep[e.g.][]{tachinami2011,wagner2019} calibrated the 1-D mixing length theory against boundary layer theory and 3-D convection models for rocky planets, and came up with different prescriptions for $l$. \citet{tachinami2011} compared the evolution result of the Earth using mixing length theory to calculations using the boundary layer theory and found $l=0.82D$ to be the best choice, which we use for all simulations in this study. Such a prescription results in a small mixing length and thus a low convective heat flow near convection boundaries, and heat transport is dominated by thermal conduction.
Calculations with other prescriptions for $l$ are left for future investigation.

In this paper, we consider whole mantle convection where the only convection boundaries are the planet surface and the CMB. However, the mixing length formulation employed in our model can be easily applied to mantle convection with additional barriers, as \citet{tachinami2011} pointed out. Several groups suggest that phase boundaries in the mantle may serve as barriers to mantle convection, such as the solid/melt boundary \citep{labrosse2007} or the upper/lower mantle phase boundary \citep{honda1993}.

To save computation time while solving equation~\ref{eqn:mantle_energy}, we pre-tabulate thermophysical quantities, including $T$, $\rho$, $\alpha$, and $c_{P,m}$ (specific heat of silicate per unit mass in liquid phase), as a function of $s$ and $P$ using the EoSs described in section~\ref{EoS}. Values of these quantities are obtained using a 2D interpolation routine in python during each thermal timestep in the evolution calculation.

Viscosity, $\eta$, and radiogenic heating, $H$, are two other important components of mantle dynamics, which we discuss in sections~\ref{section:viscosity} and~\ref{section:Qrad}. 

\subsection{Viscosity of Mantle Silicate}\label{section:viscosity}

The dynamic viscosity of mantle silicate, $\eta=\nu\rho$, is one of the most important quantities in the thermal evolution calculation of rocky planets, especially during the later stage when the entire mantle has solidified. The level of viscosity determines the cooling rate of the mantle, which then controls how fast the iron core can cool. For silicate in the liquid phase, we take $\eta_m=100\text{Pa\ s}$ \citep{abe1997}. For silicate in the solid phase, we adopt an Arrhenius formulation for temperature- and pressure-dependent viscosity \citep{ranalli2001},

\begin{equation}
    \eta_s=\frac{1}{2}\left[\frac{1}{B^{1/n_c}}\exp\left(\frac{E^*+PV^*}{nR_gT}\right)\right]\dot{\epsilon}^{(1-n_c)/n_c},
\end{equation}
where $B$, $R_g$, $n_c$, $E^*$, $V^*$ and $\dot{\epsilon}$ are the Barger coefficient, molar gas constant, creep index, activation energy, activation volume and strain rate, respectively. Values of these parameters for Mg-perovskite and olivine are listed in table~\ref{table:viscosity}. The viscosity of partially molten material should capture an abrupt change at the critical melt fraction $x_{crit}=0.4$ as the molten aggregate shifts between liquid-like and solid-like. We adopt a similar formulation to \cite{bower2018}, and model $\eta$ as

\begin{equation}
    log_{10}\eta=zlog_{10}\eta_m+(1-z)log_{10}\eta_s,
\end{equation}
where $\eta_m$ and $\eta_s$ are dynamic viscosity for liquid and solid phase silicate. $z$ is the same transition function as equation~\ref{transition function} with $y$ defined as $y=(x-x_{crit})/0.15$.

In addition to the above formulation for viscosity, our code can readily take a wide range of other viscosity parameterizations, including ones adopted by \citet{abe1995}, \citet{stamenkovic2012}, and \citet{dri14}. In this paper, we consider a single choice to demonstrate the workings of code and to explore the influence of a temperature- and pressure-dependent viscosity on the planet evolution.

\begin{deluxetable}{lccccc}\label{table:viscosity}
\tabletypesize{\footnotesize}
\tablewidth{0.99\textwidth}
\tablecaption{Parameters of viscosity for Mg-perovskite and olivine \citep{ranalli2001}}
\tablehead{
  \colhead{Material} &\colhead{B($\mathrm{Pa^{-n_c}s^{-1}}$)} & \colhead{$n_c$} & \colhead{$\mathrm{E}^*(\mathrm{10^3J\,mol^{-1}})$} &\colhead{$\mathrm{V}^*(\mathrm{10^{-6}m^2mol^{-1}})$} &\colhead{$\dot{\epsilon}$($\mathrm{s^{-1}}$)}
}
\startdata
Olivine     & 3.5$\times10^{-15}$    & 3.0    & 430 & 10 & $10^{-15}$\\
Perovskite  & 7.4$\times10^{-17}$  & 3.5 & 500 & 10 & $10^{-15}$\\
\enddata
\end{deluxetable}

\subsection{Internal Heat Production}\label{section:Qrad}

Internal heat production, $H$, is important for rocky planet evolution on geological timescales. For Earth, the heat production is generated by the decay of \ce{^{40}K}, \ce{^{232}Th}, \ce{^{235}U} and \ce{^{238}U}. The estimated amount of these elements in Earth's mantle and their half life time are compiled in table~\ref{table:radiogenic}. The total mantle radiogenic heat production is the sum over 4 elements,

\begin{equation}
    H=\sum_i q_{0,i}\exp\left(\text{ln2}\left(t_{\oplus}-t\right)/\tau_{i}\right),
\end{equation}
where $t_{\oplus}$ is the current age of Earth, $q_{0,i}$ is the current radiogenic heat production rate per unit mass in Earth's mantle and $\tau_{i}$ is the radioactive decay time of four major isotopes. Previous works have studied the effect of different concentrations of radionuclides \citep[e.g.][]{nimmo2020}. However, for this study, we assume the same abundance of radioactive elements in the mantle as those in Earth's mantle and model the elements as uniformly distributed with mass in the mantle through convection. 

\begin{deluxetable}{lccc}\label{table:radiogenic}
\tablecaption{Parameters of radiogenic elements \citep{mcd1995}}
\tablehead{
  \colhead{element}                                            &  \colhead{$\mathrm{q_0}(\mathrm{W\,kg^{-1}})$}            &        \colhead{$\tau (\mathrm{Gyr})$}           
}
\startdata
\ce{^{40}K}                   & 8.69$\times10^{-13}$                   & 1.25                \\
\ce{^{232}Th}              & 2.24$\times10^{-12}$                             & 14                         \\
\ce{^{235}U}                       &  8.48$\times10^{-14}$   & 0.704    \\
\ce{^{238}U}                     &1.97$\times10^{-12}$       &    4.47        \\
\enddata
\end{deluxetable}

\subsection{Boundary Conditions}\label{section:boundary condition}

In this study, we explore a simplified scenario where planets cool down through radiation at the planet surface. The surface heat flow is thus chosen as a gray-body radiation, 

\begin{equation} \label{eqn:Fsurf}
    F_{\text{surf}}=\sigma\left(T_{\text{surf}}^4-T_{eq}^4\right),
\end{equation}
where $F_{\text{surf}}$ is the net outgoing surface heat flux, $\sigma$ the Stefan-Boltzmann constant, $T_{\text{surf}}$ surface temperature and $T_{eq}$ the equilibrium temperature, which corresponds to the radiation from the host star. 

Our model does not explicitly differentiate between stagnant-lid and mobile-lid convection and impose the thickness of conductive thermal boundary layers. Instead, a boundary layer where thermal conduction dominates is automatically captured by the mixing length, $l$. When approaching a convection barrier (the surface or CMB), $\kappa_h$ decreases rapidly in proportion to $l^4$. Our choice of the mixing length, as described in section~\ref{section:mantle heat transport}, represents an Earth-like scenario with mobile-lid convection. \citet{wagner2019} calibrated the mixing length theory against a 3-D mantle convection simulation. They presented various parametrizations of the mixing length for stagnant-lid, sluggish-lid, and mobile-lid convection, which we plan to explore using our model in future studies.   

The surface temperature and pressure is typically determined by the atmosphere for a real planet. \citet{abe1997} discussed the thermal blanketing effect of a steam or \ce{H2}/He atmosphere, which may sustain a shallow surface magma ocean. \citet{elkins2008} explored Earth's and Mars' atmosphere growth due to magma ocean solidification, and found that the final atmospheric pressure could reach several thousand bar. However, exploring the influence of an atmosphere on the thermal evolution of rocky planets is beyond the scope of this paper. Additionally, \citet{bower2019} suggests that silicate mantle may not experience an appreciable amount of compression due to an outgassed atmosphere, as the bulk modulus of molten and solid silicate is around 100 GPa. In this paper, the surface temperature is determined by solving the energy transport equation (equation~\ref{eqn:mantle_energy}) at the planet surface and the surface pressure is set to be 1 bar. 
At the bottom surface of the mantle, namely the CMB, the heat flux is evaluated based on heat conduction through the CMB layer, the first term on the right hand side of equation~\ref{eqn:core_energy},

\begin{equation}
    F_{\text{CMB}}=-k_{LM}\left.\frac{\partial T}{\partial r}\right|_{\text{CMB}},
\end{equation}
where $k_{LM}$ is the thermal conductivity of lower mantle.

With the heat flux defined at the planet surface and CMB layer by the boundary conditions, we can calculate the net cooling rate of both the core and the mantle. Since no radiogenic heating is considered in the iron core, the net cooling rate of the core in $W$, $Q_{\text{core}}$, is 

\begin{equation}
    Q_{\text{core}}=4\pi R_c^2 F_{\text{CMB}}.
\end{equation}
The net cooling rate of the mantle is the sum of outgoing energy at the surface ($Q_{\text{surf}}$), incoming energy at the CMB ($Q_{\text{CMB}}$) and radiogenic heating ($Q_{\text{rad}}$), 

\begin{equation}\label{eqn:mantle_cooling}
    Q_{\text{mantle}}=4\pi R_{\text{pl}}^2F_{\text{surf}}-4\pi R_c^2 F_{\text{CMB}}-HM_{\text{mantle}},
\end{equation}
where $M_{\text{mantle}}$ is the mass of the mantle. The first, second and third terms in equation~\ref{eqn:mantle_cooling} represent $Q_{\text{surf}}$, $Q_{\text{CMB}}$, and $Q_{\text{rad}}$. $Q_{\text{core}}$ and $Q_{\text{mantle}}$ are positive when the core and the mantle cool down and negative when they heat up.

\subsection{Criterion to Drive a Dynamo in Rocky Planets}\label{section:dynamo_criterion}
A dynamo action within the planetary interior requires a magnetic Reynolds number that exceeds a critical value,

\begin{equation}\label{eqn:Rem}
    Re_m=\mu_0vL_c\sigma>Re_{m,crit},
\end{equation}
where $\mu_0$ is the magnetic permeability in vacuum, $v$ is the flow velocity, $L_c$ is the thickness of the magma ocean/liquid iron core, and $\sigma$ is the electrical conductivity of the convective fluid. As \citet{christensen2006} suggest, a self-sustained dynamo action requires a critical value, $Re_{m,crit}$, of order 50, which we adopt for both the iron core and the magma ocean in our model. We note, however, that the study in \citet{christensen2006} is configured to represent modern Earth's core and we advocate for dynamo studies focusing on the magma ocean.

Several groups have attempted to measure the electrical conductivity of molten silicate using dynamic and \emph{ab initio} calculations \citep{soubiran2018,stixrude2020}. They have shown that molten silicate is semi-metallic and estimated its electrical conductivity to be of the order of $10^4$ $\mathrm{S\,m^{-1}}$. Given the level of electrical conductivity and using an illustrative value for the thickness of the convective layer, $\sim10^6$ m, a minimal flow velocity of the order of 1$\mathrm{mm\,s^{-1}}$ is required so that $Re_m$ can exceed the critical value in a magma ocean. 

We evaluate $Re_m$ in each cell to determine whether and where a dynamo action can occur in the magma ocean. The flow velocity is estimated with the modified mixing length theory, $v\sim\kappa_h/l$. We use the equation provided in \citet{stixrude2020} to calculate the electrical conductivity of molten silicate, which includes ionic and spin-polarized electric contributions, 
\begin{equation}
    \sigma=\sigma_0T^{-1}\exp\left(-\frac{E^*+PV^*}{R_gT}\right).
\end{equation}
Values of $\sigma_0$, $E^*$, and $V^*$ for ionic and spin-polarized electric contributions are summarized in table~\ref{table:electrical conductivity}. The electrical conductivity of partial melts is weighted by the melt fraction.

\begin{deluxetable}{lccc}\label{table:electrical conductivity}
\tabletypesize{\footnotesize}
\tablewidth{0.99\textwidth}
\tablecaption{Parameters for electrical conductivity of molten silicate \citep{stixrude2020}}
\tablehead{
  \colhead{Contribution} &\colhead{$\sigma_0(\mathrm{Sm^{-1}})$} & \colhead{$\mathrm{E}^*(\mathrm{10^3J\,mol^{-1}})$} &\colhead{$\mathrm{V}^*(\mathrm{10^{-6}m^2mol^{-1}})$} 
}
\startdata
Spin-polarized electronic contribution     & 1.754$\times10^{9}$    & 108.6    & 0.0611\\
Ionic contribution  & 1.0811$\times10^{9}$  & 131 & 0.437\\
\enddata
\end{deluxetable}

We compute $Re_m$ for the entire liquid iron core to determine whether the dynamo can operate within and its lifetime. To translate the core convective heat flux to the flow velocity, we employ a mixing length scaling law \citep{christensen2010},
\begin{equation}\label{eqn:flow_velocity}
    v=\left(\frac{L_cF_{\text{conv}}}{\rho H_T}\right)^{1/3},
\end{equation}
where $H_T=c_P/(\alpha g)$ is the temperature scale height and $L_c$ is the thickness of the convecting liquid core. The core convective heat flux, $F_{\text{conv}}$, is calculated as the difference between the total heat flux coming out of the core ($F_{\text{CMB}}$) and the conductive heat flux along the core adiabat ($F_{\text{cond}}$), which is given as
\begin{equation}
    F_{\text{cond}}=k_c\left(\frac{\alpha gT}{c_{P}}\right)_{\text{s,CMB}}.
\end{equation}
The liquid iron core is no longer convecting if $F_{\text{CMB}}<F_{\text{cond}}$, and $F_{\text{conv}}$ is taken to be 0. We use the Wiedemann-Franz law to calculate the electrical conductivity of liquid iron,
\begin{equation}
    \sigma=\frac{k_c}{L_0T},
\end{equation}
where $k_c$ is the thermal conductivity of liquid iron and $L_0$ is the Lorenz number. We use 40 $\mathrm{Wm^{-1}K^{-1}}$ as the default value for $k_c$ \citep{monteux2011, kon2016}. However, higher values of $k_c$ under the core conditions ($>$80$\mathrm{Wm^{-1}K^{-1}}$) have been reported \citep[e.g.][]{pozzo2012,pozzo2014, ohta2016, ZhangY2020,zhangy2021,pourovskii2020, inoue2020}, and we offer a discussion on the effect of different levels of $k_c$ on the lifetime of the dynamo in the liquid core in section~\ref{section:dynamo age}. Values of $\rho$, $\alpha$, $g$, and $\sigma$ used in the calculation of $Re_m$ are mass-weighted average for the liquid iron core.

In general, unless the outer liquid iron core is very thin, so long as the liquid iron core exists and is convective, it could support a dynamo with $Re_m>Re_{m,crit}$. Using illustrative values for Earth-like and super-Earth planets for quantities in equations~\ref{eqn:Rem} and~\ref{eqn:flow_velocity}, $\sigma\sim10^6\,\mathrm{S\,m^{-1}}$, $L_c\sim10^6$~m, $\rho\sim10^4\,\mathrm{kg\,m^{-3}}$, $c_P=840\,\mathrm{kg\,J^{-1}\,K^{-1}}$, $g\sim10~\mathrm{m\,s^{-2}}$ and $\alpha\sim10^{-5}\mathrm{K^{-1}}$, the required $F_{\text{conv}}$ to sustain a dynamo in the liquid iron core is very low $\sim10^{-10}\mathrm{W\,m^{-2}}$. Since $F_{cond}>>10^{-10}\mathrm{W\,m^{-2}}$, $Re_m$ of the liquid iron core will exceed $Re_{m,crit}$ as long as it is convecting. However, in cases where the liquid iron core is thin (i.e., as the core approaches complete solidification), $Re_m$ may drop below the critical value and the dynamo might cease even while convection continues. For example, if $L_c$ decreases to 10km, the required $F_{\text{conv}}$ to sustain a dynamo increases to $\sim0.1\mathrm{W\,m^{-2}}$, which is non-negligible compared to $F_{cond}$ (typically $\sim0.05\mathrm{W\,m^{-2}}$ for $k_c=40\mathrm{Wm^{-1}K^{-1}}$).

\section{Model validation}\label{validation}
In the following, we present our simulations of magma ocean solidification and late-time evolution after the mantle fully solidifies for an 1$M_{\oplus}$ planet with CMF$=$0.326 and $T_{eq}=255K$. We compare our results to previous work \citep{lebrun2013, solomatov2015, tachinami2011} as well as observations of current-day Earth \citep{prem, davies2010} to validate our model approach. 
\subsection{Magma Ocean Solidification}\label{section:magma_validate}
\begin{figure}
\begin{center}
\includegraphics[scale=0.6]{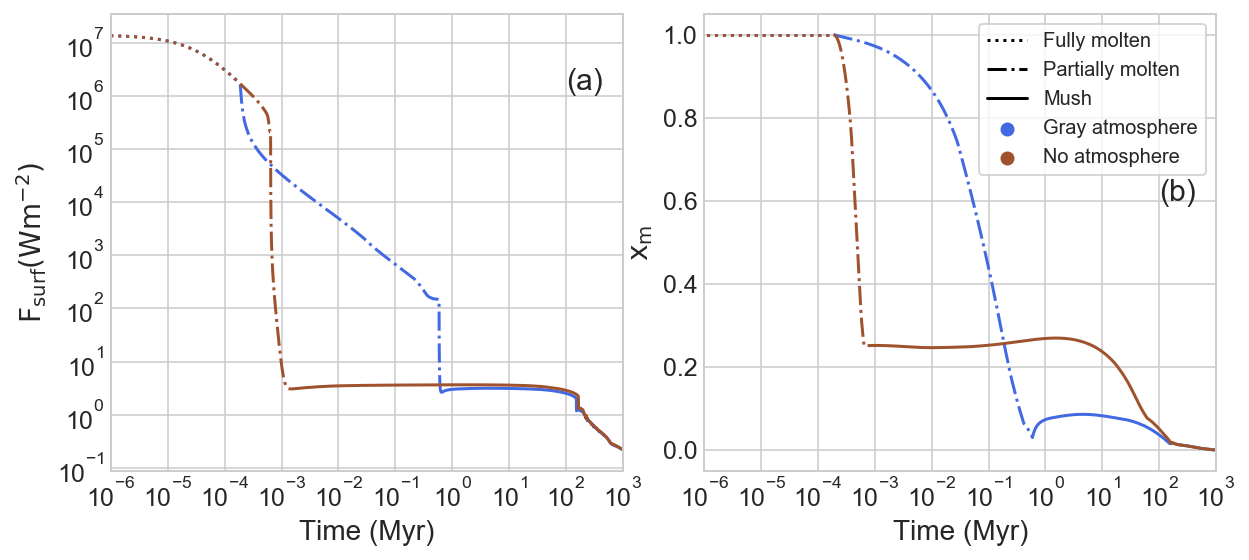}
\end{center}
\caption{Evolution of (a) surface heat flow ($F_{\text{surf}}$) and (b) fraction of total mantle mass in the liquid phase ($x_m$) for a 1~$M_{\oplus}$ planet with CMF=0.326 and $T_{eq}$=255K. The melting curve is taken to be Earth's mantle solidus \citep{hirschmann2000,andrault2011} instead of equation~\ref{eqn:pv_melt}. Brown curves are for the case with no atmosphere and blue curves are for the case with a degassed gray atmosphere with initial water content of 4.3$\times10^{-2}$wt\% and CO$_2$ content of 1.4$\times10^{-2}$wt\%. Dotted, dashdotted and solid curves represent the fully molten stage, the partially molten stage and the mush stage.  \label{fig:magma_ocean_validate}}
\end{figure}

We validate our models by comparing the magma ocean's solidification timescale with previous work \citep{lebrun2013,solomatov2015}. We simulate the magma ocean solidification of an $1M_{\oplus}$ planet with an Earth-like composition --- an iron core and silicate mantle with a core mass fraction of 0.326 --- following the procedure described in section~\ref{methodology}. 

To facilitate the comparison, we adjust some of our default model choices to match the scenarios simulated by \citet{lebrun2013}. We use Earth mantle's solidus as the melting curve instead of the melting curve of pure Mg-perovskite. We use the solidus profiles taken from \citet{andrault2011} at high pressure (P$>10$GPa) and from~\citet{hirschmann2000} at low pressure. We take an initial thermal profile comparable to that of \citet{lebrun2013}, i.e. an adiabatic profile with an initial surface temperature of 4000~K. We run simulations both with and without a degassed atmosphere. In the case that considers a degassed atmosphere, we use the same initial water ($4.3\times10^{-2}\text{wt}\%$) and CO$_2$ ($1.4\times10^{-2}\text{wt}\%$) content as \citet{lebrun2013}. The atmosphere is treated as a gray emitter and the surface heat flux is calculated using equations~7-9 in \citet{elkins2008}. The surface heat flux is modified by the emissivity of the atmosphere, which is a function of the mass of the atmosphere. The evolution of surface heat flux ($F_{\text{surf}}$) and fraction of total mantle mass in the liquid phase ($x_m=\int^{\mathrm{surf}}_{\mathrm{CMB}}xdm/M_{\mathrm{mantle}}$) with and without an overlying steam atmosphere are shown in figure~\ref{fig:magma_ocean_validate}.

Following \citet{lebrun2013}, we divide the solidification of magma ocean into three stages: (i) totally molten stage, (ii) partially molten stage, where the magma ocean starts solidifying from the CMB to the planet surface, and (iii) mush stage, where a cold thermal boundary layer starts developing at the surface and the surface temperature approaches the planet equilibrium temperature ($T_{eq}$). 

As shown in figure~\ref{fig:magma_ocean_validate}, stages (i) and (ii) are fairly short ( $\sim$1~kyrs), when there is no insulating atmosphere at the planet surface. During these stages, the convection is extremely vigorous due to the low viscosity of molten silicate, and the surface heat flow can reach as high as $\sim10^7$~$\mathrm{Wm^{-2}}$. As the mantle cools down rapidly, the magma ocean starts solidifying from the bottom and becomes partially molten near the CMB at $\sim$ 100~years. This partially molten zone quickly expands from the CMB to the surface in the next $\sim$1~kyrs and $x_m$ drops to around 20\% in this time period. After the surface starts solidifying, the convective flow near the surface becomes suppressed due a drastic increase in the local viscosity level (equation~\ref{eddy_x}). As a result, the surface temperature quickly approaches $T_{eq}$ and the mush stage begins. The mantle fully solidifies at $\sim$200~Myrs and details on the mush stage is discussed in section~\ref{section:magma ocean}. 

\begin{figure}
\begin{center}
\includegraphics[scale=0.65]{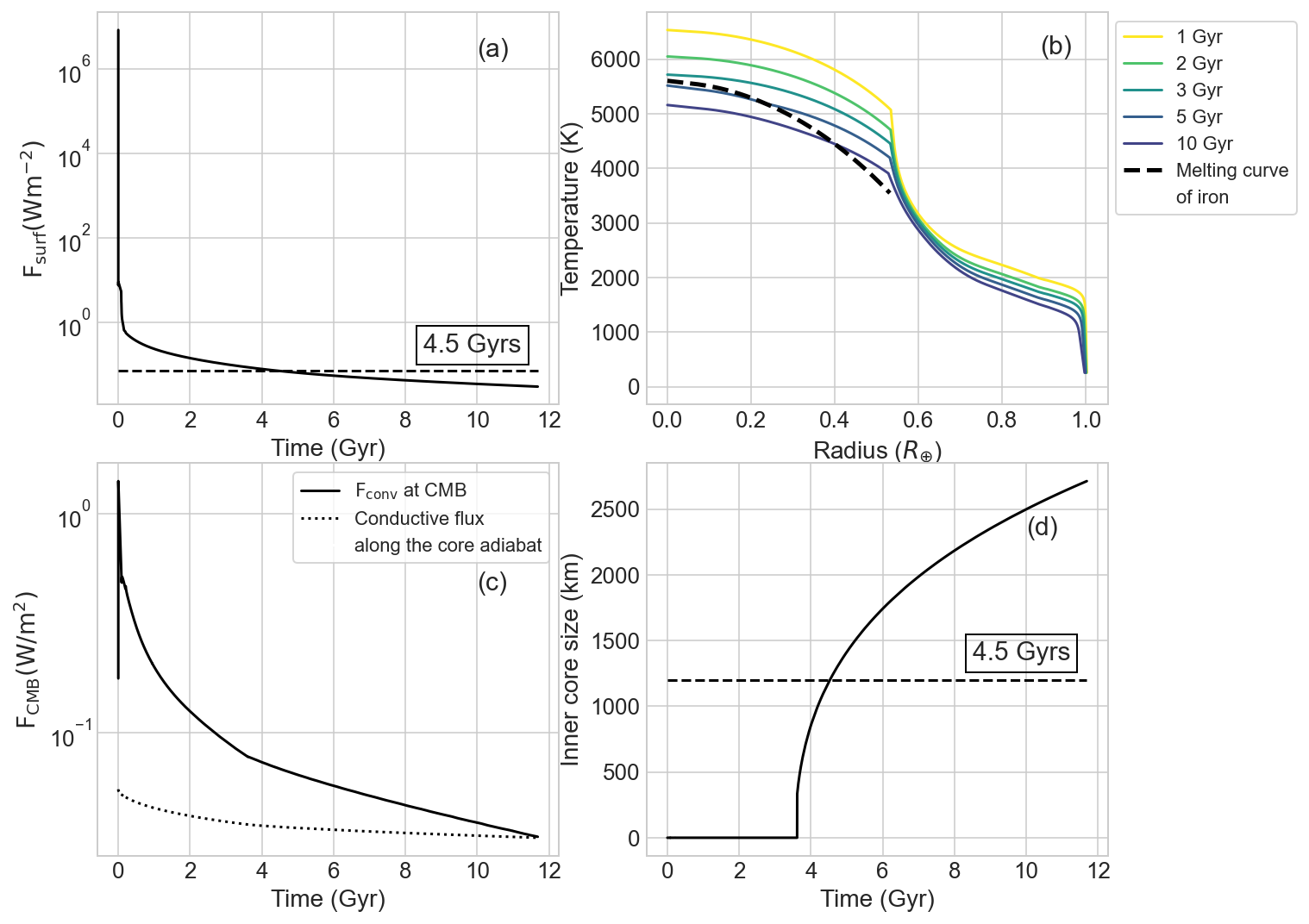}
\end{center}
\caption{Evolution of (a) surface heat flow, (b) radial temperature profiles, (c) heat flux at the CMB and (d) solid inner core size for a 1$M_{\oplus}$ planet with CMF=0.326 and $T_{eq}=255$~K. Melting temperature of iron is reduced by 20\% artificially to mimic the influence of impurities in the iron core (see text in section~\ref{section:iron_core_validate}). The surface heat flow reduces to $\sim$ 0.07 $\mathrm{W/m^2}$ (corresponding to $\sim$~36.5 TW) and the solid inner core size increases to $\sim$ 1300 km at 4.5 Gyrs.  \label{fig:validation}}
\end{figure}

Even though we use a different prescription of surface heat flux (equation~\ref{eqn:Fsurf}) from the parameterized surface flux used by \citet{lebrun2013}, the lifetimes of totally and partially molten stages are comparable, $\sim$1~kyrs. Our result also agrees qualitatively with the magma ocean crystallization timescale predicted by the scaling law by \citet{solomatov2015} within an order of magnitude ($\sim$ 400 years). In addition, our model predicts a similar level of $F_{\text{surf}}$ during all three stages of magma ocean solidification as compared to \citet{lebrun2013}, from $\sim10^7$~$\mathrm{Wm^{-2}}$ during the totally molten stage to less than $10$~$\mathrm{Wm^{-2}}$ during the mush stage.

Including a degassed atmosphere during the magma ocean solidification results in a slower cooling process than when no atmosphere is present. This increases the duration of the totally and partially molten stage to $\sim$0.6~Myrs and thus delays the mush stage. 
This result differs from that predicted by \citet{lebrun2013} ($\sim$0.2~Myrs when the atmosphere is treated as a gray emitter)  by a factor of three, but agrees to within an order of magnitude. We attribute the difference mainly to the different partition coefficients of volatiles in mantle silicate, as we assume all volatiles are degassed into the atmosphere while \citet{lebrun2013} considers non-zero partition coefficients. These illustrative simulations highlight the sensitivity of the magma ocean evolution to the choice of atmospheric boundary condition. 
Results presented in this section are qualitatively comparable to those of \citet{lebrun2013} and \citet{ solomatov2015}. This gives our model the credibility to simulate magma ocean solidification on rocky planets. 

\subsection{Iron Core Solidification}\label{section:iron_core_validate}
In this section, we present our calculation of the late-time evolution --- after the mantle is fully solidified --- of a 1$M_{\oplus}$ planet with CMF$=$0.326 and $T_{eq}=255K$. We validate our model by comparing the model predicted inner core size and surface heat flux at 4.5~Gyrs with previous work \citep{tachinami2011,dri14,lab2015} and current-day observations of Earth \citep{prem,davies2010}.

To facilitate the comparison, we fine tune our model choices to match the Earth-like case simulated by \citet{tachinami2011}. The melting temperature of iron is reduced by 20\% to account for the influence of light elements in the iron core \citep{stevenson1983}. We use an initial temperature profile for the planet based on \citet{tachinami2011}, which results in a fully molten iron core and solid silicate mantle. 

\begin{figure}
\begin{center}
\includegraphics[scale=0.67]{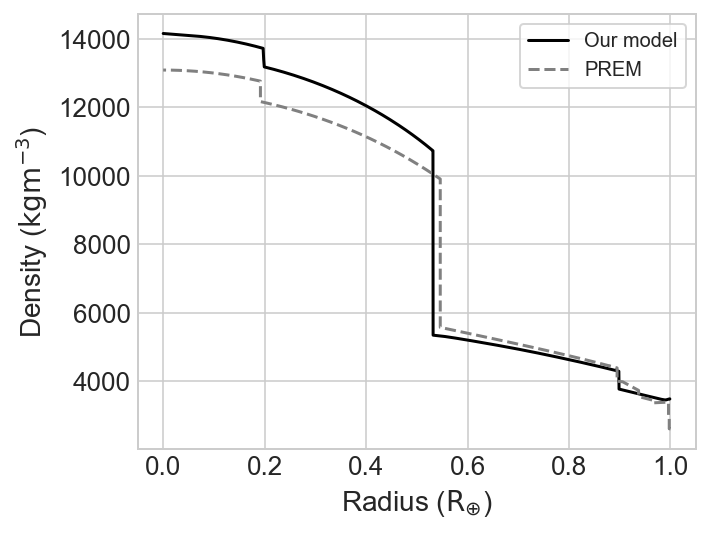}
\end{center}
\caption{Model predicted radial density profile at 4.5 Gyrs and comparison to the result from  Preliminary Reference Earth Model \citep[PREM,][]{prem}. Our model predicted a higher density in the iron core due to the lack of light impurities, which results in a slightly smaller iron core size. The model predicted planetary radius is in excellent agreement with the PREM result.   \label{fig:density_validate}}
\end{figure}

Figure~\ref{fig:validation} summarizes the time evolution of surface heat flow ($F_{\text{surf}}$), radial temperature profiles, heat flow at CMB ($F_{\text{CMB}}$) and solid inner core size ($R_{\text{ICB}}$). The cooling rate of the mantle slows down over time as the decrease in temperature enhances the viscosity and depresses the efficiency of heat transfer. The surface heat flow reduces to $\sim$0.07 $\mathrm{Wm^{-2}}$ at around 4.5 Gyrs, which agrees well with the value predicted by \citet{tachinami2011}, $\sim$0.08~$\mathrm{Wm^{-2}}$. Our result is also in decent agreement with the current observed surface heat flow of Earth, $\sim$ 0.09 $\mathrm{Wm^{-2}}$ \citep{davies2010}. 

Figures~\ref{fig:validation} (c) and (d) show the history of $F_{\text{CMB}}$ and $R_{\text{ICB}}$, respectively. The onset of inner core solidification happens around 3.5~Gyrs. This puts the $\sim$1 Gyr age of the solid inner core predicted by our model within the 0.5-2 Gyrs range indicated by many previous thermal models of Earth \citep[e.g.,][]{tachinami2011, dri14,lab2015, ZhangY2020}. The solid inner core grows to $\sim$ 1300 km at 4.5 Gyrs, which is comparable to the results of \citet{tachinami2011} and \citet{dri14} and the observed value of current day Earth \citep{prem}. The formation of the solid inner core also releases latent heat, which act as an additional heat source in the iron core and slow the core cooling. As a result, the decline in $F_{\text{CMB}}$ slows after $\sim$ 3.5 Gyrs. Our models predict that thermal convection in the liquid outer core will eventually shut off around 11.5 Gyrs. 

Our model obtains an interior structure comparable to that of present-day Earth using the Henyey code. Figure~\ref{fig:density_validate} shows the radial density profile of the planet at 4.5 Gyrs with comparison to results of Preliminary Reference Earth Model \citep[PREM,][]{prem}. The comparison shows decent agreement and the mismatch in the planetary radius is less than 0.1$\%$, which further validates our modeling approach. The discrepancy in the core density between PREM and our model is due to the presence of elements lighter than iron in Earth's core and our choice to simulate a distilled pure-iron core composition. As a result, the model predicted core radius at 4.5 Gyrs is about 92 km smaller than the seismologically observed value. Additionally, our model does not compute the increase in the concentration of light elements in the liquid outer core due to exsolving light elements from the solid inner core. This would further reduce the melting temperature of iron as the solid core grows. A consistent inclusion of light elements in the iron core is left for future studies. 

\begin{figure}
\begin{center}
\includegraphics[scale=0.6]{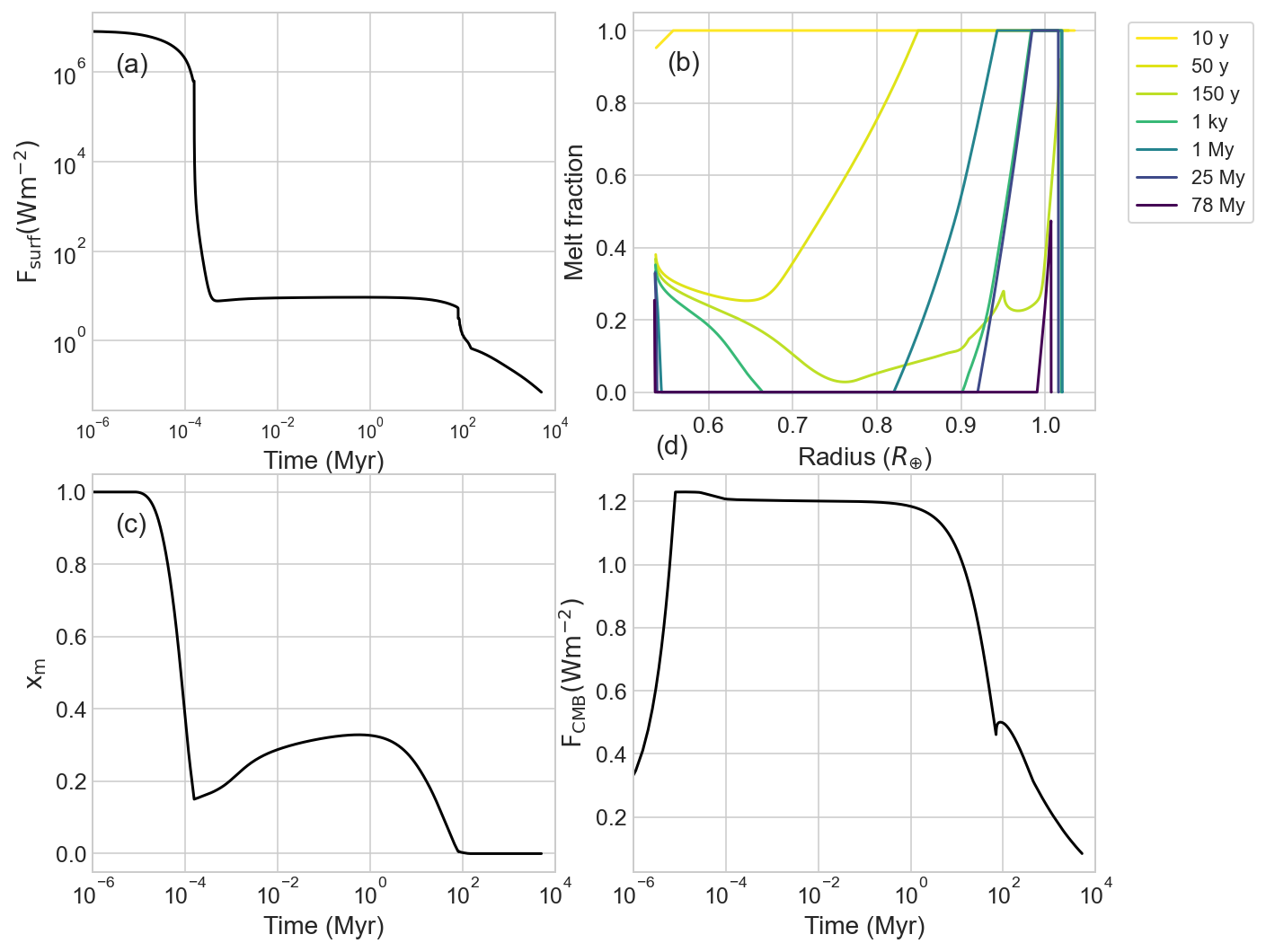}
\end{center}
\caption{Evolution of (a) surface heat flow, (b) melt fraction distribution, (c) fraction of the total mantle mass in the liquid phase and (d) heat flux at the CMB for a 1~$M_{\oplus}$ planet with CMF=0.326 and $T_{eq}$=255K. The planetary radius and core radius start off at 6600~km and 3424~km, and decrease to 6400~km and 3409~km respectively. The sub-surface magma ocean reaches its maximum size of 1277km at an age of $\sim$1 million years and fully solidifies by $\sim$~150 million years \label{fig:magma_evol_Earth}}
\end{figure}

\section{Thermal evolution results and the lifetime of magnetic field}\label{result}

We applied our model to a grid of 12 planets with 2 different planetary masses (1 and 3 $M_{\oplus}$), 2 CMFs (an Earth-like value of 0.326 and an iron-enhanced value of 0.7), and 3 equilibrium temperatures (255K, 1250K and 2500K). Our goals are to explore how these factors influence the planetary thermal evolution and to assess magma oceans and liquid iron cores as potential dynamo source regions. Additionally, the results demonstrate the various capabilities of our model.  We first discuss the solidification history of and dynamo action within magma oceans in 12 planets in section~\ref{section:magma ocean}. Then in section~\ref{section:iron core}, we comment on the thermal history of the planet following the solidification of magma oceans with a focus on the cooling history of and dynamo activity within iron cores.  

\subsection{Thermal Evolution of Magma Ocean \label{section:magma ocean}}

\subsubsection{Earth-like Case}

We now discuss how the magma ocean solidifies for an Earth-mass ($M=1M_{\oplus}$) planet with a core mass fraction (CMF) of 0.326 and an equilibrium temperature ($T_{eq}$) of 255K (the baseline case). Figure~\ref{fig:magma_evol_Earth}b shows the evolution of the radial melt fraction profile of the baseline case, and figures~\ref{fig:magma_evol_Earth}a, c and d show the time evolution of the surface heat flux ($F_{\text{surf}}$), fraction of the total mantle mass in the liquid phase ($x_m$), and the heat flux across the CMB ($F_{\text{CMB}}$). 

The fully and partially molten stages of the magma ocean evolution are similar to those of the validation case in section~\ref{section:magma_validate} but shorter. The magma ocean starts solidifying and becomes partially molten starting from the CMB layer at $\sim$10~years. The partially molten zone expands to the surface in the next 150 years and the mush stage starts earlier than in the validation case. This is a direct consequence of using the melting temperature of pure Mg-perovskite, which is hotter than the solidus temperature of Earth's mantle. 

During the mush stage in the evolution (from around 150 years to 150 Myrs), the outgoing heat flux at the surface, $F_{\text{surf}}$, drops down to less than 10 $\text{W}\text{m}^{-2}$, as shown in figure~\ref{fig:magma_evol_Earth}a, which significantly slows the cooling rate of the mantle compared to the previous stages. The dramatic decrease in $F_{\text{surf}}$ is triggered by the reduction in the melt fraction near the surface. As the melt fraction near the surface drops below $x_{crit}=0.4$, the viscosity level near the surface increases drastically, which leads to a suppressed convective heat flow near the surface boundary (equation~\ref{eddy_x}). As a result, a cold solid boundary forms at the surface, and the surface temperature drops quickly and approaches $T_{eq}$, which leads to the dramatic decrease in $F_{\text{surf}}$.

\begin{figure}
\begin{center}
\includegraphics[scale=0.6]{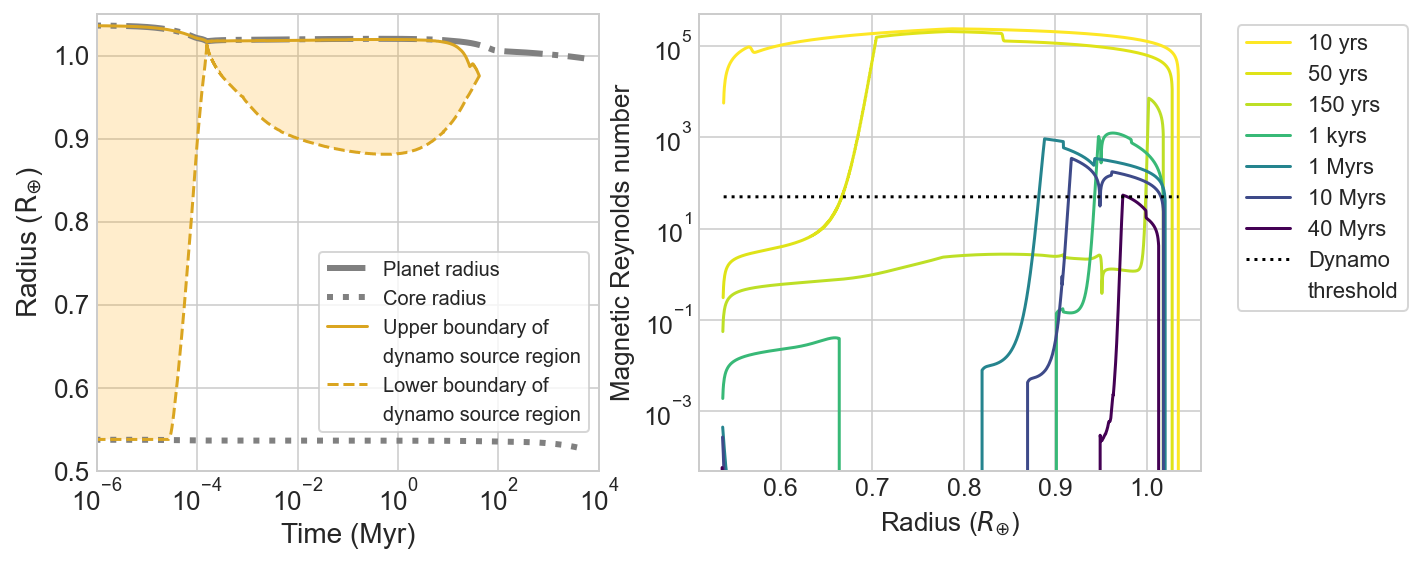}
\end{center}
\caption{Time evolution of (a) dynamo source region and (b) magnetic Reynolds number profiles of the magma ocean for 1$M_{\oplus}$ planet with CMF=0.326. The dotted line in (b) indicates the threshold above which the region can potentially support a dynamo. The dynamo source region starts off as the entire fully molten mantle and is confined to the sub-surface magma ocean only $\sim$150 years after the start. The size of the region while confined to the sub-surface magma ocean reaches its maximum of 890km at $\sim$ 1 Myrs and vanishes $\sim$ 40 million years. \label{fig:magma_dynamo_Earth}}
\end{figure}

Within the mush stage, a sub-surface magma ocean develops as figure~\ref{fig:magma_evol_Earth}b indicates. A layer in the middle of the mantle first becomes fully solid around $\sim$200 years, as heat continues to be transported from the deep interior of the mantle toward the surface via convection. With a low surface heat flux limiting the heat lost rate, the region just below the surface accumulates the heat coming from the deep interior, which cannot be radiated into space as quickly as it is convected from the interior. Thus, the melt fraction just below the surface starts increasing again and a sub-surface magma ocean develops, causing $x_m$ to increase (figure~\ref{fig:magma_evol_Earth}c). At 1~Myr, the sub-surface magma ocean reaches its maximum size, after which it starts to shrink. The mantle eventually fully solidifies at 150~Myrs. We note that the behavior of the magma ocean during these short-lived transient stages would occur only if the mantle is fully molten at the end of the planet formation or after the last giant impact.

As explained in section~\ref{section:dynamo_criterion}, a dynamo may be sustained in the magma ocean where $Re_m$ exceeds the critical value of 50. Figure~\ref{fig:magma_dynamo_Earth}a shows the evolution of the dynamo source region in the magma ocean and figure~\ref{fig:magma_dynamo_Earth}b shows the evolution of the magnetic Reynolds number profile. At the very beginning of our simulation, the entire liquid mantle could potentially generate a magnetic field, with a low viscosity (100 Pa s) and a magnetic Reynolds number ($>1000$) well above the dynamo threshold ($\sim 50$) throughout the entire mantle. As the mantle becomes partially molten, the melt fraction descends below $x_{crit}$ starting from the CMB and moving upward in the first 200 years. In the region with $x<x_{crit}$, the magnetic Reynolds number, which depends on the convective velocity, is limited by the viscous drag force (with a $\sim$10 orders of magnitude increase in the viscosity level compared to $x>x_{crit}$, equation~\ref{stokes_velocity}) and declines below the dynamo threshold. The thickness of the dynamo source region decreases from over 3000~km to only 20~km near the very surface in this time period. As the sub-surface magma ocean develops, the melt fraction in this region increases again causing the viscosity level to drop and thus the magnetic Reynolds number to climb back above the dynamo threshold. The dynamo source is then confined to the region in the sub-surface magma ocean where the melt fraction is high enough to allow liquid-like convection. The thickness of the dynamo source region reaches its maximum of 890km around 1~Myr and then decreases to 0 at 40~Myrs. 

In addition to the convective velocity, the dynamo lifetime at the top of the sub-surface magma ocean is also limited by the decreasing electrical conductivity of silicate melts toward lower pressure levels \citep{stixrude2020}. As figure~\ref{fig:magma_dynamo_Earth} shows, the dynamo starts shutting off from its upper boundary after 10 Myrs. At 10 Myrs, the electrical conductivities of silicate melts at the upper and lower boundaries of the dynamo source region are $\sim4000\ \mathrm{Sm^{-1}}$ and $\sim12000\ \mathrm{Sm^{-1}}$. As a result, the lower level of $\sigma$ near the planet surface leads $Re_m$ to fall below the critical value.

\subsubsection{Exoplanets}\label{MO solidification}

\begin{figure}
\begin{center}
\includegraphics[scale=0.6]{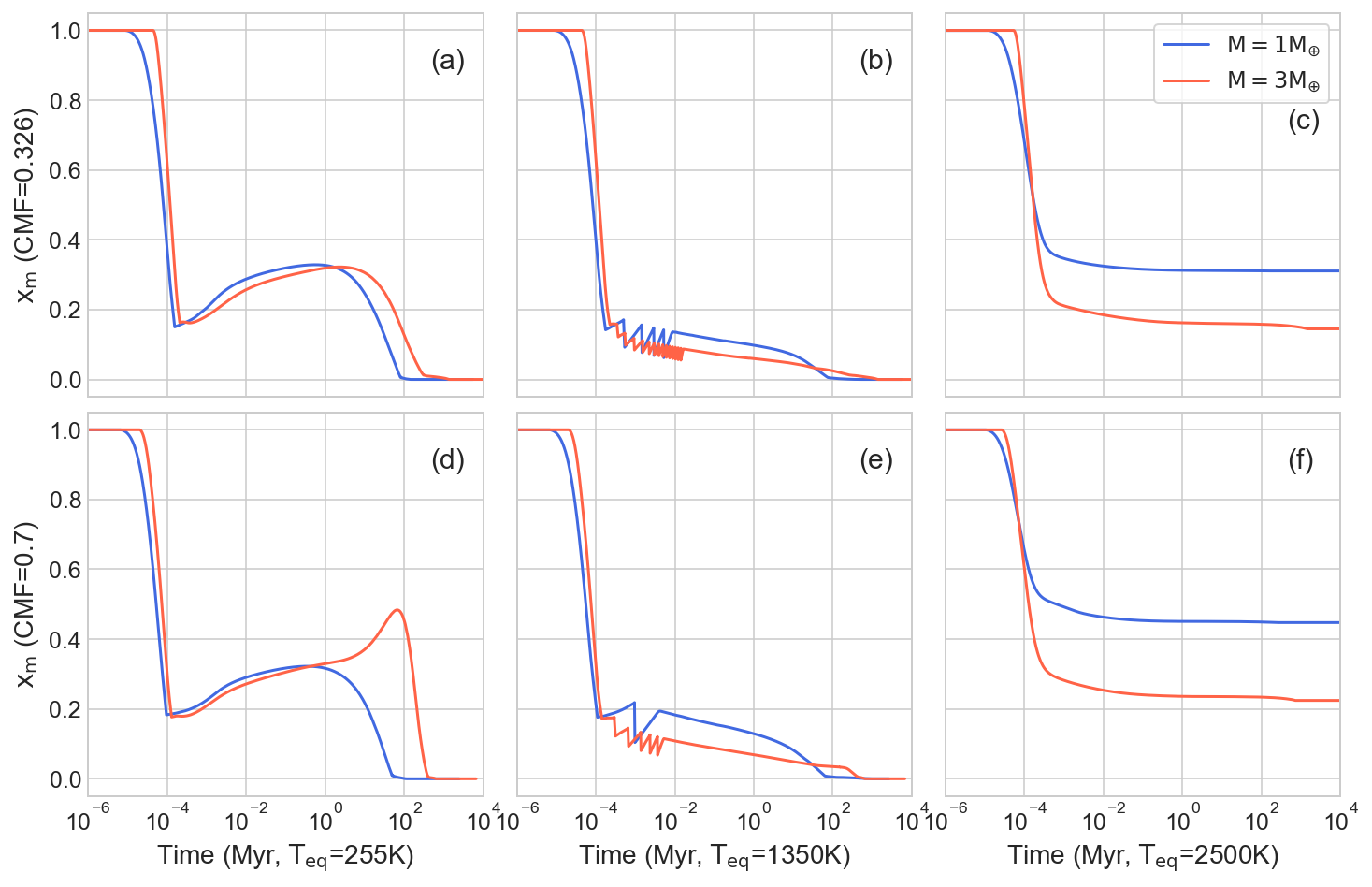}
\end{center}
\caption{Time evolution of the fraction of the total mantle mass in liquid phase ($x_m$) in planets with CMF=0.326 (top row), CMF=0.7 (bottom row), $T_{eq}=$255K (left column), $T_{eq}=$1350K (middle column) and $T_{eq}=$2500K (right column). Blue and red curves indicate planets with 1 and 3 $M_{\oplus}$, respectively. $T_{eq}$ has the major influence on the evolution of the magma ocean. The lifetime and the depth of the magma ocean have a slight dependence on the planetary mass and CMF, which together determine the mass of the mantle. \label{fig:melt_exoplanet}}
\end{figure}

In this section, we explore how planet mass, CMF, and equilibrium temperature affect the evolution of magma oceans in cooling rocky planets. We consider planets with masses of 1 and 3 $M_{\oplus}$, CMFs of 0.326 and 0.7, and equilibrium temperatures of 255K, 1350K and 2500K to demonstrate the versatility of our model. Figure~\ref{fig:melt_exoplanet} shows the evolution of overall melt fraction of the 12 cases explored. 

The equilibrium temperature, $T_{eq}$, determines the regimes of the evolution history of magma oceans, given the surface boundary adopted in this study. We chose three different levels of $T_{eq}$ representing three regimes of behaviors of magma oceans: a low level (255~K) far below the melting temperature of silicate at the surface, a medium level (1350~K) that is near but just below the same melting temperature, and a high level (2500~K) above that melting temperature. As figure~\ref{fig:melt_exoplanet} shows, the evolution of $x_m$ follows three different trends with different equilibrium temperatures. 

In the 4 cases of a low equilibrium temperature ($T_{eq}=255\mathrm{K}$), the evolution of the magma ocean follows a similar trend as in the baseline case (figures~\ref{fig:melt_exoplanet}a and d). The fully and partially molten stages only last a few hundred years and the overall fraction of the total mantle mass in the liquid phase quickly drops to $\sim$0.2 at the end of the partially molten stage. As a cold thermal boundary layer develops at the surface, the cooling rate of the mantle decreases substantially. A sub-surface magma ocean forms due to a lower cooling rate at the surface and solid-state convection in the deeper interior continuing transporting heat upwards. As the sub-surface magma ocean widens and  eventually shrinks over the next few hundreds of million years, the overall melt fraction increases and then reduces to 0.

\begin{figure}
\begin{center}
\includegraphics[scale=0.6]{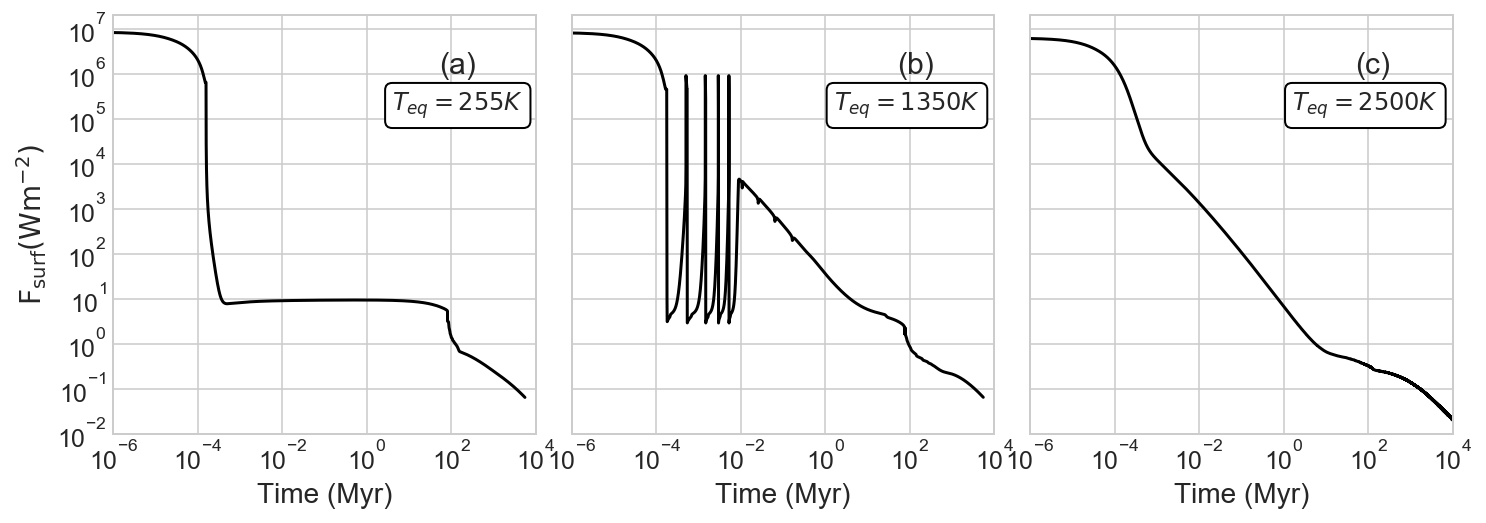}
\end{center}
\caption{Time evolution of the surface heat flux of 1$M_{\oplus}$ planet with CMF=0.326 and (a) $T_{eq}=$255K, (b) 1350K and (c) 2500K. \label{fig:Fsurf_1E}}
\end{figure}

Compared to cases with $T_{eq}=255\mathrm{K}$, the magma ocean on planets with an equilibrium temperature of 1350~K go through an additional episodic stage after the cold thermal boundary first forms at the surface (figures~\ref{fig:melt_exoplanet}b and e). In this episodic stage, as the sub-surface magma ocean widens, heat transported to the surface by the sub-surface magma ocean can remelt the surface together with a $T_{eq}$ sufficiently close to the melting temperature of silicate at the surface. This can lead to a drastic increase in the surface heat flux back to $10^6\mathrm{W\,m^{-2}}$ (figure~\ref{fig:Fsurf_1E}b).  With an enhanced level of surface heat flux, the overall melt fraction starts to drop again and the cold thermal boundary layer eventually redevelops. Depending on the initial heat content of the mantle, this process can repeat multiple times until the overall melt fraction finally reduces to 0. During this stage, the episodic surface melting enhances the surface heat flux periodically and allows for a more efficient time-average cooling rate for the magma ocean despite of a higher stellar incident flux compared to the cases with the lower $T_{eq}$ of 255K. The enhanced $F_{\text{surf}}$ also limits the size of the sub-surface magma ocean and the potential dynamo source region. Our results present a counterintuitive scenario where more irradiated planets can potentially cool more quickly than less irradiated ones. We checked whether the episodic behavior is an artifact due to the size of timesteps or spatial resolution in individual cells. The size of timesteps decreases during the episodic stage due to rapid changes in $x_m$ and $F_{\text{surf}}$, and is much smaller than the period of the oscillations ($\sim$~10 years vs $\sim$~1000 years). However, our model does not resolve the surface layer down to $\sim$cm-m thickness. As the surface layer remelts due to the widening of sub-surface magma ocean and high $T_{eq}$, the layer could become thin enough such that foundering may occur. Additional studies that treat surface foundering are required to predict whether the non-monotonic behavior in $x_m$ and $F_{\text{surf}}$ may persist. 

The surface magma ocean on planets with $T_{eq}=2500\mathrm{K}$, our highest considered equilibrium temperature, persist for the age of the universe. This is because $T_{eq}$ is above the melting temperature of mantle silicate at the surface of the planet. As figures ~\ref{fig:melt_exoplanet}c and f show, the overall mantle melt fraction in each 2500~K case quickly drops to steady-state levels in the first few hundred years and then stays roughly constant for the next 10 billion years. 

In the cases of the low (255~K) and medium (1350~K) levels of $T_{eq}$, the lifetime of the magma ocean increases with increasing mantle mass, which is determined by both planet mass and CMF. With $T_{eq}=255$K, the lifetime of the magma ocean for 1$M_{\oplus}$ planet with a CMF of 0.7 is 0.12 Gyr, whereas the lifetime increases to 1.38 Gyr for 3$M_{\oplus}$ planet with a CMF of 0.326. In the high $T_{eq}=2500$K cases, the fraction of the mantle in the liquid phase, $x_m$, achieved after the surface magma ocean reaches a steady state decreases with increasing mantle mass. This trend is simply the result of greater mantle mass in the solid phase below the surface magma ocean. The 3$M_{\oplus}$ planet with a CMF of 0.326 has the smallest overall mantle melt fraction at 10 billion years ($x_m\sim0.15$) out of the 4 cases, and the 1$M_{\oplus}$ planet with a CMF of 0.7 has the greatest melt fraction of 0.45.

\begin{figure}
\begin{center}
\includegraphics[scale=0.5]{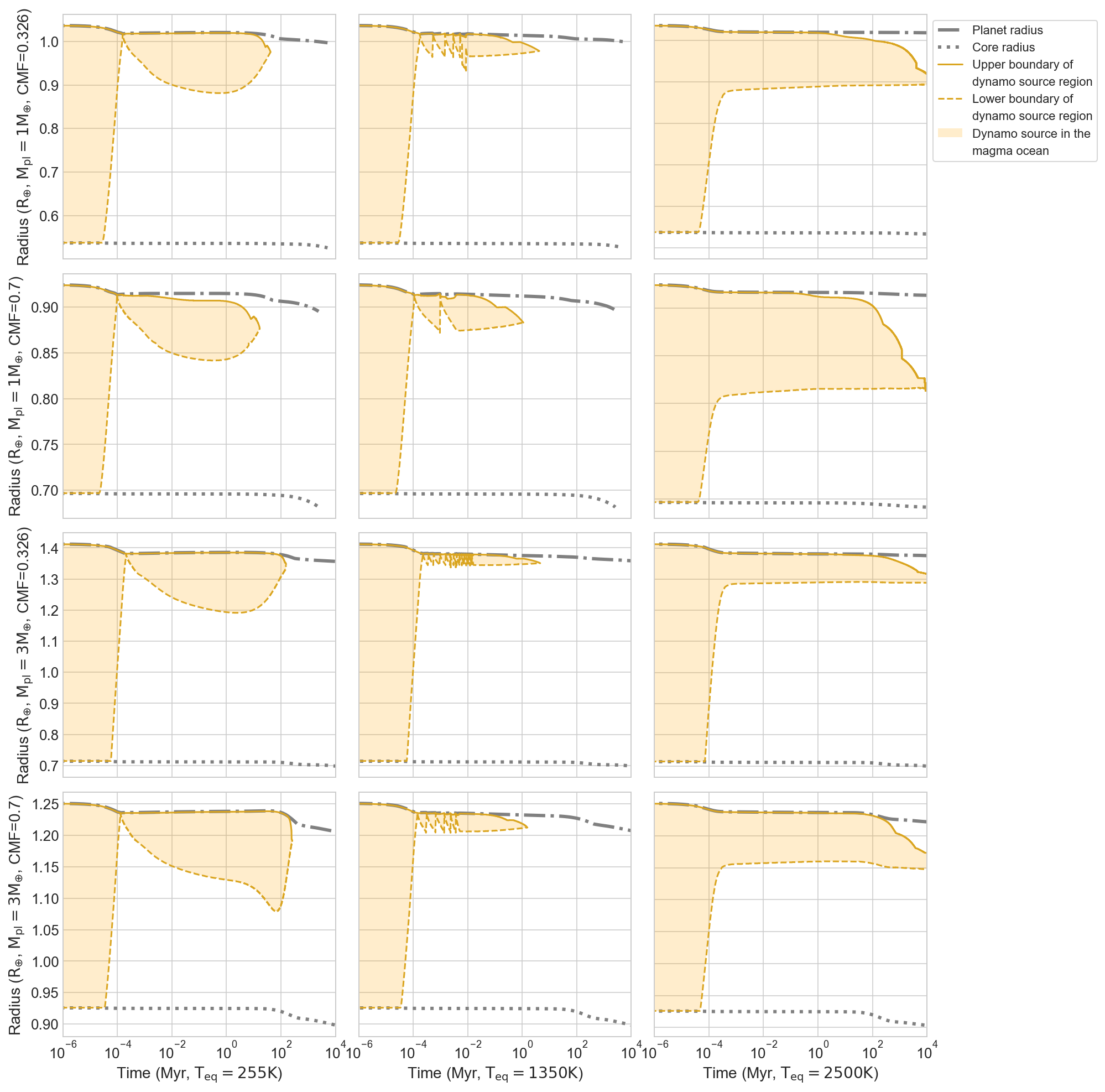}
\end{center}
\caption{Evolution of the potential dynamo source regions in the magma ocean for all 12 cases. The three columns are for cases with $T_{eq}=$255~K, 1350~K and 2500~K. The four rows are for cases with different combinations of $M_{pl}$ and CMF. The shaded area indicates the dynamo source region in the magma ocean. Solid and dashed lines indicate the upper and lower boundaries of the dynamo source region in the magma ocean. Dotted and dashdotted lines indicate radii of the planet and the iron core.
\label{fig:dynamo_exoplanet}}
\end{figure}

Similar to the baseline model, in all cases the dynamo source region in the magma ocean is determined by $Re_m$. Results are summarized in figure~\ref{fig:dynamo_exoplanet}. 

For cases with $T_{eq}=255K$ (left column of figure~\ref{fig:dynamo_exoplanet}), the entire fully molten mantle can be the potential dynamo source region initially. As the temperature reaches the melting curve from the bottom of the mantle upward and a cold thermal boundary quickly develops at the surface, the dynamo source region shrinks and is quickly confined ($\sim$1 kyr) to the very top layers in the mantle. As the sub-surface magma ocean forms and the eventually solidifies, the dynamo source region widens and diminishes. The dynamo source regions in the cases with $T_{eq}=1350K$ show similar overall trends as the low equilibrium temperature cases (figure~\ref{fig:dynamo_exoplanet}, middle column). The $T_{eq}=1350K$ cases have an additional episodic stage due to the remelt of the cold thermal boundary layer at the surface, which results in a thinner dynamo source region compared to the low $T_{eq}$ cases. Lastly, in cases where $T_{eq}=2500K$, the dynamo source region is quickly reduced to the surface magma ocean in only $\sim$1 kyr. 

In all 12 combinations of mass, CMF, and $T_{eq}$ that we explored, the top of the magma ocean eventually stops contributing to the dynamo, and the upper boundary of the dynamo source region falls below the planet surface. This is due to the pressure dependence of the electrical conductivity of silicate melt (which decreases toward lower pressures). As the planet ages and the convective heat flux declines, molten ($x>x_{crit}$) regions at low pressures near the planet surface are not able to keep $Re_m$ above the critical value.

\subsection{Thermal Evolution of Solid Phase Mantle and Iron Core}\label{section:iron core}

In this section, we explore the thermal evolution of all 12 cases mentioned in section~\ref{section:magma ocean} following the solidification of the magma ocean. Our focus is to investigate the possibility of the liquid outer iron core being a potential dynamo source. We end the simulation once the entire iron core fully solidifies.

\subsubsection{Earth-like Case}
\begin{figure}
\begin{center}
\includegraphics[scale=0.6]{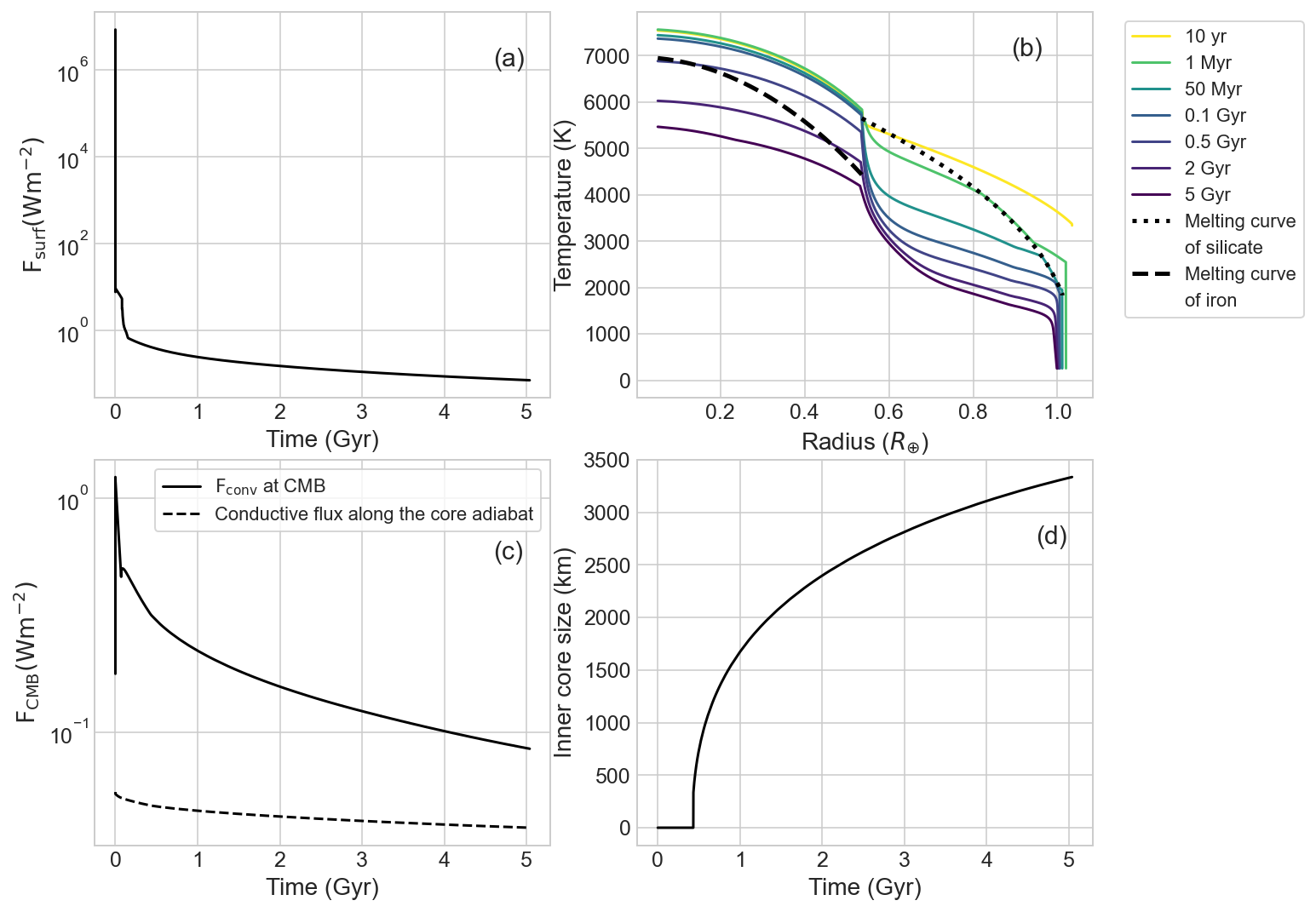}
\end{center}
\caption{Evolution of (a) surface heat flow, (b) radial temperature distribution, (c) heat flux at the CMB, and (d) radius of the solid inner core for a 1$M_{\oplus}$ planet with CMF=0.326 and a pure iron core/pure Mg silicate mantle composition. The dotted curve in (b) represents the melting curve of the silicate mantle. The dashed curve in (c) shows the conductive flux along the core adiabat with $k_c=40\ \mathrm{Wm^{-1}K^{-1}}$. In order for the liquid core to be convective and thus to be a potential dynamo source region, $F_{\text{CMB}}$ needs to exceed the conductive flux. The iron core fully solidifies at $\sim$ 5 Gyrs, and the shorter age of the liquid core compared to the validation case (figure~\ref{fig:validation}) is caused by the high melting temperature of pure iron without impurities.  \label{fig:1ME_mantlecore}}
\end{figure}

Here we discuss the thermal evolution of our baseline case: i.e., a planet with $M_{pl}=1M_{\oplus}$, $T_{eq}=255K$ and $\mathrm{CMF=0.326}$. Figure~\ref{fig:1ME_mantlecore} shows the evolution of the radial temperature distribution ($T(r)$), solid inner core size ($R_{\text{ICB}}$), as well as the surface heat flux ($F_{\text{surf}}$) and heat flux at the CMB ($F_{\text{CMB}}$), which indicate the cooling rate of the entire planet and iron core respectively. The cooling rate of the mantle is extremely rapid at first, due to the low viscosity of molten and partially molten silicate. This leads to a rapid planetary contraction early on, and the pressure level at the CMB increases by $\sim$10 GPa from 127 to 137 GPa in the first $\sim$0.2 Myrs. The increase in the pressure level in the iron core heats it up (equation~\ref{eqn:core_energy}). As the mantle solidifies and cools down in the next $\sim$5 Gyrs, the viscosity in the mantle increases, which reduces the convective heat flux in the mantle. Thus the mantle becomes less efficient at transporting heat to the surface, which is also reflected by the decreasing surface heat flux (figure~\ref{fig:1ME_mantlecore}a).

The cooling rate of the iron core is directly controlled by heat conduction through the CMB layer. As the mantle quickly cools down, the temperature difference across the CMB layer quickly builds up, causing $F_{\text{CMB}}$ to grow in the first $\sim$1 Myr. As the iron core cools down and subsequently loses heat into the mantle, the temperature difference across the CMB layer decreases over time, which lowers $F_{\text{CMB}}$ (figure~\ref{fig:1ME_mantlecore}c) and slows down the cooling rate of the iron core. 

At around $\sim$0.5 Gyr, the iron core starts solidifying from the center of the planet. This early onset of inner core solidification compared to the validation case (section~\ref{validation}) is a result of adopting a pure iron composition in the core. The presence of impurities in the iron-dominated core would reduce the melting temperature, delaying the start of inner core solidification and lengthening the lifetime of the liquid outer core. As the solid inner core grows, latent heat is released, which work as additional internal heat sources and slow down the cooling rate of the iron core and the rate of decline of $F_{\text{CMB}}$ ( figure~\ref{fig:1ME_mantlecore}c). The slower cooling rate combined with the spherical geometry of the iron core causes the growth rate of the solid inner core radius to decrease (figure~\ref{fig:1ME_mantlecore}d). Due in part to the latent heat released, $F_{\text{CMB}}$ remains above the conductive flux along the core adiabat (indicated by the dashed curve in figure~\ref{fig:1ME_mantlecore}c) for the remainder of the evolution.

\subsubsection{Equilibrium Temperature Dependence of the Thermal Evolution History \label{section:influence of Teq}}

\begin{figure}
\begin{center}
\includegraphics[scale=0.67]{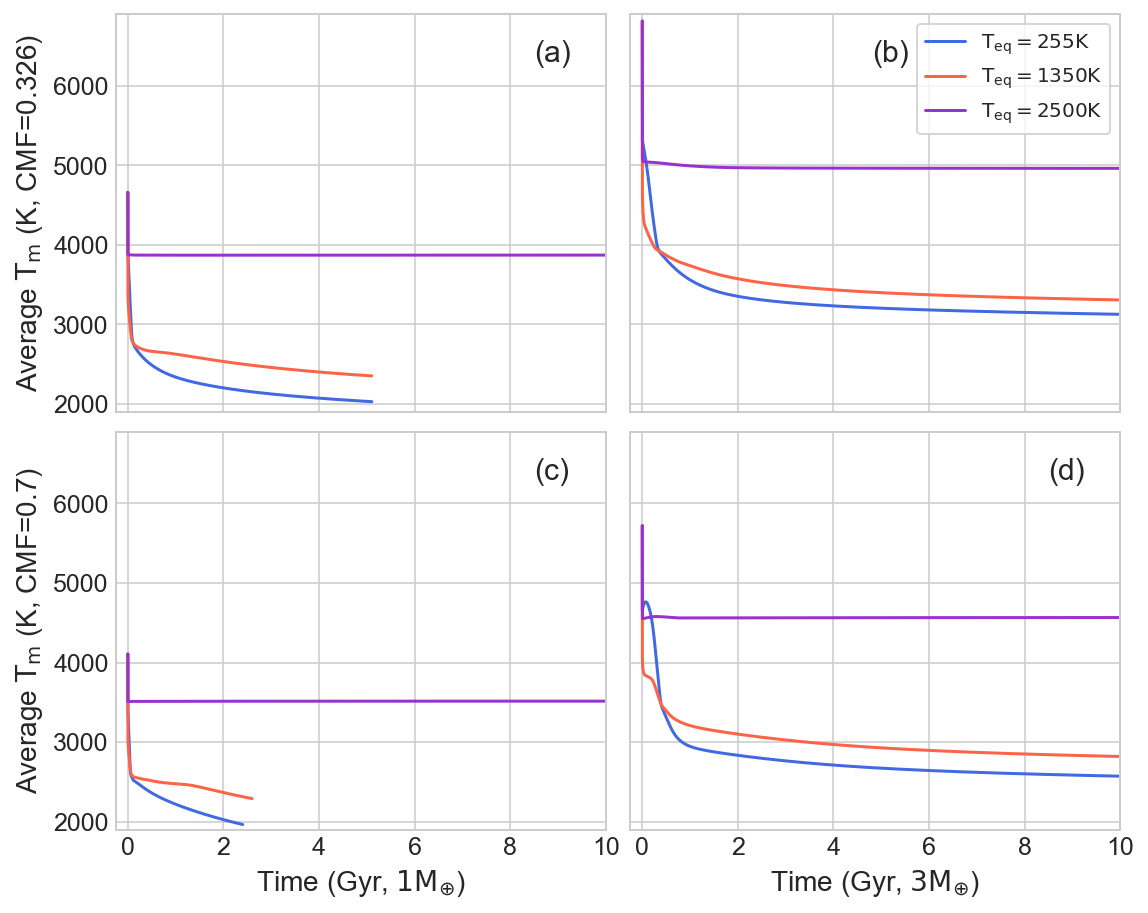}
\end{center}
\caption{Evolution of mass-averaged mantle temperature of planets with 1$M_{\oplus}$ (left column), 3$M_{\oplus}$ (right column), CMF=0.326 (top row) and CMF=0.7 (bottom row). Blue, red and violet curves indicate cases with $T_{eq}$ of 255K, 1350K and 2500K, respectively.  In cases where $T_{eq}$=2500K, the mantle reaches a steady state after $\sim$1000 years and hardly cools in the next $\sim$ 10 Gyrs. \label{fig:Tm}}
\end{figure}

\begin{figure}
\begin{center}
\includegraphics[scale=0.65]{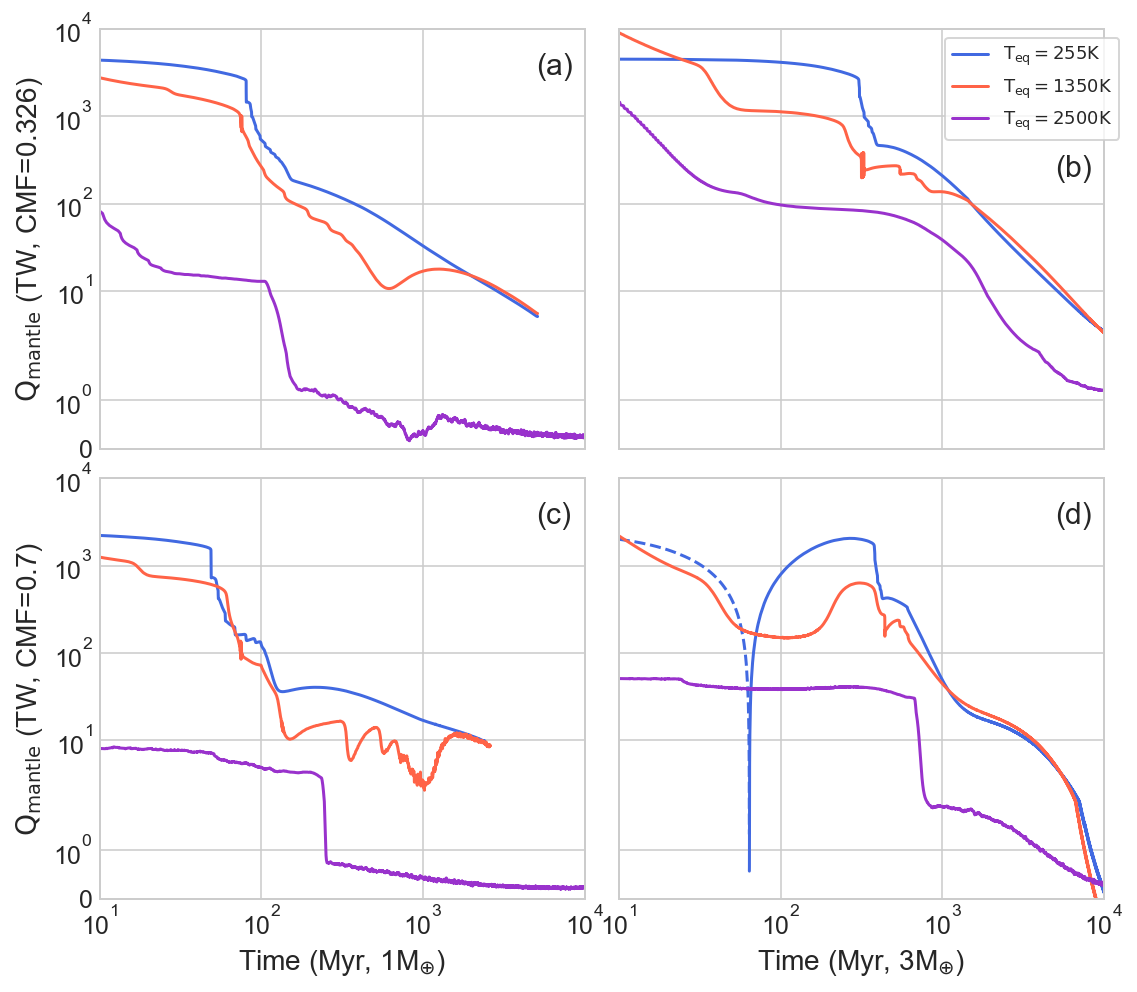}
\end{center}
\caption{Evolution of mantle cooling rate,  $Q_{\mathrm{mantle}}$ (equation~\ref{eqn:mantle_cooling}), with 1$M_{\oplus}$ (left column), 3$M_{\oplus}$ (right column), CMF=0.326 (top row) and CMF=0.7 (bottom row). This plot shows the zoomed in portion between 10~Myr and 10~Gyr to emphasize the behavior of $Q_{\mathrm{mantle}}$ during this stage. Blue, red and violet curves indicate cases with $T_{eq}$ of 255K, 1350K and 2500K, respectively. Dashed curves in panel (d) indicates a negative $Q_{\mathrm{mantle}}$ (i.e. the mantle heats up). In cases where $T_{eq}$=2500K, the mantle reaches a steady state where the $Q_{\mathrm{mantle}}$ approaches a zero level in less than a few Gyrs. The vertical blue curve near 100~Myr in panel (d) indicates the mantle switches from heating up to cooling down.  \label{fig:mantle_cooling_rate}}
\end{figure}

In this section, we discuss how $T_{eq}$ influences the cooling history of the mantle and the core. The evolution of mass-averaged mantle temperature, mantle cooling rate, heat flux at the CMB and solid inner core size is shown in figures~\ref{fig:Tm} to~\ref{fig:Ric}.

The planet equilibrium temperature affects the evolution of the mantle in two different regimes. In cases with $T_{eq}=2500K$, the mantle quickly reaches a steady state where the net cooling rate ($Q_{\text{mantle}})$ approaches zero and the average mantle temperature stays almost constant, as shown by the 2500K evolution track (purple lines) in figures~\ref{fig:Tm} and~\ref{fig:mantle_cooling_rate}. For the 3$M_{\oplus}$ planet with CMF=0.326, which has the most massive mantle out of all planets in our model grid, the mantle reaches the steady state in $\sim$2 Gyrs, while the other 3 combinations of planet mass and CMF reach the steady state in less than 1 Gyr. In comparison, when $T_{eq}=255K$ or 1350K, the net cooling rate of the mantle never reaches the zero level before the iron core fully solidifies, and thus the average mantle temperature continues to decrease until the end of the evolution. 

The evolution history of the iron core (both $F_{\text{CMB}}$ and $R_{\text{ICB}}$) is nearly identical in the low and intermediate $T_{eq}$ cases (255~K and 1350~K), for each of the 4 combinations of planet mass and CMF that we explore. This is reflected in Figures~\ref{fig:F_cmb} and \ref{fig:Ric} where the 255~K evolution tracks (blue lines) are barely visible under the 1350~K tracks (red lines). For example, in the case of $M=1M_{\oplus}$ and CMF=0.326, the onset of inner core solidification occurs at $\sim$0.5 Gyr and the completion of the core solidification occurs at $\sim$5 Gyr in both the $T_{eq}=255$~K and 1350~K cases. Thus, planet equilibrium temperature --- so long as it is below the low-pressure melting temperature of silicates --- has a minimal influence on the thermal history of the iron core and the possibility of a dynamo therein.

Increasing the planet equilibrium temperature to $T_{eq}=2500K$ does significantly impact the thermal history of the iron core. For these cases, the mantles quickly reach the steady state and hardly cools for the remainder of the evolution. The mantles thus act as blankets overlaying and insulating the iron cores. As a result, the cooling rates of iron cores are slower compared to cases with the low and intermediate equilibrium temperatures, as indicated by the 2500~K evolution track (purple lines) in figure~\ref{fig:F_cmb}. The liquid iron cores can survive for more than 10 Gyrs. 

\subsubsection{Mass Dependence of the Thermal Evolution History}
\begin{figure}
\begin{center}
\includegraphics[scale=0.63]{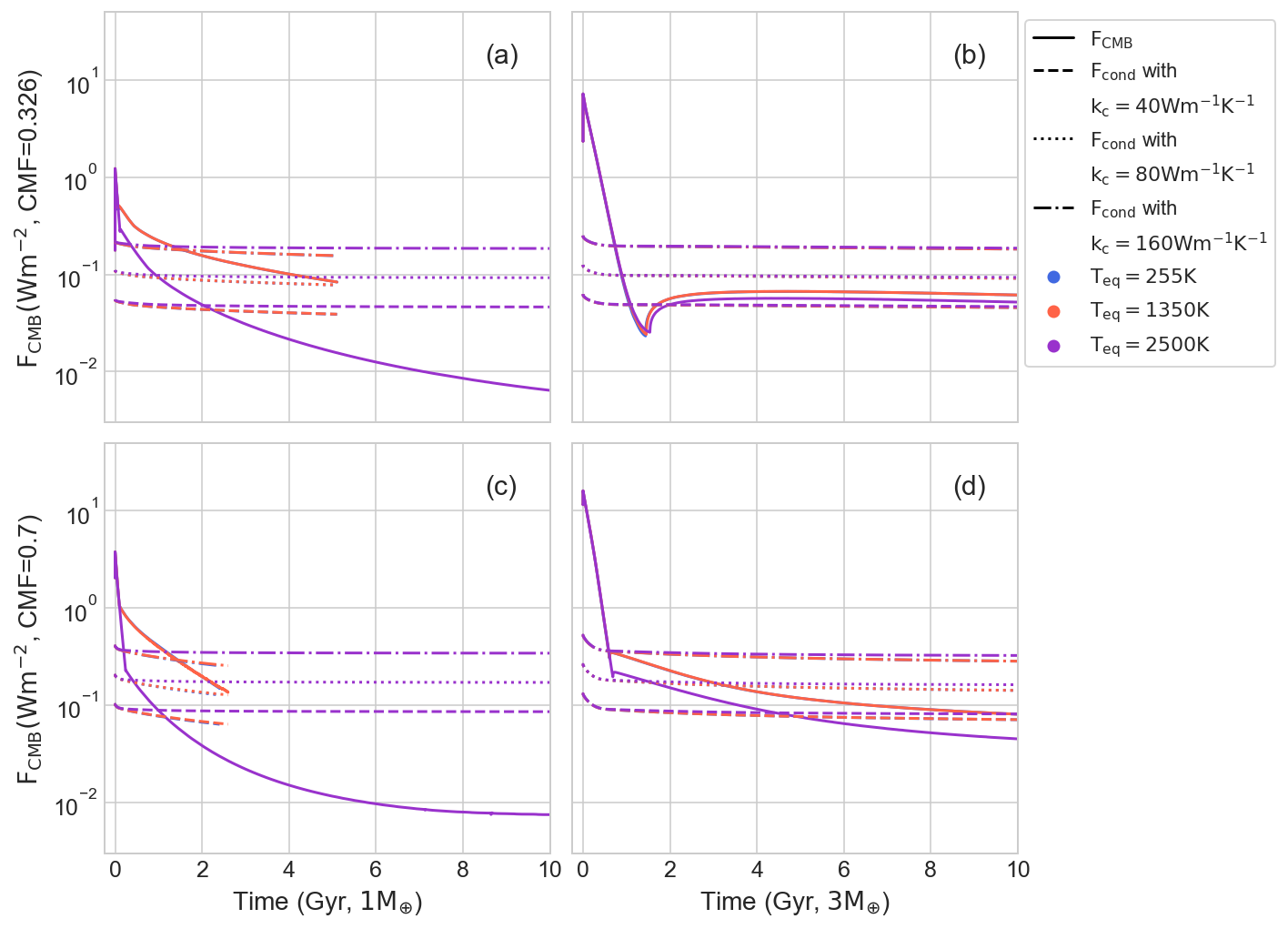}
\end{center}
\caption{Time evolution of heat flux at CMB, $F_{\text{CMB}}$ for planets with 1$M_{\oplus}$ (left column), 3$M_{\oplus}$ (right column), CMF=0.326 (top row) and CMF=0.7 (bottom row). Blue, red and violet curves indicate cases with a $T_{eq}$ of 255K, 1350K, and 2500K. Dashed, dotted and dashdotted curves are the conductive heat flux along the core adiabat calculated with $k_c=\mathrm{40Wm^{-1}K^{-1}}$ (the fiducial choice), $\mathrm{80Wm^{-1}K^{-1}}$ and $\mathrm{160Wm^{-1}K^{-1}}$, which we adopt as the threshold flux for the liquid core to be convecting. 255K (blue curves) and 1350K evolution tracks (red curves) almost overlap each other and $T_{eq}$ has almost no influence on the cooling history of the iron core for cases with the low and intermediate $T_{eq}$. In contrast, increasing $T_{eq}$ to 2500K suppresses the cooling rate of the iron core. 
\label{fig:F_cmb}}
\end{figure}

The thermal history of planets with 1 and 3 $M_{\oplus}$ differ in many aspects. Here, we focus on the influence of planetary mass on the cooling history of the iron core, which has a direct impact on the plausibility of dynamo action in the liquid outer core. 

\begin{figure}
\begin{center}
\includegraphics[scale=0.67]{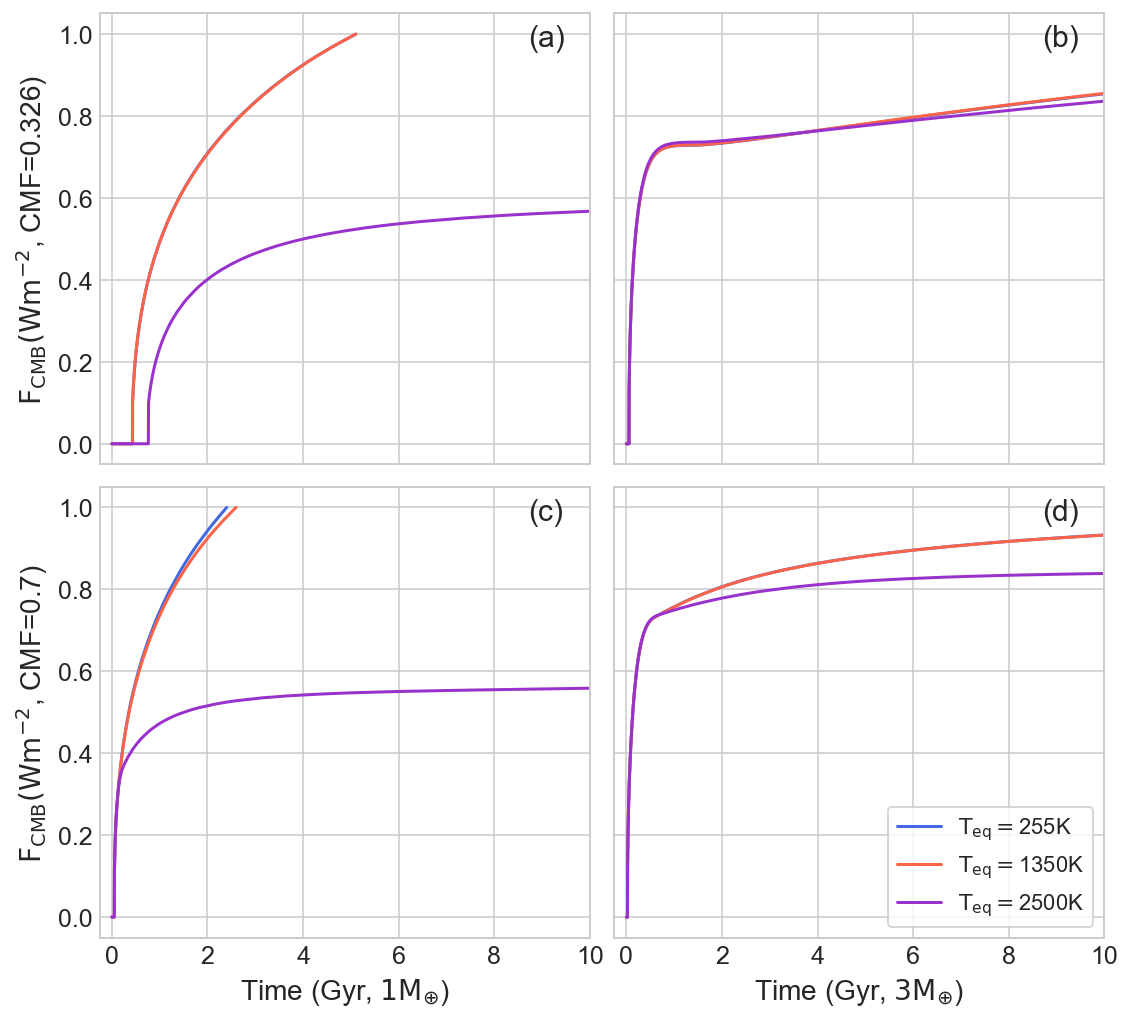}
\end{center}
\caption{Time evolution of solid inner core size for planets with 1$M_{\oplus}$ (left column), 3$M_{\oplus}$ (right column), CMF=0.326 (top row) and CMF=0.7 (bottom row). Blue, red and violet curves indicate cases with a $T_{eq}$ of 255K, 1350K and 2500K. \label{fig:Ric}}
\end{figure}

In the first few hundreds of Myrs after the onset of core solidification, the inner cores of the more massive planets grow more quickly than the inner cores of the lower mass planets, as shown in figure~\ref{fig:Ric}. This is because the centers of more massive planets have greater pressure levels and the melting curve of iron ($dT_{si,m}/dP$) has a smaller gradient at higher pressures. In addition, once the bottom layer in the mantle becomes partially molten, the temperature difference across the CMB layer is greater for more massive planets in our simulations ($\sim$240K for 1$M_{\oplus}$ planets compared to 1880K for 3$M_{\oplus}$ planets). 
This causes $F_{\text{CMB}}$ of more massive planets to peak at a higher level in the first $\sim$Myr, allowing the iron cores to cool faster during this stage. The shape of the melting curve of iron combined with a higher $F_{\text{CMB}}$ in the first $\sim$Myr allow the solid inner core in the $3M_{\oplus}$ planets to grow faster than in the $1M_{\oplus}$ planets. For example, with CMF$=$0.7, $R_{\text{ICB}}/R_c$ grows to $\sim$ 0.6 at $\sim$100 Myrs after the onset of inner core solidification for the planet with $3M_{\oplus}$, but to only 0.3 for the planet with $1M_{\oplus}$ 

Despite the fast initial growth of the inner core, at low and intermediate equilibrium temperatures ($T_{eq}=255K$ and $1350K$) the overall lifetime of the liquid outer core is extended for more massive planets compared to lower mass planets, as figure~\ref{fig:Ric} shows. The reason is that the viscosity at the bottom of the mantle in a more massive planet is greater due to the higher pressure level, resulting in a lower convective heat flow. This causes $F_{\text{CMB}}$ of the more massive 3$M_{\oplus}$ planet to eventually (after~600 Myr) reduce to a lower level than the case of 1$M_{\oplus}$ planet (figure~\ref{fig:F_cmb}). In addition, the iron core of the more massive planet has a greater heat capacity ($\propto M_c$, the core mass), while the core surface area only increases weakly with planetary mass \citep[$\propto M_c^{0.53}$ for planets with Earth-like composition,][]{noack2020}. Thus, it takes longer for the cores of more massive planets to fully solidify. As figure~\ref{fig:Ric} shows, in both the $T_{eq}$= 255K and 1350K cases the core of the 1$M_{\oplus}$ planets with CMF=0.7 and 0.326 fully solidify after $\sim$ 2.5 and 5 Gyrs, respectively. In comparison, for 3$M_{\oplus}$ planets with both CMFs, the liquid outer core can survive more than 10 Gyrs.

For cases with $T_{eq}$=2500K, the effect of high irradiation on the core evolution is reduced with increased planetary mass. This results from a greater separation between the planetary surface and CMB for more massive planets. As figures~\ref{fig:F_cmb} and~\ref{fig:Ric} show, the evolution tracks of high $T_{eq}$ cases (purple lines) are more similar to those of lower $T_{eq}$ cases (red/blue lines) at higher planetary masses. 

\subsubsection{Core Mass Fraction of the Thermal Evolution History}
The influence of CMF on the evolution of the planet falls into two separate regimes depending on whether the planet equilibrium temperature is below or above the low-pressure melting temperature of silicate.

CMF has a major impact on the cooling rate of the iron core on a billion year timescale in cases with low and intermediate $T_{eq}$ (255K or 1350K). For these cases, $F_{\text{CMB}}$ is several times higher for planets with a higher CMF than those with a lower CMF, given the same planetary mass (figure~\ref{fig:F_cmb}). As a result, the solid inner core in a planet with a higher CMF grows faster than in a planet with a lower CMF, despite the fact that the planet with a higher CMF has a more massive iron core. For instance, the core mass of a 1$M_{\oplus}$ planet with CMF=0.7 is around twice as massive as that of a planet with the same mass but a CMF of 0.326, and yet the iron core fully solidifies after only $\sim$ 2.5 Gyrs in the case where CMF=0.7 compared to $\sim$ 5 Gyrs in the cases where CMF=0.326.

\begin{figure}
\begin{center}
\includegraphics[scale=0.6]{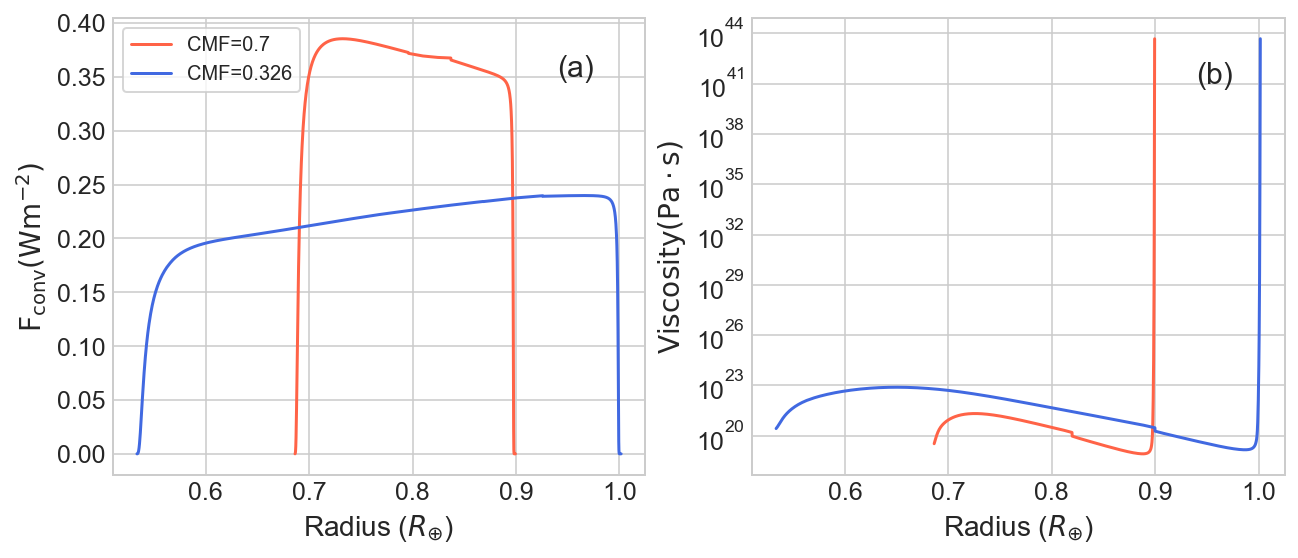}
\end{center}
\caption{The convective heat flow (a) and the viscosity profiles (b) in the mantles of planets with $1M_{\oplus}$, $T_{eq}$=255K, and two values of the CMF at an age of 1 Gyr. The viscosity is several orders of magnitude greater in the lower CMF case compared to the higher CMF case, due to the higher pressures encountered in the mantle. This results in a lower convective heat flux and a suppressed efficiency of heat transport through the mantle in the case of CMF=0.326, compared to CMF=0.7. \label{fig:Fconv_mantle_1ME}}
\end{figure}

The varying cooling rates of the core is a consequence of the blanketing effect of the mantle, namely the ability of the mantle allowing the core to cool. Given the same planet mass, a planet with a greater CMF has a less massive mantle with less blanketing effect on the core, which thus allows a faster growth of the solid inner core. The main contributing factor to the diminished blanketing effect of the less massive mantle is its viscosity level. Figure~\ref{fig:Fconv_mantle_1ME}b shows the radial viscosity profiles of 1$M_{\oplus}$ planets with $T_{eq}$=255K and two CMFs (0.326 and 0.7) at 1 Gyr. The pressure level at the CMB in the low CMF case is 139~GPa whereas it is 76~GPa in the high CMF case. The viscosity level in the mantle of the low CMF case is around 2 orders of magnitude greater due to a higher pressure level in the mantle. The influence of viscosity on the convective heat flow in the mantle dominates over other factors, such as the change in the mixing length, and results in a reduced convective heat low in the mantle in the case of CMF=0.326 (figure~\ref{fig:Fconv_mantle_1ME}a). This depresses the efficiency of convective heat transport through the mantle, and further limits the cooling rate of the iron core. As a result, the liquid iron core persists longer for planets with lower CMFs, when $T_{eq}$ is below the low-pressure melting temperature of silicate. 

When $T_{eq}$ increases to 2500K, the impact of high irradiation on the iron core is lessened for planets with lower CMFs (shown in figures~\ref{fig:F_cmb} and~\ref{fig:Ric}. The effect is similar to mass dependence of evolution of planets with high $T_{eq}$. Compared to cases with CMF=0.7, the mantle of planets with CMF=0.326 acts as a greater barrier between stellar irradiation and the iron core, thus reducing the impact of high irradiation on the iron core. For instance, for 3$M_{\oplus}$ planets with CMF$=0.326$, $R_{\text{ICB}}$ at 10 Gyrs differs by only $\sim$2\% between cases with high and low $T_{eq}$. In comparison, if CMF$=0.7$ and $M_{pl}=3M_{\oplus}$, the difference increases to $\sim$10\% between cases with high and low $T_{eq}$.

\subsubsection{Lifetime of the Dynamo in the Liquid Iron Core}\label{section:dynamo age}

\begin{figure}
\begin{center}
\includegraphics[scale=0.48]{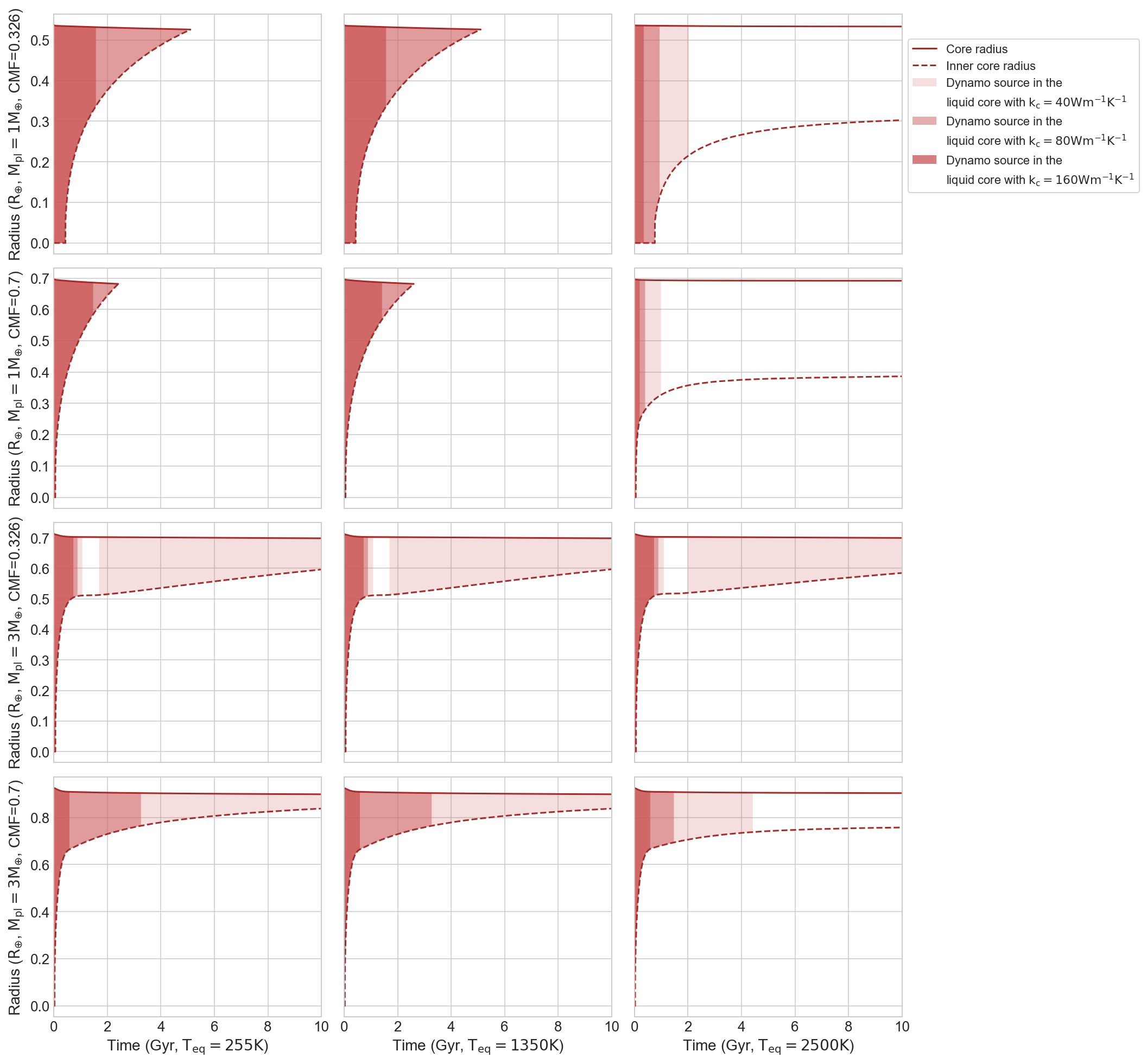}
\end{center}
\caption{Evolution of the potential dynamo source regions in the liquid outer core for all 12 cases. The three columns are for cases with $T_{eq}=$255~K, 1350~K and 2500~K. The four rows are for cases with different combinations of $M_{pl}$ and CMF. The red shaded area indicates the dynamo source region in the liquid iron core (for the fiducial thermal conductivity of liquid iron of 40 $\mathrm{Wm^{-1}K^{-1}}$). Two darker red shadings indicates areas wherein the liquid outer core would still be convecting even if it had higher thermal conductivities of 80 $\mathrm{Wm^{-1}K^{-1}}$ and 160 $\mathrm{Wm^{-1}K^{-1}}$. Solid and dashed lines indicate radii of the iron core and the solid inner core, which are also the upper and lower boundaries of the dynamo source region in the liquid outer core. \label{fig:dynamo_region}}
\end{figure}

The lifetime of the possible dynamo source in the liquid iron core is closely related to the lifetime of thermal convection therein, i.e. the time period for which the liquid iron core exists and stays convective. In cases where the liquid iron core does not fully solidify, so long as the liquid iron core is convecting, its magnetic Reynolds number will exceed $Re_{m,crit}$. This was argued based on order-of-magnitude reasoning in Section~\ref{section:dynamo_criterion} and is born out in our simulations (Figure~\ref{fig:dynamo_region}). On the other hand, in cases where the entire core solidifies before thermal convection stops (i.e., $F_{\text{CMB}}>F_{\text{cond}}$ at the time of complete solidification), the dynamo may shut off shortly before the solidification of the convective liquid core as $L_c$ decreases and reduces $Re_m$. The difference between the lifetimes of the dynamo and of the thermal convection in the liquid core is small. For example, in the baseline model ($M_{pl}=1M_{\oplus}$, CMF=0.326, $T_{eq}=255\mathrm{K}$ and $k_c=40\mathrm{Wm^{-1}K^{-1}}$), the lifetime of the dynamo is 4.96~Gyr while the liquid convecting core lasts for 5.1 Gyrs (a barely perceptible difference on the scale of Figure~\ref{fig:dynamo_region}).

For 1$M_{\oplus}$ planets with $T_{eq}=255K$ or 1350K, the dynamo shuts off because the entire iron core solidifies. As figure~\ref{fig:F_cmb} shows, $F_{\text{CMB}}$ stays above the conductive flux along the core adiabat until the iron core fully solidifies for both CMFs simulated. Thus the dynamo shuts off shortly before the convective core fully solidifies at $\sim$2.5 Gyrs for cases with CMF=0.7 and $\sim$5 Gyrs for cases with CMF=0.326.

The limiting factor on the dynamo lifetime for 3$M_{\oplus}$ planet with $T_{eq}=255K$ or 1350K is the early shut down of convection in the liquid core before it fully solidifies. This is because the cooling rate of the iron core, $F_{\text{CMB}}$, is lower in these planets due to the higher viscosity at the bottom of the mantle compared to the less massive cases. Thus, convection in the liquid core shuts down prior to the full solidification of the iron core. The lifetime of the dynamo is $\sim$10 Gyrs for cases with CMF=0.7, and $\sim$12 Gyrs for cases with CMF=0.326.

The lifetime of the dynamo in the liquid iron core is generally shortened as $T_{eq}$ increases to 2500K. As figure~\ref{fig:F_cmb} shows, $F_{\text{CMB}}$ is reduced with the high $T_{eq}$ compared to the lower $T_{eq}$ cases for all combinations of planet mass and CMF explored. Thus, convection as well as the dynamo in the liquid iron core shuts down prior to the solidification of the entire core in all cases simulated with $T_{eq}=2500K$. The dynamo can only exist for $\sim$ 1 Gyr and 2 Gyrs for 1$M_{\oplus}$ planets with CMF=0.7 and 0.326, and $\sim$4.5 Gyrs and over 10 Gyrs for 3$M_{\oplus}$ planets with the same levels of CMFs. 

The dynamo lifetime depends sensitively on the thermal conductivity of liquid iron, $k_c$, at the temperature and pressure conditions in the iron core. Estimates of $k_c$ at Earth's core condition from first-principles computations and experimental results range from as low as 18$\mathrm{W\, m^{-1}\,K^{-1}}$ \citep[e.g.,][]{stacey2007,kon2016} to over 100 $\mathrm{W\, m^{-1}\,K^{-1}}$ \citep[e.g.,][]{dekoker2012, ohta2016, ZhangY2020}. The large range of reported values for thermal conductivity of iron has large implications for the lifetime of thermal convection in the cores of rocky planets. If $k_c$ is lower than our fiducial choice of 40$\mathrm{W\, m^{-1}\,K^{-1}}$, the conductive heat flux along the core adiabat will be lower than the level shown in figure~\ref{fig:F_cmb}. This would generally extend the lifetime of the dynamo action in the liquid iron core (except in cases where complete core solidification is the limiting factor). On the other hand, if $k_c$ is greater than the fiducial choice, the lifetime of the dynamo action in the liquid iron core would typically be shortened. As shown in figure~\ref{fig:F_cmb}, thermal convection in the liquid core shuts off prior to the solidification of the entire core and the lifetime of the dynamo action in the iron core is shortened in all cases. Further refinements and verification in the direct measurements of thermal conductivity will reduce the uncertainty in the prediction of the lifetime of dynamo-generated magnetic fields for rocky planets. 

\section{Discussion and conclusion}\label{section:conclusion}

\subsection{Overview of Model Limitations}
In this paper, we performed a Gedanken experiment to predict possible thermal evolution tracks of rocky planets consisting of a pure iron core and a pure silicate mantle. While we explored the possible influence of $M_{pl}$, CMF and $T_{eq}$ on exoplanet thermal evolution and dynamo source regions in this work, some of the results may be sensitive to the precise chemical compositions and surface boundary conditions of the planet. We describe three broad caveats in our model here.

\subsubsection{Equilibrium Temperature and Surface Boundary Conditions}
Planets with intermediate and high levels of $T_{eq}$ explored in this paper have ultra-short orbital periods and are expected to likely be tidally locked. These planets have a sub-stellar hemisphere facing their host stars permanently and an anti-stellar hemisphere facing the other way. Our 1D model is inappropriate for tidally locked planets, as differences and heat transport between the sub- and anti-stellar hemispheres need to be accounted for to properly simulate their evolution. \citet{gelman2011} suggests that a magma pond could form on the star-facing side of highly-irradiated tidally locked planets instead of a global magma ocean. A 2D/3D thermal model may determine whether the episodic behavior of magma ocean can occur and to properly calculate the evolution of the localized magma pond on the star-facing side.

We do not consider an atmosphere degassed from magma ocean solidification in simulations of the grid of 12 planets, since the mantle is assumed to be pure silicate. However, the example in the validation case (section~\ref{section:magma_validate}) and previous studies \citep[e.g.][]{abe1997,elkins2008} found that a thin atmosphere may form as the magma ocean solidifies even with an initial water content as low as 100 ppm, and is able to retain enough heat to keep the planetary surface molten for much of solidification of the volume of the magma ocean. Consequently, the solidification of the initial magma ocean may slow down and the subsequent sub-surface magma ocean seen in our simulations may not be able to form. 

While cases with high and intermediate $T_{eq}$ may not represent physical planets, our simulation demonstrates the ability of the code to treat rocky planets with a range of surface temperatures. For instance, low-mass low-density planets could be comprised of metal, silicate and a volatile envelope \citep{chiang2013,lopez2014}, with a range of temperature and pressure conditions at the boundary of rocky interior and envelope. Our code can be applied to calculate energy transport of the rocky interior of such planets, and to simulate the concomitant thermal evolution of the rocky interior and envelope when coupled with a radiative/convective transfer model for the envelope.

\subsubsection{Effect of Impurities on Mantle Dynamics}
By assuming a pure silicate mantle composition (\ce{MgSiO3} and \ce{Mg2SiO4}), we ignore the effect of Fe and trace elements in the mantle on mantle dynamics and magma ocean solidification. 

Our model adopts a melting curve for the mantle because of the assumed pure composition, in contrast to a solidus and liquidus for realistic mantle mixtures. There are two major ways the melting curve could affect the mantle evolution. In our model, the bottom of the mantle starts solidifying, the local temperature is fixed on the melting curve and does not decrease until the local zone becomes fully solid. This causes a discontinuity in the time derivative of $F_{\text{CMB}}$ during the early stage where the magma ocean undergoes solidification (see figure~\ref{fig:F_cmb}). Moreover, the mantle solidus would be lower than the melting temperature of Mg-perovskite because of impurities, resulting in a lengthened lifetime of the fully and partially molten stages of a magma ocean.

As magma ocean solidifies, Fe and trace elements could exsolve into the magma ocean and affect the mantle dynamics. The new solid that forms at the solid/liquid boundary at later times during the solidification has a higher concentration in Fe and the trace elements, which would influence its density. This may lead to a gravitationally unstable density profile with denser material at the top compared to the bottom of the solid mantle. Consequently, cold dense material may sink to the bottom of the mantle and light material may float up. The mass transfer may transport additional heat from the deep interior to the surface in addition to convective heat transfer considered in our model. We need a realistic compositional model coupled with our thermal evolution model to study the feedback between mantle dynamics and chemical compositions and to examine whether the magma ocean solidification regimes mentioned in section~\ref{MO solidification} could still develop. 

\subsubsection{Effect of Impurities on Core Dynamics}\label{section:limitation core impurities}
Light elements, such as sulfur, silicon, and oxygen, are likely to be present in the iron core of a rocky planet \citep[e.g.][]{birch1952,mcdonough2003}, which could lower the core's melting temperature. Previous thermal models of terrestrial planets \citep[e.g.][]{stevenson1983,tachinami2011,stixrude2014} considered a depression in the melting temperature of pure iron to simulate core solidification. Such a depression would prolong the age of the liquid core and potentially the age of the dynamo within. In section~\ref{validation}, we present the evolution results for the baseline model ($M_{pl}=1M_{\oplus}$, CMF=0.326 and $T_{eq}=255\mathrm{K}$) with the melting temperature of the core lowered by 20\% to account for a depression owing to possible impurities. The liquid iron core can survive more than 10 Gyrs and the dynamo shuts off at $\sim$11.5 Gyrs as the liquid core becomes conductive, which is much longer than the age of the liquid iron core and the dynamo region without considering the depression in the melting temperature. 

In addition, when light impurities are present in the core, they could exsolve into the liquid outer part during inner core solidification. The exsolution of light elements into the liquid core results in compositional buoyancy and helps to sustain the dynamo in the outer core. For instance, the present day geodynamo is powered by both thermal and compositional convection with the compositional convection making a greater contribution \citep{lister1995,gubbins2003}. In section~\ref{section:dynamo age}, we demonstrate that thermal convection in the liquid core in all 12 cases explored may shut off earlier if the thermal conductivity of the core is higher than the fiducial choice, $40\mathrm{Wm^{-1}K^{-1}}$. In these scenarios, compositional convection may continue to drive the dynamo if impurities are present in the core and there is ongoing inner core solidification. To refine the prediction on the age of dynamo in cores of rocky planets, a self-consistent treatment of impurities and compositional convection in the evolution model is required. 

\subsection{Summary of Results}

We developed a 1D thermal evolution model coupled with a Henyey solver to study the effect of planetary mass, core mass fraction and equilibrium temperature on the thermal history of and possible dynamo source regions in exoplanets. We extended the modified mixing length theory and applied it to model the convective heat flow within a 2-phase region for the idealized case of a pure substance in the mantle. Figures ~\ref{fig:dynamo_exoplanet} and~\ref{fig:dynamo_region} compile the evolution of dynamo source regions in both the silicate mantle and iron core for all 12 cases explored in this paper. Here, we offer a summary on the thermal evolution and dynamo source region results.
\begin{enumerate}
    \item $T_{eq}$ has a modest effect on the evolution of the core so long as $T_{eq}$ is below the melting temperature at the surface. In all combinations of $M_{pl}$ and CMF, the core evolution is almost identical with low (255~K) and intermediate (1350~K) levels of $T_{eq}$. When $T_{eq}$ exceeds the surface melting temperature, the mantle reaches a steady state quickly in a few hundred to thousand years and slows the growth of the inner core.  
    \item The mantle mass determines the thermal isolating effect of the mantle on the core. A less massive mantle tends to have a lower viscosity and greater convective heat flow owing to the lower mantle pressure. As a result, planets with greater CMFs have a greater core cooling rate, $Q_{\text{core}}$, given the same planet mass. 
    \item More massive planets have a depressed core heat flux, $F_{\text{CMB}}$. The temperature- and pressure-dependent viscosity employed in our model predicts the silicates to be more viscous at the base of the mantle of more massive planets, resulting in lower $F_{\text{CMB}}$. 
    \item More massive planets have longer lived liquid iron cores. $F_{\text{CMB}}$ is depressed for the more massive cases, which leads to a depressed $Q_{\text{core}}$. The total heat capacity of planetary cores increases more strongly with planet mass than core surface area. 
    As a result, the liquid iron core in more massive planets can last longer. 
    \item The lifetime of the dynamo in the liquid iron core is limited by different factors for 1 and 3$M_{\oplus}$ planets. For 1$M_{\oplus}$ planets, the liquid core stays convective until it fully solidifies, which shuts off the dynamo. For 3$M_{\oplus}$ planets, the liquid iron core becomes conductive before the iron core fully solidifies, which shuts off the dynamo.
    \item Varying the thermal conductivity of iron, $k_c$, within its uncertainty range may change the lifetime of the dynamo in the liquid iron core. For example, the lifetime for a 3$M_{\oplus}$ planet with CMF=0.326 changes from over 10~Gyrs to $\sim$1~Gyrs and $\sim$0.8~Gyrs when $k_c$ increases from 40 to 80~$\mathrm{Wm^{-1}K^{-1}}$ and 160~$\mathrm{Wm^{-1}K^{-1}}$. However, varying k$_c$ only affects the dynamo lifetime significantly if convection shuts off before the core solidifies; for 1$M_{\oplus}$ planet with CMF of 0.326 and $T_{eq}$ of 255K, the change in dynamo lifetime is negligible for increases in $k_c$ from our fiducial value up to 70$\mathrm{Wm^{-1}K^{-1}}$ (4.96~Gyrs vs 4.97~Gyrs).
    \item $T_{eq}$ determines the solidification regimes of the magma ocean. Planets with low and intermediate levels of $T_{eq}$ (255~K and 1350~K) may develop a sub-surface magma ocean after the boundary between solid mantle and magma ocean progresses to the surface in a few hundred to thousand years. The entire mantle solidifies in a few $\sim$100 Myrs. If $T_{eq}$ is sufficiently close to the surface melting temperature, the sub-surface magma ocean may experience a episodic stage where its depth widens and shrinks for multiple times before it fully solidifies. A surface magma ocean may persist for the age of the universe if $T_{eq}$ is greater than surface melting temperature.
    \item The dynamo source region in the magma ocean is confined to the region where the melt fraction is above the $x_{crit}$. The liquid/solid assemblages in such regions behave like a liquid and therefore could have a convective velocity great enough to support a dynamo. In comparison, partially molten silicate with low melt fraction is solid-like. The viscous drag force dramatically suppresses the convective velocity and the dynamo shuts off. Melt fraction greater than $x_{crit}$ is a necessary but not sufficient criterion for $Re_m\geq50$ in the magma ocean.
    \item The electrical conductivity of liquid silicate decreases as pressure decreases and can limit the radial extent of the dynamo source region in surface magma oceans. With a low level of electrical conductivity near the planet surface ($\sim10^3\mathrm{Sm^{-1}}$), $Re_m$ at the top of the surface magma ocean eventually falls below the dynamo threshold as the surface heat flux decreases over time. Dynamos in long-lasting surface magma oceans on planets with $T_{eq}$ greater than surface melting temperature may eventually shut off.
\end{enumerate}

Our model is a powerful tool to constrain the thermal structures of planets and their lifetimes of potential dynamos. It has the benefits of employing local values of thermophysical quantities and accounting for heating/cooling due to planet contraction/expansion self-consistently. Though not important for old rocky planets \citep{gubbins2003}, planet contraction/expansion is important for sub-Neptunes. Our Henyey-style code can be extended to properly simulate such planets in the future. The model treats convective heat flow with both high and low levels of viscosity in the mantle. As a result, we observe a non-monotonic behavior in the evolution of $x_m$, which does not occur in models that ignore solid state convection \citep[e.g.][]{elkins2008}. In addition, we calculate the evolution tracks of planets starting from an initial hot state with molten cores and mantles. We find that the liquid iron cores in more massive planets could survive longer than low mass planets, which is in contrast to results by snap-shot models \citep[e.g.][]{val2006, val2007}. 

The thermal and dynamo histories of rocky planets are subtle and complex. In this work, we have presented an initial sparse grid of simulations sampling a single end-member choice of core and mantle chemical composition, and surface boundary condition. Looking forward, the model can be applied to map out the dynamo lifetime as a function of planet mass and CMF. In addition, the model can be readily extended to enable a broader exploration of planet parameter space by including additional EOSs for other bulk compositions, investigating various atmospheric boundary conditions, and considering different convection regimes (e.g., mobile-/sluggish-lid convection). Ultimately, we aim to develop the code presented here into a premier modeling tool to help guide and interpret observational searches for strong planetary-scale magnetic fields on low-mass exoplanets. 

\bibliography{ms}

\begin{thebibliography}{}
\expandafter\ifx\csname natexlab\endcsname\relax\def\natexlab#1{#1}\fi
\providecommand{\url}[1]{\href{#1}{#1}}
\providecommand{\dodoi}[1]{doi:~\href{http://doi.org/#1}{\nolinkurl{#1}}}
\providecommand{\doeprint}[1]{\href{http://ascl.net/#1}{\nolinkurl{http://ascl.net/#1}}}
\providecommand{\doarXiv}[1]{\href{https://arxiv.org/abs/#1}{\nolinkurl{https://arxiv.org/abs/#1}}}

\bibitem[{{Abe}(1995)}]{abe1995}
{Abe}, Y. 1995, Journal of Physics of the Earth, 43, 515

\bibitem[{{Abe}(1997)}]{abe1997}
---. 1997, Physics of the Earth and Planetary Interiors, 100, 27,
  \dodoi{10.1016/S0031-9201(96)03229-3}

\bibitem[{{Adams} {et~al.}(2008){Adams}, {Seager}, \&
  {Elkins-Tanton}}]{adams2008}
{Adams}, E.~R., {Seager}, S., \& {Elkins-Tanton}, L. 2008, \apj, 673, 1160,
  \dodoi{10.1086/524925}

\bibitem[{{Alibay} {et~al.}(2017){Alibay}, {Kasper}, {Lazio}, \&
  {Neilsen}}]{sunrise}
{Alibay}, F., {Kasper}, J.~C., {Lazio}, J., \& {Neilsen}, T.~L. 2017, in AGU
  Fall Meeting Abstracts, Vol. 2017, A41I--2410

\bibitem[{{Al'Tshuler} {et~al.}(1987){Al'Tshuler}, {Brusnikin}, \&
  {Kuz'menkov}}]{altshuler1987}
{Al'Tshuler}, L.~V., {Brusnikin}, S.~E., \& {Kuz'menkov}, E.~A. 1987, Journal
  of Applied Mechanics and Technical Physics, 28, 129,
  \dodoi{10.1007/BF00918785}

\bibitem[{{Anderson} \& {Duba}(1997)}]{anderson1997}
{Anderson}, O.~L., \& {Duba}, A. 1997, \jgr, 102, 22,659,
  \dodoi{10.1029/97JB01641}

\bibitem[{{Andrault} {et~al.}(2011){Andrault}, {Bolfan-Casanova}, {Nigro},
  {Bouhifd}, {Garbarino}, \& {Mezouar}}]{andrault2011}
{Andrault}, D., {Bolfan-Casanova}, N., {Nigro}, G.~L., {et~al.} 2011, Earth and
  Planetary Science Letters, 304, 251, \dodoi{10.1016/j.epsl.2011.02.006}

\bibitem[{{Belonoshko} {et~al.}(2005){Belonoshko}, {Skorodumova}, {Rosengren},
  {Ahuja}, {Johansson}, {Burakovsky}, \& {Preston}}]{belonoshko2005}
{Belonoshko}, A.~B., {Skorodumova}, N.~V., {Rosengren}, A., {et~al.} 2005,
  \prl, 94, 195701, \dodoi{10.1103/PhysRevLett.94.195701}

\bibitem[{{Birch}(1952{\natexlab{a}})}]{bir52}
{Birch}, F. 1952{\natexlab{a}}, \jgr, 57, 227, \dodoi{10.1029/JZ057i002p00227}

\bibitem[{{Birch}(1952{\natexlab{b}})}]{birch1952}
---. 1952{\natexlab{b}}, \jgr, 57, 227, \dodoi{10.1029/JZ057i002p00227}

\bibitem[{{Bonati} {et~al.}(2021){Bonati}, {Lasbleis}, \& {Noack}}]{Bonati2020}
{Bonati}, I., {Lasbleis}, M., \& {Noack}, L. 2021, Journal of Geophysical
  Research (Planets), 126, e06724, \dodoi{10.1029/2020JE006724}

\bibitem[{{Bower} {et~al.}(2019){Bower}, {Kitzmann}, {Wolf}, {Sanan}, {Dorn},
  \& {Oza}}]{bower2019}
{Bower}, D.~J., {Kitzmann}, D., {Wolf}, A.~S., {et~al.} 2019, \aap, 631, A103,
  \dodoi{10.1051/0004-6361/201935710}

\bibitem[{{Bower} {et~al.}(2018){Bower}, {Sanan}, \& {Wolf}}]{bower2018}
{Bower}, D.~J., {Sanan}, P., \& {Wolf}, A.~S. 2018, Physics of the Earth and
  Planetary Interiors, 274, 49, \dodoi{10.1016/j.pepi.2017.11.004}

\bibitem[{{Burke} \& {Franklin}(1955)}]{Bur55}
{Burke}, B.~F., \& {Franklin}, K.~L. 1955, \jgr, 60, 213,
  \dodoi{10.1029/JZ060i002p00213}

\bibitem[{{Chiang} \& {Laughlin}(2013)}]{chiang2013}
{Chiang}, E., \& {Laughlin}, G. 2013, \mnras, 431, 3444,
  \dodoi{10.1093/mnras/stt424}

\bibitem[{{Christensen}(2010)}]{christensen2010}
{Christensen}, U.~R. 2010, \ssr, 152, 565, \dodoi{10.1007/s11214-009-9553-2}

\bibitem[{{Christensen} \& {Aubert}(2006)}]{christensen2006}
{Christensen}, U.~R., \& {Aubert}, J. 2006, Geophysical Journal International,
  166, 97, \dodoi{10.1111/j.1365-246X.2006.03009.x}

\bibitem[{{C{\^o}t{\'e}} {et~al.}(2008){C{\^o}t{\'e}}, {Vo{\v{c}}adlo}, \&
  {Brodholt}}]{cote2008}
{C{\^o}t{\'e}}, A.~S., {Vo{\v{c}}adlo}, L., \& {Brodholt}, J.~P. 2008, \grl,
  35, L05306, \dodoi{10.1029/2007GL032788}

\bibitem[{{Davies} \& {Davies}(2010)}]{davies2010}
{Davies}, J.~H., \& {Davies}, D.~R. 2010, Solid Earth, 1, 5,
  \dodoi{10.5194/se-1-5-2010}

\bibitem[{{de Koker} {et~al.}(2012){de Koker}, {Steinle-Neumann}, \&
  {Vl{\v{c}}ek}}]{dekoker2012}
{de Koker}, N., {Steinle-Neumann}, G., \& {Vl{\v{c}}ek}, V. 2012, Proceedings
  of the National Academy of Science, 109, 4070,
  \dodoi{10.1073/pnas.1111841109}

\bibitem[{{Dewaele} {et~al.}(2006){Dewaele}, {Loubeyre}, {Occelli}, {Mezouar},
  {Dorogokupets}, \& {Torrent}}]{Dewaele2006}
{Dewaele}, A., {Loubeyre}, P., {Occelli}, F., {et~al.} 2006, \prl, 97, 215504,
  \dodoi{10.1103/PhysRevLett.97.215504}

\bibitem[{{Dorogokupets} {et~al.}(2017){Dorogokupets}, {Dymshits}, {Litasov},
  \& {Sokolova}}]{Dorogokupets2017}
{Dorogokupets}, P.~I., {Dymshits}, A.~M., {Litasov}, K.~D., \& {Sokolova},
  T.~S. 2017, Scientific Reports, 7, 41863, \dodoi{10.1038/srep41863}

\bibitem[{{Driscoll} \& {Bercovici}(2014)}]{dri14}
{Driscoll}, P., \& {Bercovici}, D. 2014, Physics of the Earth and Planetary
  Interiors, 236, 36, \dodoi{10.1016/j.pepi.2014.08.004}

\bibitem[{{Dziewonski} \& {Anderson}(1981)}]{prem}
{Dziewonski}, A.~M., \& {Anderson}, D.~L. 1981, Physics of the Earth and
  Planetary Interiors, 25, 297, \dodoi{10.1016/0031-9201(81)90046-7}

\bibitem[{{Elkins-Tanton}(2008)}]{elkins2008}
{Elkins-Tanton}, L.~T. 2008, Earth and Planetary Science Letters, 271, 181,
  \dodoi{10.1016/j.epsl.2008.03.062}

\bibitem[{{Ellingson} {et~al.}(2009){Ellingson}, {Clarke}, {Cohen}, {Craig},
  {Kassim}, {Pihlstrom}, {Rickard}, \& {Taylor}}]{ellingson2009}
{Ellingson}, S.~W., {Clarke}, T.~E., {Cohen}, A., {et~al.} 2009, IEEE
  Proceedings, 97, 1421, \dodoi{10.1109/JPROC.2009.2015683}

\bibitem[{{Fei} {et~al.}(2004){Fei}, {van Orman}, {Li}, {van Westrenen},
  {Sanloup}, {Minarik}, {Hirose}, {Komabayashi}, {Walter}, \&
  {Funakoshi}}]{fei04}
{Fei}, Y., {van Orman}, J., {Li}, J., {et~al.} 2004, Journal of Geophysical
  Research (Solid Earth), 109, B02305, \dodoi{10.1029/2003JB002562}

\bibitem[{{Finlay} {et~al.}(2010){Finlay}, {Maus}, {Beggan}, {Bondar},
  {Chambodut}, {Chernova}, {Chulliat}, {Golovkov}, {Hamilton}, \&
  {Hamoudi}}]{Finlay2010}
{Finlay}, C.~C., {Maus}, S., {Beggan}, C.~D., {et~al.} 2010, Geophysical
  Journal International, 183, 1216, \dodoi{10.1111/j.1365-246X.2010.04804.x}

\bibitem[{{Gelman} {et~al.}(2011){Gelman}, {Elkins-Tanton}, \&
  {Seager}}]{gelman2011}
{Gelman}, S.~E., {Elkins-Tanton}, L.~T., \& {Seager}, S. 2011, \apj, 735, 72,
  \dodoi{10.1088/0004-637X/735/2/72}

\bibitem[{{Gubbins} {et~al.}(2003){Gubbins}, {Alf{\`e}}, {Masters}, {Price}, \&
  {Gillan}}]{gubbins2003}
{Gubbins}, D., {Alf{\`e}}, D., {Masters}, G., {Price}, G.~D., \& {Gillan},
  M.~J. 2003, Geophysical Journal International, 155, 609,
  \dodoi{10.1046/j.1365-246X.2003.02064.x}

\bibitem[{{Gunell} {et~al.}(2018){Gunell}, {Maggiolo}, {Nilsson}, {Stenberg
  Wieser}, {Slapak}, {Lindkvist}, {Hamrin}, \& {De Keyser}}]{gunell2018}
{Gunell}, H., {Maggiolo}, R., {Nilsson}, H., {et~al.} 2018, \aap, 614, L3,
  \dodoi{10.1051/0004-6361/201832934}

\bibitem[{{Hallinan} {et~al.}(2013){Hallinan}, {Sirothia}, {Antonova},
  {Ishwara-Chand ra}, {Bourke}, {Doyle}, {Hartman}, \& {Golden}}]{hallinan2013}
{Hallinan}, G., {Sirothia}, S.~K., {Antonova}, A., {et~al.} 2013, \apj, 762,
  34, \dodoi{10.1088/0004-637X/762/1/34}

\bibitem[{{Henyey} {et~al.}(1959){Henyey}, {Wilets}, {B{\"o}hm}, {Lelevier}, \&
  {Levee}}]{hen59}
{Henyey}, L.~G., {Wilets}, L., {B{\"o}hm}, K.~H., {Lelevier}, R., \& {Levee},
  R.~D. 1959, \apj, 129, 628, \dodoi{10.1086/146661}

\bibitem[{{Hess}(1990)}]{hess1990}
{Hess}, P.~C. 1990, in Lunar and Planetary Science Conference, Vol.~21, Lunar
  and Planetary Science Conference, 501

\bibitem[{{Hirschmann}(2000)}]{hirschmann2000}
{Hirschmann}, M.~M. 2000, Geochemistry, Geophysics, Geosystems, 1, 1042,
  \dodoi{10.1029/2000GC000070}

\bibitem[{{Honda} {et~al.}(1993){Honda}, {Yuen}, {Balachandar}, \&
  {Reuteler}}]{honda1993}
{Honda}, S., {Yuen}, D.~A., {Balachandar}, S., \& {Reuteler}, D. 1993, Science,
  259, 1308, \dodoi{10.1126/science.259.5099.1308}

\bibitem[{{Huang} {et~al.}(2022){Huang}, {Rice}, \& {Steffen}}]{huang2022}
{Huang}, C., {Rice}, D.~R., \& {Steffen}, J.~H. 2022, \mnras, 513, 5256,
  \dodoi{10.1093/mnras/stac1133}

\bibitem[{{Inoue} {et~al.}(2020){Inoue}, {Suehiro}, {Ohta}, {Hirose}, \&
  {Ohishi}}]{inoue2020}
{Inoue}, H., {Suehiro}, S., {Ohta}, K., {Hirose}, K., \& {Ohishi}, Y. 2020,
  Earth and Planetary Science Letters, 543, 116357,
  \dodoi{10.1016/j.epsl.2020.116357}

\bibitem[{{Kao} {et~al.}(2018){Kao}, {Hallinan}, {Pineda}, {Stevenson}, \&
  {Burgasser}}]{kao18}
{Kao}, M.~M., {Hallinan}, G., {Pineda}, J.~S., {Stevenson}, D., \& {Burgasser},
  A. 2018, The Astrophysical Journal Supplement Series, 237, 25,
  \dodoi{10.3847/1538-4365/aac2d5}

\bibitem[{{Katsura} {et~al.}(2009){Katsura}, {Shatskiy}, {Manthilake}, {Zhai},
  {Fukui}, {Yamazaki}, {Matsuzaki}, {Yoneda}, {Ito}, \& {Kuwata}}]{Katsura2009}
{Katsura}, T., {Shatskiy}, A., {Manthilake}, M.~A. G.~M., {et~al.} 2009,
  Physics of the Earth and Planetary Interiors, 174, 86,
  \dodoi{10.1016/j.pepi.2008.08.002}

\bibitem[{{Kivelson} {et~al.}(1996){Kivelson}, {Khurana}, {Russell}, {Walker},
  {Warnecke}, {Coroniti}, {Polanskey}, {Southwood}, \&
  {Schubert}}]{kivelson1996}
{Kivelson}, M.~G., {Khurana}, K.~K., {Russell}, C.~T., {et~al.} 1996, \nat,
  384, 537, \dodoi{10.1038/384537a0}

\bibitem[{{Kon{\^o}pkov{\'a}} {et~al.}(2016){Kon{\^o}pkov{\'a}}, {McWilliams},
  {G{\'o}mez-P{\'e}rez}, \& {Goncharov}}]{kon2016}
{Kon{\^o}pkov{\'a}}, Z., {McWilliams}, R.~S., {G{\'o}mez-P{\'e}rez}, N., \&
  {Goncharov}, A.~F. 2016, \nat, 534, 99, \dodoi{10.1038/nature18009}

\bibitem[{{Labrosse}(2015)}]{lab2015}
{Labrosse}, S. 2015, Physics of the Earth and Planetary Interiors, 247, 36,
  \dodoi{10.1016/j.pepi.2015.02.002}

\bibitem[{{Labrosse} {et~al.}(2007){Labrosse}, {Hernlund}, \&
  {Coltice}}]{labrosse2007}
{Labrosse}, S., {Hernlund}, J.~W., \& {Coltice}, N. 2007, \nat, 450, 866,
  \dodoi{10.1038/nature06355}

\bibitem[{{Lebrun} {et~al.}(2013){Lebrun}, {Massol}, {Chassefi{\`e}Re},
  {Davaille}, {Marcq}, {Sarda}, {Leblanc}, \& {Brandeis}}]{lebrun2013}
{Lebrun}, T., {Massol}, H., {Chassefi{\`e}Re}, E., {et~al.} 2013, Journal of
  Geophysical Research (Planets), 118, 1155, \dodoi{10.1002/jgre.20068}

\bibitem[{{Lister} \& {Buffett}(1995)}]{lister1995}
{Lister}, J.~R., \& {Buffett}, B.~A. 1995, Physics of the Earth and Planetary
  Interiors, 91, 17, \dodoi{10.1016/0031-9201(95)03042-U}

\bibitem[{{Lopez} \& {Fortney}(2014)}]{lopez2014}
{Lopez}, E.~D., \& {Fortney}, J.~J. 2014, \apj, 792, 1,
  \dodoi{10.1088/0004-637X/792/1/1}

\bibitem[{{Lundin} {et~al.}(2007){Lundin}, {Lammer}, \& {Ribas}}]{lundin2007}
{Lundin}, R., {Lammer}, H., \& {Ribas}, I. 2007, \ssr, 129, 245,
  \dodoi{10.1007/s11214-007-9176-4}

\bibitem[{{Mao} {et~al.}(2005){Mao}, {Meng}, {Shen}, {Prakapenka}, {Campbell},
  {Heinz}, {Shu}, {Caracas}, {Cohen}, \& {Fei}}]{andy2005}
{Mao}, W.~L., {Meng}, Y., {Shen}, G., {et~al.} 2005, Proceedings of the
  National Academy of Science, 102, 9751, \dodoi{10.1073/pnas.0503737102}

\bibitem[{{McDonough}(2003)}]{mcdonough2003}
{McDonough}, W.~F. 2003, Treatise on Geochemistry, 2, 568,
  \dodoi{10.1016/B0-08-043751-6/02015-6}

\bibitem[{{McDonough} \& {Sun}(1995)}]{mcd1995}
{McDonough}, W.~F., \& {Sun}, S.~s. 1995, Chemical Geology, 120, 223,
  \dodoi{10.1016/0009-2541(94)00140-4}

\bibitem[{{Monteux} {et~al.}(2011){Monteux}, {Jellinek}, \&
  {Johnson}}]{monteux2011}
{Monteux}, J., {Jellinek}, A.~M., \& {Johnson}, C.~L. 2011, Earth and Planetary
  Science Letters, 310, 349, \dodoi{10.1016/j.epsl.2011.08.014}

\bibitem[{{Mosenfelder} {et~al.}(2009){Mosenfelder}, {Asimow}, {Frost},
  {Rubie}, \& {Ahrens}}]{mosenfelder2009}
{Mosenfelder}, J.~L., {Asimow}, P.~D., {Frost}, D.~J., {Rubie}, D.~C., \&
  {Ahrens}, T.~J. 2009, Journal of Geophysical Research (Solid Earth), 114,
  B01203, \dodoi{10.1029/2008JB005900}

\bibitem[{{Ness} {et~al.}(1986){Ness}, {Acuna}, {Behannon}, {Burlaga},
  {Connerney}, {Lepping}, \& {Neubauer}}]{ness1986}
{Ness}, N.~F., {Acuna}, M.~H., {Behannon}, K.~W., {et~al.} 1986, Science, 233,
  85, \dodoi{10.1126/science.233.4759.85}

\bibitem[{{Ness} {et~al.}(1989){Ness}, {Acuna}, {Burlaga}, {Connerney},
  {Lepping}, \& {Neubauer}}]{ness1989}
{Ness}, N.~F., {Acuna}, M.~H., {Burlaga}, L.~F., {et~al.} 1989, Science, 246,
  1473, \dodoi{10.1126/science.246.4936.1473}

\bibitem[{{Ness} {et~al.}(1975){Ness}, {Behannon}, {Lepping}, \&
  {Whang}}]{Ness75}
{Ness}, N.~F., {Behannon}, K.~W., {Lepping}, R.~P., \& {Whang}, Y.~C. 1975,
  \nat, 255, 204, \dodoi{10.1038/255204a0}

\bibitem[{{Nimmo} {et~al.}(2020){Nimmo}, {Primack}, {Faber}, {Ramirez-Ruiz}, \&
  {Safarzadeh}}]{nimmo2020}
{Nimmo}, F., {Primack}, J., {Faber}, S.~M., {Ramirez-Ruiz}, E., \&
  {Safarzadeh}, M. 2020, \apjl, 903, L37, \dodoi{10.3847/2041-8213/abc251}

\bibitem[{{Noack} \& {Lasbleis}(2020)}]{noack2020}
{Noack}, L., \& {Lasbleis}, M. 2020, \aap, 638, A129,
  \dodoi{10.1051/0004-6361/202037723}

\bibitem[{{Oganov} {et~al.}(2001){Oganov}, {Brodholt}, \& {Price}}]{Oganov2001}
{Oganov}, A.~R., {Brodholt}, J.~P., \& {Price}, G.~D. 2001, \nat, 411, 934,
  \dodoi{10.1038/35082048}

\bibitem[{{Ohta} {et~al.}(2016){Ohta}, {Kuwayama}, {Hirose}, {Shimizu}, \&
  {Ohishi}}]{ohta2016}
{Ohta}, K., {Kuwayama}, Y., {Hirose}, K., {Shimizu}, K., \& {Ohishi}, Y. 2016,
  \nat, 534, 95, \dodoi{10.1038/nature17957}

\bibitem[{{Papuc} \& {Davies}(2008)}]{papuc2008}
{Papuc}, A.~M., \& {Davies}, G.~F. 2008, \icarus, 195, 447,
  \dodoi{10.1016/j.icarus.2007.12.016}

\bibitem[{{Pourovskii} {et~al.}(2020){Pourovskii}, {Mravlje}, {Pozzo}, \&
  {Alf{\`e}}}]{pourovskii2020}
{Pourovskii}, L.~V., {Mravlje}, J., {Pozzo}, M., \& {Alf{\`e}}, D. 2020, Nature
  Communications, 11, 4105, \dodoi{10.1038/s41467-020-18003-9}

\bibitem[{{Pozzo} {et~al.}(2012){Pozzo}, {Davies}, {Gubbins}, \&
  {Alf{\`e}}}]{pozzo2012}
{Pozzo}, M., {Davies}, C., {Gubbins}, D., \& {Alf{\`e}}, D. 2012, \nat, 485,
  355, \dodoi{10.1038/nature11031}

\bibitem[{{Pozzo} {et~al.}(2014){Pozzo}, {Davies}, {Gubbins}, \&
  {Alf{\`e}}}]{pozzo2014}
---. 2014, Earth and Planetary Science Letters, 393, 159,
  \dodoi{10.1016/j.epsl.2014.02.047}

\bibitem[{{Ranalli}(2001)}]{ranalli2001}
{Ranalli}, G. 2001, Journal of Geodynamics, 32, 425,
  \dodoi{10.1016/S0264-3707(01)00042-4}

\bibitem[{{Rogers} \& {Seager}(2010)}]{rogers2010}
{Rogers}, L.~A., \& {Seager}, S. 2010, \apj, 712, 974,
  \dodoi{10.1088/0004-637X/712/2/974}

\bibitem[{{Rosenfeld} \& {Tarazona}(1998)}]{rosenfeld1998}
{Rosenfeld}, Y., \& {Tarazona}, P. 1998, Molecular Physics, 95, 141,
  \dodoi{10.1080/00268979809483145}

\bibitem[{{Sasaki} \& {Nakazawa}(1986{\natexlab{a}})}]{sas86}
{Sasaki}, S., \& {Nakazawa}, K. 1986{\natexlab{a}}, Journal of Geophysical
  Research, 91, 9231, \dodoi{10.1029/JB091iB09p09231}

\bibitem[{{Sasaki} \& {Nakazawa}(1986{\natexlab{b}})}]{sasaki1986}
---. 1986{\natexlab{b}}, \jgr, 91, 9231, \dodoi{10.1029/JB091iB09p09231}

\bibitem[{{Senshu} {et~al.}(2002){Senshu}, {Kuramoto}, \&
  {Matsui}}]{senshu2002}
{Senshu}, H., {Kuramoto}, K., \& {Matsui}, T. 2002, Journal of Geophysical
  Research (Planets), 107, 5118, \dodoi{10.1029/2001JE001819}

\bibitem[{{Smith} {et~al.}(1980){Smith}, {Davis}, {Jones}, {Coleman},
  {Colburn}, {Dyal}, \& {Sonett}}]{smith1980}
{Smith}, E.~J., {Davis}, L., {Jones}, D.~E., {et~al.} 1980, Science, 207, 407,
  \dodoi{10.1126/science.207.4429.407}

\bibitem[{Solomatov(2015)}]{solomatov2015}
Solomatov, V. 2015, in Treatise on Geophysics (Second Edition), second edition
  edn., ed. G.~Schubert (Oxford: Elsevier), 81--104,
  \dodoi{https://doi.org/10.1016/B978-0-444-53802-4.00155-X}

\bibitem[{{Soubiran} \& {Militzer}(2018)}]{soubiran2018}
{Soubiran}, F., \& {Militzer}, B. 2018, Nature Communications, 9, 3883,
  \dodoi{10.1038/s41467-018-06432-6}

\bibitem[{{Spera} {et~al.}(2011){Spera}, {Ghiorso}, \& {Nevins}}]{spera2011}
{Spera}, F.~J., {Ghiorso}, M.~S., \& {Nevins}, D. 2011, \gca, 75, 1272,
  \dodoi{10.1016/j.gca.2010.12.004}

\bibitem[{{Stacey} \& {Davis}(2004)}]{stacey04}
{Stacey}, F.~D., \& {Davis}, P.~M. 2004, Physics of the Earth and Planetary
  Interiors, 142, 137, \dodoi{10.1016/j.pepi.2004.02.003}

\bibitem[{{Stacey} \& {Davis}(2008)}]{stacey08}
---. 2008, {Physics of the Earth}

\bibitem[{{Stacey} \& {Loper}(2007)}]{stacey2007}
{Stacey}, F.~D., \& {Loper}, D.~E. 2007, Physics of the Earth and Planetary
  Interiors, 161, 13, \dodoi{10.1016/j.pepi.2006.12.001}

\bibitem[{{Stamenkovi{\'c}} {et~al.}(2011){Stamenkovi{\'c}}, {Breuer}, \&
  {Spohn}}]{sta11}
{Stamenkovi{\'c}}, V., {Breuer}, D., \& {Spohn}, T. 2011, \icarus, 216, 572,
  \dodoi{10.1016/j.icarus.2011.09.030}

\bibitem[{{Stamenkovi{\'c}} {et~al.}(2012){Stamenkovi{\'c}}, {Noack}, {Breuer},
  \& {Spohn}}]{stamenkovic2012}
{Stamenkovi{\'c}}, V., {Noack}, L., {Breuer}, D., \& {Spohn}, T. 2012, \apj,
  748, 41, \dodoi{10.1088/0004-637X/748/1/41}

\bibitem[{{Stanley} \& {Glatzmaier}(2010)}]{sta10}
{Stanley}, S., \& {Glatzmaier}, G.~A. 2010, \ssr, 152, 617,
  \dodoi{10.1007/s11214-009-9573-y}

\bibitem[{{Stevenson}(2010)}]{stevenson2010}
{Stevenson}, D.~J. 2010, \ssr, 152, 651, \dodoi{10.1007/s11214-009-9572-z}

\bibitem[{{Stevenson} {et~al.}(1983){Stevenson}, {Spohn}, \&
  {Schubert}}]{stevenson1983}
{Stevenson}, D.~J., {Spohn}, T., \& {Schubert}, G. 1983, \icarus, 54, 466,
  \dodoi{10.1016/0019-1035(83)90241-5}

\bibitem[{{Stixrude}(2014)}]{stixrude2014}
{Stixrude}, L. 2014, Philosophical Transactions of the Royal Society of London
  Series A, 372, 20130076, \dodoi{10.1098/rsta.2013.0076}

\bibitem[{{Stixrude} {et~al.}(2020){Stixrude}, {Scipioni}, \&
  {Desjarlais}}]{stixrude2020}
{Stixrude}, L., {Scipioni}, R., \& {Desjarlais}, M.~P. 2020, Nature
  Communications, 11, 935, \dodoi{10.1038/s41467-020-14773-4}

\bibitem[{{Tachinami} {et~al.}(2011){Tachinami}, {Senshu}, \&
  {Ida}}]{tachinami2011}
{Tachinami}, C., {Senshu}, H., \& {Ida}, S. 2011, \apj, 726, 70,
  \dodoi{10.1088/0004-637X/726/2/70}

\bibitem[{{Turner} {et~al.}(2020){Turner}, {Zarka}, {Grie{\ss}meier}, {Lazio},
  {Cecconi}, {Enriquez}, {Girard}, {Jayawardhana}, {Lamy}, {Nichols}, \& {de
  Pater}}]{turner2020}
{Turner}, J.~D., {Zarka}, P., {Grie{\ss}meier}, J.-M., {et~al.} 2020, arXiv
  e-prints, arXiv:2012.07926.
\newblock \doarXiv{2012.07926}

\bibitem[{{Unterborn} \& {Panero}(2019)}]{unterborn2019}
{Unterborn}, C.~T., \& {Panero}, W.~R. 2019, Journal of Geophysical Research
  (Planets), 124, 1704, \dodoi{10.1029/2018JE005844}

\bibitem[{{Valencia} {et~al.}(2006){Valencia}, {O'Connell}, \&
  {Sasselov}}]{val2006}
{Valencia}, D., {O'Connell}, R.~J., \& {Sasselov}, D. 2006, \icarus, 181, 545,
  \dodoi{10.1016/j.icarus.2005.11.021}

\bibitem[{{Valencia} {et~al.}(2007){Valencia}, {Sasselov}, \&
  {O'Connell}}]{val2007}
{Valencia}, D., {Sasselov}, D.~D., \& {O'Connell}, R.~J. 2007, \apj, 656, 545,
  \dodoi{10.1086/509800}

\bibitem[{{van Haarlem} {et~al.}(2013){van Haarlem}, {Wise}, {Gunst}, {Heald},
  {McKean}, {Hessels}, {de Bruyn}, {Nijboer}, {Swinbank}, {Fallows},
  {Brentjens}, {Nelles}, {Beck}, {Falcke}, {Fender}, {H{\"o}randel},
  {Koopmans}, {Mann}, {Miley}, {R{\"o}ttgering}, {Stappers}, {Wijers},
  {Zaroubi}, {van den Akker}, {Alexov}, {Anderson}, {Anderson}, {van Ardenne},
  {Arts}, {Asgekar}, {Avruch}, {Batejat}, {B{\"a}hren}, {Bell}, {Bell}, {van
  Bemmel}, {Bennema}, {Bentum}, {Bernardi}, {Best}, {B{\^\i}rzan}, {Bonafede},
  {Boonstra}, {Braun}, {Bregman}, {Breitling}, {van de Brink}, {Broderick},
  {Broekema}, {Brouw}, {Br{\"u}ggen}, {Butcher}, {van Cappellen}, {Ciardi},
  {Coenen}, {Conway}, {Coolen}, {Corstanje}, {Damstra}, {Davies}, {Deller},
  {Dettmar}, {van Diepen}, {Dijkstra}, {Donker}, {Doorduin}, {Dromer}, {Drost},
  {van Duin}, {Eisl{\"o}ffel}, {van Enst}, {Ferrari}, {Frieswijk}, {Gankema},
  {Garrett}, {de Gasperin}, {Gerbers}, {de Geus}, {Grie{\ss}meier}, {Grit},
  {Gruppen}, {Hamaker}, {Hassall}, {Hoeft}, {Holties}, {Horneffer}, {van der
  Horst}, {van Houwelingen}, {Huijgen}, {Iacobelli}, {Intema}, {Jackson},
  {Jelic}, {de Jong}, {Juette}, {Kant}, {Karastergiou}, {Koers}, {Kollen},
  {Kondratiev}, {Kooistra}, {Koopman}, {Koster}, {Kuniyoshi}, {Kramer},
  {Kuper}, {Lambropoulos}, {Law}, {van Leeuwen}, {Lemaitre}, {Loose}, {Maat},
  {Macario}, {Markoff}, {Masters}, {McFadden}, {McKay-Bukowski}, {Meijering},
  {Meulman}, {Mevius}, {Middelberg}, {Millenaar}, {Miller-Jones}, {Mohan},
  {Mol}, {Morawietz}, {Morganti}, {Mulcahy}, {Mulder}, {Munk}, {Nieuwenhuis},
  {van Nieuwpoort}, {Noordam}, {Norden}, {Noutsos}, {Offringa}, {Olofsson},
  {Omar}, {Orr{\'u}}, {Overeem}, {Paas}, {Pandey-Pommier}, {Pandey}, {Pizzo},
  {Polatidis}, {Rafferty}, {Rawlings}, {Reich}, {de Reijer}, {Reitsma},
  {Renting}, {Riemers}, {Rol}, {Romein}, {Roosjen}, {Ruiter}, {Scaife}, {van
  der Schaaf}, {Scheers}, {Schellart}, {Schoenmakers}, {Schoonderbeek},
  {Serylak}, {Shulevski}, {Sluman}, {Smirnov}, {Sobey}, {Spreeuw}, {Steinmetz},
  {Sterks}, {Stiepel}, {Stuurwold}, {Tagger}, {Tang}, {Tasse}, {Thomas},
  {Thoudam}, {Toribio}, {van der Tol}, {Usov}, {van Veelen}, {van der Veen},
  {ter Veen}, {Verbiest}, {Vermeulen}, {Vermaas}, {Vocks}, {Vogt}, {de Vos},
  {van der Wal}, {van Weeren}, {Weggemans}, {Weltevrede}, {White}, {Wijnholds},
  {Wilhelmsson}, {Wucknitz}, {Yatawatta}, {Zarka}, {Zensus}, \& {van
  Zwieten}}]{haarlem2013}
{van Haarlem}, M.~P., {Wise}, M.~W., {Gunst}, A.~W., {et~al.} 2013, \aap, 556,
  A2, \dodoi{10.1051/0004-6361/201220873}

\bibitem[{{Vinet} {et~al.}(1989){Vinet}, {Rose}, {Ferrante}, \&
  {Smith}}]{vinet1989}
{Vinet}, P., {Rose}, J.~H., {Ferrante}, J., \& {Smith}, J.~R. 1989, Journal of
  Physics Condensed Matter, 1, 1941, \dodoi{10.1088/0953-8984/1/11/002}

\bibitem[{{Vitense}(1953)}]{vitense53}
{Vitense}, E. 1953, \zap, 32, 135

\bibitem[{{Wagner} {et~al.}(2019){Wagner}, {Plesa}, \& {Rozel}}]{wagner2019}
{Wagner}, F.~W., {Plesa}, A.~C., \& {Rozel}, A.~B. 2019, Geophysical Journal
  International, 217, 75, \dodoi{10.1093/gji/ggy543}

\bibitem[{{Winterhalter} {et~al.}(2006){Winterhalter}, {Kuiper}, {Majid},
  {Chandra}, {Lazio}, {Zarka}, {Naudet}, {Bryden}, {Gonzalez}, \&
  {Treumann}}]{winter2006}
{Winterhalter}, D., {Kuiper}, T., {Majid}, W., {et~al.} 2006, in Planetary
  Radio Emissions VI, 595--602

\bibitem[{{Wolf} \& {Bower}(2018)}]{wolf2018}
{Wolf}, A.~S., \& {Bower}, D.~J. 2018, Physics of the Earth and Planetary
  Interiors, 278, 59, \dodoi{10.1016/j.pepi.2018.02.004}

\bibitem[{{Yukutake}(2000)}]{yukutake2000}
{Yukutake}, T. 2000, Physics of the Earth and Planetary Interiors, 121, 103,
  \dodoi{10.1016/S0031-9201(00)00163-1}

\bibitem[{{Zarka}(2007)}]{zarka2007}
{Zarka}, P. 2007, \planss, 55, 598, \dodoi{10.1016/j.pss.2006.05.045}

\bibitem[{Zhang {et~al.}(2015)Zhang, Liu, Liu, \& Cai}]{zhang2015}
Zhang, W.-J., Liu, Z.-Y., Liu, Z.-L., \& Cai, L.-C. 2015, Physics of the Earth
  and Planetary Interiors, 244, 69 ,
  \dodoi{https://doi.org/10.1016/j.pepi.2014.10.011}

\bibitem[{Zhang {et~al.}(2020)Zhang, Hou, Liu, Zhang, Prakapenka, Greenberg,
  Fei, Cohen, \& Lin}]{ZhangY2020}
Zhang, Y., Hou, M., Liu, G., {et~al.} 2020, Phys. Rev. Lett., 125, 078501,
  \dodoi{10.1103/PhysRevLett.125.078501}

\bibitem[{Zhang {et~al.}(2022)Zhang, Luo, Hou, Driscoll, Salke, Minár,
  Prakapenka, Greenberg, Hemley, Cohen, \& Lin}]{zhangy2021}
Zhang, Y., Luo, K., Hou, M., {et~al.} 2022, Proceedings of the National Academy
  of Sciences, 119, e2119001119, \dodoi{10.1073/pnas.2119001119}

\end{thebibliography}
\bibliographystyle{aasjournal}

\appendix
We give a brief overview on the modified mixing length formulation, which provides an estimation on the convective heat flow within the planetary interior. The original formulation \citep{abe1995} was derived for thermal convection driven by thermal profiles with super-adiabatic temperature gradients. The formula for $F_{\text{conv}}$ for viscous fluid and inviscid fluid are given by \citet{sasaki1986} and \citet{vitense53}. Here, we derive the same formula in terms of the entropy gradient, since entropy is a natural coordinate for convecting systems of both pure solid/liquid and partially molten aggregates. We further develop similar formulae for convection driven by super-adiabatic melt fraction gradients.

\section{Fundamentals of the modified mixing length formulation}\label{section:mlt}
The general idea of the mixing length theory is that a thermal fluid parcel in the planetary interior can move in between regions with high and low heat content due to a buoyancy force. The thermal parcel can move for a characteristic distance of $l$ before it merges with the surroundings. For thermal convection, the buoyancy force is generated by a temperature difference between the fluid parcel and the surroundings, and the heat flux can be written as 
\begin{equation}\label{mlt}
    F_{\text{conv}}=\frac{1}{2}\rho c_P v \Delta T,
\end{equation}
where $\rho$ is the density, $c_P$ is the specific heat capacity per unit mass, $v$ is the fluid velocity. $\Delta T$ is the temperature difference between the fluid parcel and the surroundings, which is generated as the fluid parcel moves for $l$, and is estimated as $\Delta T=l[(\partial T/\partial r)_s-\partial T/\partial r]$, where $(\partial T/\partial r)_s$ is the adiabatic temperature gradient. The numerical factor 1/2 comes from the fact that about half of the matter rises and the other half descends at any location with convective instability. 

In cases where the viscous drag force is significant, the fluid velocity is constrained by the Stoke's flow, and is given by the Stoke's velocity,

\begin{equation}\label{stokes_velocity}
    v=\frac{2\Delta \rho l^2g}{9\rho\nu},
\end{equation}
where $g$ is the local gravitational acceleration, $\nu$ is the kinematic viscosity, and $\Delta\rho$ is the density difference between the surrounding fluid and the fluid parcel, and approximated as $\Delta \rho=\rho\alpha\Delta T$. Convective instability occurs when $\Delta\rho$ is positive. Plugging equation~\ref{stokes_velocity} into equation~\ref{mlt}, and adding an additional numerical factor of 1/2 to  match the original formulae provided in \citet{sasaki1986} and \citet{abe1995}, we obtain

\begin{equation}\label{mlt_T_high_nu}
    F_{\text{conv}}=\frac{\rho c_P\alpha gl^4}{18\nu}\left(\left.\frac{\partial T}{\partial r}\right\vert_s-\frac{\partial T}{\partial r}\right)^2.
\end{equation}
We note that the numerical factor of 1/2 is inconsequential as equation~\ref{mlt_T_high_nu} is calibrated against a 1-D boundary layer theory model for rocky planets to obtain the proper mixing length \citep{tachinami2011} (but nonetheless should be included to ensure proper calibration). The effective eddy diffusivity, $\kappa_h\sim vl$, is obtained from equation~\ref{eq:Fconv_kh} as 

\begin{equation}\label{kappa_T_high_nu}
    \kappa_h=\frac{\alpha  gl^4}{18\nu}\left(\left.\frac{\partial T}{\partial r}\right|_s-\frac{\partial T}{\partial r}\right).
\end{equation}
Both temperature gradients are defined to be negative. Convective instability does not occur when $(\partial T/\partial r)_s\leq\partial T/\partial r$, in which case $F_{\text{conv}}$ and $\kappa_h$ are both 0. 

Now we want to express the temperature gradients in terms of the entropy gradient, $\partial s/\partial r$. $\partial T/\partial r$ can be expressed as 

\begin{equation}\label{eqn:dTdr differnce}
    \begin{split}
        \frac{\partial T}{\partial r}&=\left.\frac{\partial T}{\partial P}\right\vert_s\frac{\partial P}{\partial r}+\left.\frac{\partial T}{\partial s}\right\vert_P\frac{\partial s}{\partial r}\\
        &=\left.\frac{\partial T}{\partial r}\right\vert_s+\frac{T}{c_P}\frac{\partial s}{\partial r}.
    \end{split}
\end{equation}
Now we can rewrite $\Delta\rho$, $\Delta T$, $F_{\text{conv}}$ and $\kappa_h$ in terms of the entropy gradient, 

\begin{equation}
    \Delta T=-l\frac{T}{c_P}\frac{\partial s}{\partial r},
\end{equation}
\begin{equation}\label{eqn:delta rho}
    \Delta\rho=-\rho\alpha l\frac{T}{c_P}\frac{\partial s}{\partial r},
\end{equation}
\begin{equation}\label{eqn:F_conv}
    F_{\text{conv}}=\frac{\rho\alpha gT^2l^4}{18\nu c_P}\left(\frac{\partial s}{\partial r}\right)^2,
\end{equation}
\begin{equation}\label{eqn:kappa_h}
    \kappa_h=-\frac{\alpha gTl^4}{18c_P\nu}\frac{\partial s}{\partial r}.
\end{equation}
Convection occurs when the entropy gradient is negative, and $F_{\text{conv}}$ and $\kappa_h$ are calculated using equations~\ref{eqn:F_conv} and~\ref{eqn:kappa_h}. Otherwise, both $F_{\text{conv}}$ and $\kappa_h$ are 0. 

In cases where viscosity is low and viscous drag force is insignificant, then the flow velocity is the free fall velocity estimated based on the energy exchange between the gravitational potential energy and kinematic energy of the fluid parcel. The sum of the buoyancy force and gravity on the fluid parcel per unit volume is $f=g\Delta\rho$. The fluid parcel can rise (i.e. convective instability occurs) when $f$ is positive. The work done on the fluid parcel per unit volume to move it through a distance $l$ is 

\begin{equation}
    W(l)=\int^{l}_0 g\Delta\rho dl'=g\int^{l}_0\Delta\rho dl'=\frac{1}{2}g\Delta\rho l.
\end{equation}
Assuming $W(l)$ could all be transformed into kinetic energy of the parcel, $E_k=\rho v^2/2$, then the fluid velocity is 

\begin{equation}\label{free fall velocity}
    v=\left(\frac{g\Delta\rho l}{\rho}\right)^{0.5},
\end{equation}
and $\Delta \rho$ is the same density difference between the fluid parcel and the surrounding fluid as in equation~\ref{stokes_velocity}. Plugging this velocity into equation~\ref{mlt} and including the numerical factor of 1/2 to match the formula in \citet{vitense53} and \cite{abe1995}, we have 

\begin{equation}\label{mlt_T_low_nu}
    F_{\text{conv}}=\rho c_P\sqrt{\frac{\alpha gl^4}{16}\left(\left.\frac{\partial T}{\partial r}\right|_s-\frac{\partial T}{\partial r}\right)^3},
\end{equation}
and 

\begin{equation}\label{kappa_T_low_nu}
    \kappa_h=\sqrt{\frac{\alpha gl^4}{16}\left(\left.\frac{\partial T}{\partial r}\right|_s-\frac{\partial T}{\partial r}\right)}.
\end{equation}
We now use equation~\ref{eqn:dTdr differnce} to rewrite equations~\ref{mlt_T_low_nu} and~\ref{kappa_T_low_nu} in terms of the entropy gradient instead of the temperature gradient, we have 

\begin{equation}
    F_{\text{conv}}=\rho T\sqrt{-\frac{\alpha gTl^4}{16c_P}\left(\frac{\partial s}{\partial r}\right)^3},
\end{equation}
and 

\begin{equation}
    \kappa_h=\sqrt{-\frac{\alpha gTl^4}{16c_P}\frac{\partial s}{\partial r}}.
\end{equation}
The convection criterion for inviscid fluid remains the same as that for viscous fluid. 

The transition criterion between the low and high viscosity regimes is determined based on Stokes velocity and the free fall velocity of the fluid parcel. The convective heat flux and $\kappa_h$ are limited by the either one of the slower velocity. We determine the transition by equating equations~\ref{stokes_velocity} and~\ref{free fall velocity}, 

\begin{equation}\label{criterion}
    \frac{\Delta\rho gl^3}{18\rho\nu^2}=\frac{9}{8}.
\end{equation}
Plugging equation~\ref{eqn:delta rho} into equation~\ref{criterion}, we obtain the transition criterion

\begin{equation}\label{criterion_s}
    -\frac{\alpha gTl^4}{18c_P\nu^2}\frac{\partial s}{\partial r}=\frac{9}{8}.
\end{equation}
When the quantity on the left side of equation~\ref{criterion_s} is greater than 9/8, the convective heat flow is then in the low viscosity regime, otherwise the flow is in the high viscosity regime. 

\section{Modified mixing length formulation for solid and liquid mixtures} \label{section:mlt_x}
In this section, we aim to extend the modified mixing length formulation described in appendix~\ref{section:mlt} and develop a formula to estimate the convective heat flow within a mixture of solid and liquid phases. Herein, we consider the case where the two phases are well mixed and do not separate (no rain-out); the solid condensate (fluid melt) is completely entrained with the fluid (solid). For a pure substance, the temperature of a solid-liquid mixture is fixed on the melting curve. The density of the solid and liquid mixture is 

\begin{equation}
    \rho=\left(\frac{x}{\rho_m}+\frac{1-x}{\rho_s}\right)^{-1},
\end{equation}
where $\rho_m$ and $\rho_s$ are the densities of the liquid and solid phases on the melting curve.

The buoyancy force that drives the convection is generated by a difference in the melt fraction between the adiabatically displaced fluid parcel and the surroundings. The fluid parcel carries latent heat with it as it moves up and down due to the convection, and the convective heat flux is 

\begin{equation}\label{mlt_x}
    F_{\text{conv}}=\frac{1}{2}\rho L v \Delta x,
\end{equation}
where $L$ is the latent heat of fusion and $\Delta x$ is the difference in melt fraction between the fluid parcel and the surrounding fluid. Similar to $\Delta T$ in thermal convection, $\Delta x$ can be approximated as $\Delta x=l\left[(\partial x/\partial r)_s-\partial x/\partial r\right]$, with $(\partial x/\partial r)_s$ being the adiabatic melt fraction gradient. 

In the case where viscous drag force is significant, the flow velocity is given by equation~\ref{stokes_velocity}. Including the factor of 1/2 the same way as equation~\ref{mlt_T_high_nu}, the convective heat flux is 

\begin{equation}\label{mlt_dxdr}
    F_{\text{conv}}=\frac{L\Delta\rho l^3g}{18\nu}\left(\left.\frac{\partial x}{\partial r}\right\vert_s-\frac{\partial x}{\partial r}\right),
\end{equation}
where $\Delta \rho$ is the density difference between the surrounding material and the fluid parcel. In analogy to the thermal expansion coefficient, we define an expansion coefficient due to changes in melt fraction, $\alpha_x$,

\begin{equation}
    \alpha_x \equiv -\frac{1}{\rho}\left.\frac{\partial \rho}{\partial x}\right\vert_P = \rho\left(\frac{1}{\rho_m}-\frac{1}{\rho_s}\right),
\end{equation}
so that $\Delta \rho = \rho \alpha_x \Delta x$. 
Substituting this expression back into equation~\ref{mlt_dxdr}, we have

\begin{equation}\label{mlt_x_high_nu}
    F_{\text{conv}}=\frac{\rho\alpha_x Lg l^4}{18\nu}\left(\left.\frac{\partial x}{\partial r}\right\vert_s-\frac{\partial x}{\partial r}\right)^2,
\end{equation}
and $\kappa_h$ is obtained using equation~\ref{eq:Fconv_kh} 

\begin{equation}\label{kappa_x_high_nu}
    \kappa_h=\frac{\alpha_x g l^4}{18\nu}\left(\left.\frac{\partial x}{\partial r}\right\vert_s-\frac{\partial x}{\partial r}\right).
\end{equation}

The next goal is to rewrite the melt fraction gradient in terms of the entropy gradient. The specific entropy of a solid and liquid mixture is 

\begin{equation}
    s=xs_m+(1-x)s_s,
\end{equation}
where $s_m$ and $s_s$ are the specific entropy of the liquid and solid components on the melting curve. We expand $\partial s/\partial r$ and express it as 

\begin{equation}\label{dsdr_full}
    \frac{\partial s}{\partial r}=\left[x\left(\frac{\mathrm{d} s_m}{\mathrm{d} P}-\frac{\mathrm{d} s_s}{\mathrm{d} P}\right)+\frac{\mathrm{d} s_s}{\mathrm{d} P}\right]\frac{\partial P}{\partial r}+(s_m-s_s)\frac{\partial x}{\partial r}.
\end{equation}
Rearranging the equation, we obtain

\begin{equation}
    \frac{\partial x}{\partial r}=-\frac{1}{s_m-s_s}\left[x\left(\frac{\mathrm{d} s_m}{\mathrm{d} P}-\frac{\mathrm{d} s_s}{\mathrm{d} P}\right)+\frac{\mathrm{d} s_s}{\mathrm{d} P}\right]\frac{\partial P}{\partial r}+\frac{1}{s_m-s_s}\frac{\partial s}{\partial r}.
\end{equation}
The first term on the right side of the equation is the adiabatic melt fraction gradient and the second term is the super-adiabatic component. Given that $s_m-s_s=L/T_m$ with $T_m$ being the melting temperature, we have 

\begin{equation}
    \left.\frac{\partial x}{\partial r}\right|_s-\frac{\partial x}{\partial r}=-\frac{T_m}{L}\frac{\partial s}{\partial r}.
\end{equation}
Putting this into equations~\ref{mlt_x_high_nu} and~\ref{kappa_x_high_nu}, we have 

\begin{equation}
    F_{\text{conv}}=\frac{\rho\alpha_x g T_m^2l^4}{18L\nu}\left(\frac{\partial s}{\partial r}\right)^2,
\end{equation}
and 

\begin{equation}
    \kappa_h=-\frac{\alpha_xg T_ml^4}{18 L\nu}\left(\frac{\partial s}{\partial r}\right).
\end{equation}
The convective instability can occur when the entropy gradient is negative. 

In the case where the viscous drag force is insignificant and the convective heat flow is limited by the free fall velocity (equation~\ref{free fall velocity}). Again, with the numerical factor 1/2, the convective heat flow and $\kappa_h$ are 

\begin{equation}
    F_{\text{conv}}=\rho T_m\sqrt{-\frac{\alpha_xgT_ml^4}{16 L}\left(\frac{\partial s}{\partial r}\right)^3},
\end{equation}
and 

\begin{equation}
    \kappa_h=\sqrt{-\frac{\alpha_xgT_ml^4}{16 L}\frac{\partial s}{\partial r}}.
\end{equation}
The transition criterion between the low and high viscosity regimes for convection driven by a super-adiabatic melf fraction gradient is still given by equation~\ref{criterion}. Substituting the expression for $\Delta \rho$ due to melt fraction differences between the parcel and surroundings into equation~\ref{criterion}, we have

\begin{equation}\label{criterion_s_x}
    -\frac{\alpha_xgT_ml^4}{18 L \nu^2}\frac{\partial s}{\partial r}=\frac{9}{8}.
\end{equation}
When the quantity on the left side of equation~\ref{criterion_s_x} is greater than 9/8, the convective heat flow is then in the low viscosity regime, otherwise the flow is in the high viscosity regime. 

\end{document}